\documentclass[1p,number,times,sort&compress]{elsarticle}
\usepackage[usenames]{color} 
\usepackage[normalem]{ulem} 
\usepackage{graphicx}
\usepackage{subfigure}
\usepackage{amssymb}
\usepackage{amsmath}
\usepackage{epstopdf}
\usepackage{float}
\usepackage{booktabs} 
\usepackage{microtype}
\usepackage[font=footnotesize]{caption}
\usepackage[colorlinks,breaklinks]{hyperref}
\newcommand{\Rn}{\mathbb{R}^n}

\newcommand{\Lee}[1]{\textcolor{blue}{#1}}

\begin{document} 


\title{Constructing Reference Metrics on Multicube\\
  Representations of Arbitrary Manifolds}

\author[caltech,ucsd,MSC]{Lee Lindblom} 
\author[caltech]{Nicholas W.~Taylor}
\author[aeigolm,FreiUniv]{Oliver Rinne}

\address[caltech]{Theoretical Astrophysics 350-17, California
  Institute of Technology, Pasadena, CA 91125, USA} 
\address[ucsd]{Center
  for Astrophysics and Space Sciences, University of California at San
  Diego,\\ 9500 Gilman Drive, La Jolla, CA 92093, USA}
\address[MSC]{Mathematical Sciences Center, Tsinghua University,
Beijing 100084, China}
\address[aeigolm]{Max Planck Institute for Gravitational Physics
  (Albert Einstein Institute), Am M\"uhlenberg 1, 14476 Potsdam,
  Germany} 
\address[FreiUniv]{Department of Mathematics and Computer
  Science, Freie Universit\"at Berlin, Arnimallee 2-6, 14195 Berlin,
  Germany}

\date{\today}
 
\begin{abstract}
Reference metrics are used to define the differential structure on
multicube representations of manifolds, i.e., they provide a simple
and practical way to define what it means globally for tensor fields
and their derivatives to be continuous.  This paper introduces a
general procedure for constructing reference metrics automatically on
multicube representations of manifolds with arbitrary topologies.  The
method is tested here by constructing reference metrics for compact,
orientable two-dimensional manifolds with genera between zero and
five.  These metrics are shown to satisfy the Gauss-Bonnet identity
numerically to the level of truncation error (which converges toward
zero as the numerical resolution is increased).  These reference
metrics can be made smoother and more uniform by evolving them with
Ricci flow.  This smoothing procedure is tested on the two-dimensional
reference metrics constructed here.  These smoothing evolutions (using
volume-normalized Ricci flow with DeTurck gauge fixing) are all shown
to produce reference metrics with constant scalar curvatures (at the
level of numerical truncation error).
\end{abstract}

\begin{keyword}
 topological manifolds \sep differential structure \sep numerical
   methods \sep Ricci flow
\end{keyword} 

\maketitle

\section{Introduction}
\label{s:Introduction}

The problem of developing methods for solving partial differential
equations numerically on manifolds with nontrivial topologies has been
studied in recent years by a number of researchers.  The most widely
studied approach, the surface finite element method, was developed
originally by Gerhard Dziuk and collaborators~\cite{Dziuk1988,
  Dziuk1991, DeckelnickEtAl2005, DziukElliott2007, DemlowDziuk2007}.
This method can be applied to manifolds having isometric embeddings as
codimension one surfaces in $\Rn$.  Triangular (or higher dimensional
simplex) meshes on these surfaces are used to define discrete
differential operators using fairly standard finite element methods.
The topological structures of these manifolds are encoded in the
simplicial meshes, while their differential structures and geometries
are inherited by projection from the enveloping Euclidean $\Rn$.  The
surface finite element method has been used in a number of
applications on surfaces, including various evolving surface
problems~\cite{Demlow2009, DziukElliott2012} and harmonic map flows on
surfaces with nontrivial topologies~\cite{Bartels2005, Bartels2010,
  BartelsProhl2007}. The method is somewhat restrictive in that it
only applies to manifolds that can be embedded isometrically as
codimension one surfaces in $\Rn$.

The surface finite element method has been generalized in different
ways to allow the possibility of studying problems on larger classes
of manifolds, which need not be embedded surfaces in $\Rn$.  For
instance, Michael Holst and collaborators~\cite{Holst2001,
  HolstStern2012a, HolstStern2012b} have developed methods for
defining discrete representations of differential forms on simplicial
representations of manifolds with arbitrary topologies.  The
differential structure of a manifold in this approach is determined by
explicitly specifying the set of coordinate overlap maps that cover
the interfaces between neighboring simplices.  The geometry of the
manifold (needed for example to define the covariant Laplace-Beltrami
operator, or the dual transformations of differential forms) is
determined in this approach by a metric on the manifold that must also
be explicitly supplied.  Oliver Sander and
collaborators~\cite{Sander2010, Sander2012, GrohsEtAl2014, Sander2015,
  Hardering2015} have introduced a different generalization of the
surface finite element method.  Their approach, called the geodesic
finite element method, uses the geometry of the manifold (which must
be specified explicitly) to construct discrete differential operators
that conform more precisely to the manifold.  The usual interpolation
rule along straight coordinate lines in the reference element is
replaced with geodesic interpolation in a curved manifold.  The global
topology and the differentiable structures must be specified
explicitly for each manifold.  These approaches are very general, but
they are somewhat cumbersome to use in practice since they require a
great deal of detailed information to be explicitly provided in order
to determine the differential and geometrical structures for each
manifold studied.

Multicube representations of manifolds~\cite{Lindblom2013} provide a
framework for the development of simpler methods for solving PDEs
numerically on manifolds with arbitrary topologies.  This approach,
which we review in the following paragraphs, has several significant
advantages over the finite element methods discussed above.  For one,
the multicube method represents a manifold as a mesh of
non-overlapping cubes (or hypercubes) rather than simplices.  This
makes it simpler to introduce natural bases for vector and tensor
fields on these manifolds.  The cubic structure is also better suited
for spectral numerical methods, which converge significantly faster
than finite element methods of any (fixed) order.  Another distinct
advantage of the multicube approach is that the differential
structures on multicube manifolds can be determined by a smooth
reference metric.  Therefore one need not specify the differential
structure explicitly as would be required by the earlier
generalizations of the surface finite element method.  In our previous
work involving the multicube method we specified the needed reference
metrics analytically for the few simple manifolds that we
studied~\cite{Lindblom2013, Lindblom2014}.  In more complicated
cases, however, the problem of finding an appropriate smooth reference
metric is more difficult.  The main purpose of this paper is to
develop methods for generating the needed reference metrics
automatically.

The multicube representation of a manifold $\Sigma$ consists of a
collection of non-intersecting $n$-dimensional cubic regions ${\cal
  B}_A\subset \Rn$ for ${A}=1,2, ..., N_R$, together with a set of
one-to-one invertible maps $\Psi^{A\alpha}_{B\beta}$ that determine
how the boundaries of these regions are to be connected together.  The
maps $\partial_\alpha {\cal B}_A=
\Psi^{A\alpha}_{B\beta}(\partial_\beta{\cal B}_B)$ define these
connections by identifying points on the boundary face
$\partial_\beta{\cal B}_B$ of region ${\cal B}_B$ with points on the
boundary face $\partial_\alpha{\cal B}_A$ of region ${\cal B}_A$
(cf. Ref.~\cite{Lindblom2013} and \ref{s:2dMulticubeManifolds}).  
It is convenient to choose all these
cubic regions in $\Rn$ to have the same coordinate size $L$, the same
orientation, and to locate them so that regions intersect (if at all)
in $\Rn$ only at faces that are identified by the
$\Psi^{A\alpha}_{B\beta}$ maps.  Since the regions do not overlap, the
global Cartesian coordinates of $\Rn$ can be used to identify points
in $\Sigma$.  Tensor fields on $\Sigma$ can be represented by their
components in the tensor bases associated with these global Cartesian
coordinates.

The Cartesian components of smooth tensor fields on a multicube
manifold are smooth functions of the global Cartesian coordinates
within each region ${\cal B}_A$, but these components may not be
smooth (or even continuous) across the interface boundaries
$\partial_\alpha{\cal B}_A$ between regions.  Smooth tensor fields
must instead satisfy more complicated interface continuity conditions
defined by certain Jacobians, $J^{A\alpha i}_{B\beta j}$, that
determine how vectors $v^i$ and covectors $w_i$ transform across
interface boundaries: $v^i_A=J^{A\alpha i}_{B\beta j}\,v^j_B$ and
$w_{Ai}=J_{A\alpha i}^{*B\beta j}\,w_{Bj}$.  As discussed in
Ref.~\cite{Lindblom2013}, the needed Jacobians are easy to construct
given a smooth, positive-definite reference metric $\tilde g_{ij}$ on
$\Sigma$.

A smooth reference metric also makes it possible to define what it
means for tensor fields to be $C^1$, i.e., to have continous
derivatives across interface boundaries.  Tensors are $C^1$ if their
covariant gradients (defined with respect to the smooth connection
determined by the reference metric) are continuous.  At interface
boundaries, the continuity of these gradients (which are themselves
tensors) is defined by the Jacobians $J^{A \alpha i}_{B \beta j}$ in
the same way it is defined for any tensor field.

A reference metric is therefore an extremely useful (if not essential)
tool for defining and enforcing continuity of tensor fields and their
derivatives on multicube representations of manifolds.  Unfortunately
there is (at present) no straightforward way to construct these
reference metrics on manifolds with arbitrary topologies.  The
examples given to date in the literature have been limited to
manifolds with simple topologies where explicit formulas for smooth
metrics were already known~\cite{Lindblom2013}.  The purpose of this
paper is to present a general approach for constructing suitable
reference metrics for arbitrary manifolds.  The goal is to develop a
method that can be implemented automatically by a code using as input
only the multicube structure of the manifold, i.e., from a knowledge
of the collection of regions $\mathcal{B}_A$ and the way these regions
are connected together by the interface maps
$\Psi^{A\alpha}_{B\beta}$.

In this paper we develop, implement, and test a method for
constructing positive-definite (i.e., Riemannian) $C^1$ reference
metrics for compact, orientable two-dimensional manifolds with
arbitrary topologies.  While $C^\infty$ reference metrics might
theoretically be preferable, $C^1$ metrics are all that are required
to define the continuity of tensor fields and their derivatives.  We
show in~\ref{s:Uniqueness} that any $C^1$ reference metric provides
the same definitions of continuity of tensor fields and their
derivatives across interface boundaries as a $C^\infty$ reference
metric.  This level of smoothness is all that is needed to provide the
appropriate interface boundary conditions for the solutions of the
systems of second-order PDEs most commonly used in mathematical
physics.  For all practicable purposes, therefore, $C^1$ reference
metrics are all that are generally required.

\begin{figure}[!htb]
\centering
\includegraphics[width=0.36\textwidth]{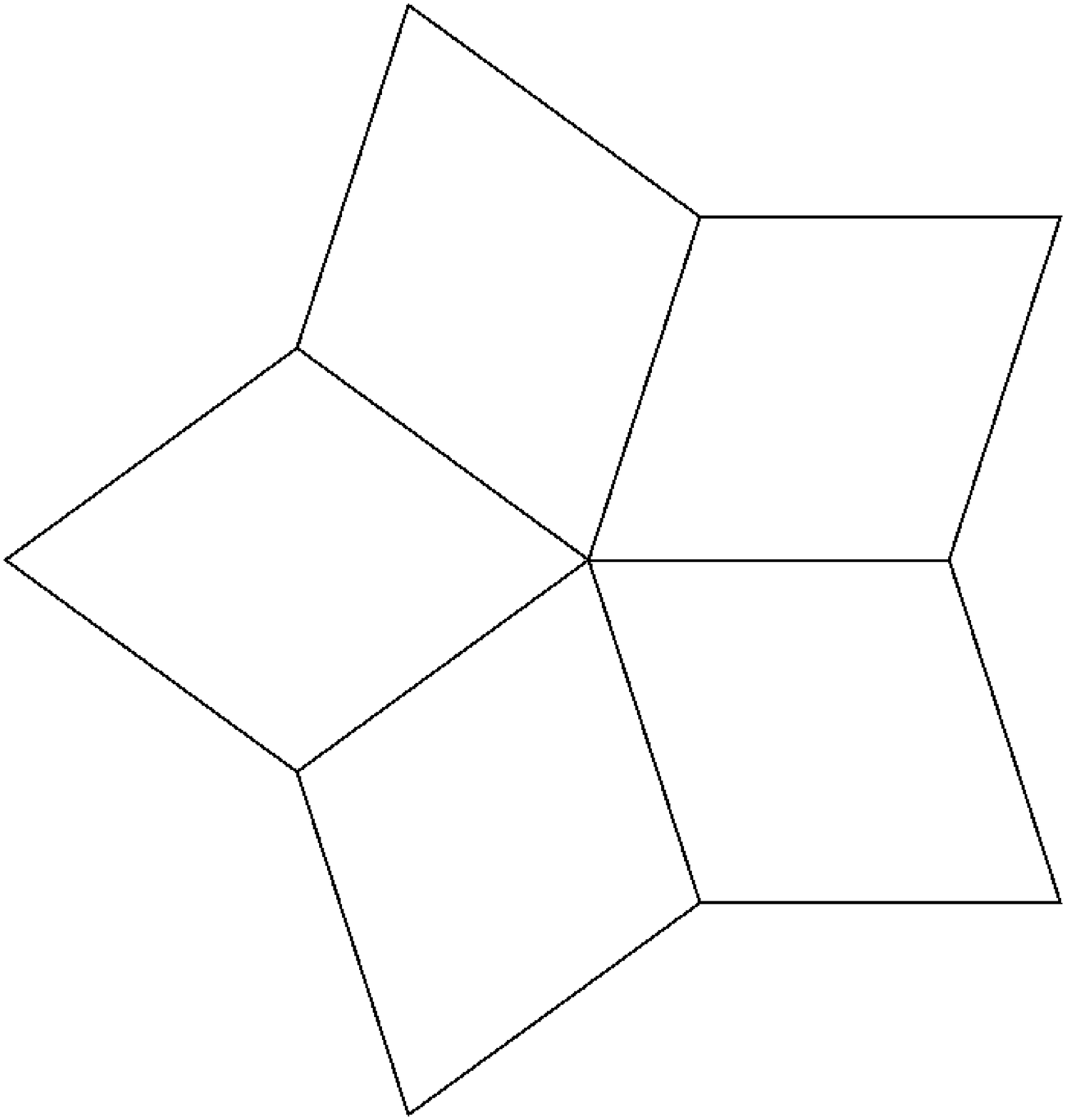}
\caption{\label{f:StarShapedRegions} Two-dimensional star-shaped
  domain $\mathcal{S}_I$ whose center $\mathcal{V}_I$ is a vertex
  point where five regions $\mathcal{B}_A$ intersect.}
\end{figure}

Our method of constructing a reference metric $\tilde g_{ij}$ on
$\Sigma$ is built on a collection of star-shaped domains ${\cal S}_I$
with ${I}=1,2,...,N_S$ that surround the vertex points
$\mathcal{V}_I$, which make up the corners of the multicube regions.
The star-shaped domain ${\cal S}_I$ is composed of
copies of all the regions $\mathcal{B}_A$ that intersect at the vertex
point $\mathcal{V}_I$.  The interface boundaries of the regions that
include the vertex $\mathcal{V}_I$ are to be connected together within
$\mathcal{S}_I$ using the same interface boundary maps
$\Psi^{A\alpha}_{B\beta}$ that define the multicube structure.
Figure~\ref{f:StarShapedRegions} illustrates a two-dimensional example
of a star-shaped domain $\mathcal{S}_I$ whose center $\mathcal{V}_I$
is a vertex point where five regions intersect.  A region
$\mathcal{B}_A$ would be represented multiple times in a particular
$\mathcal{S}_I$ if more than one of its vertices is identified by the
interface boundary maps with the vertex point $\mathcal{V}_I$ at the
center of $\mathcal{S}_I$.  For example, consider a one-region
representation of $T^2$.  The single $\mathcal{S}_I$ in this case
consists of four copies of the single region $\mathcal{B}_A$, glued
together so that each of the vertices of the original region coincides
with the center of $\mathcal{S}_I$.  The interior of each star-shaped
domain $\mathcal{S}_I$ has the topology of an open ball in $\Rn$, and
together they form a set of overlapping domains that cover the
manifold: $\cup_I{\cal S}_I=\Sigma$.

A smooth reference metric is constructed on each star-shaped
domain ${\cal S}_I$ by introducing local Cartesian
coordinates on it that have smooth transition maps with the global
multicube coordinates of each region ${\cal B}_A$ that it contains.
Let $e^I_{ij}$ denote the flat Euclidean metric within ${\cal S}_I$,
i.e., the tensor whose components are the unit matrix when written in
terms of the local Cartesian coordinates of ${\cal S}_I$.  These
metrics are manifestly free of singularities within each
$\mathcal{S}_I$, and they can be transformed from the local
star-shaped domain coordinates into the global
multicube coordinates in each $\mathcal{B}_A$ using the smooth
transition maps that relate them.

These smooth metrics on the star-shaped domains $\mathcal{S}_I$ can be
combined to form a global metric on $\Sigma$ by introducing a
partition of unity $u_I({\vec x})$.  These functions must be positive,
$u_I({\vec x})>0$, for points ${\vec x}$ in the interior of ${\cal
  S}_I$; they must vanish, $u_I({\vec x})=0$, for points outside
${\cal S}_I$; and they are normalized so that $1=\sum_I u_I(\vec x)$
at every point $\vec x$ in $\Sigma$.  Using these functions, the
tensor $\bar g_{ij}({\vec x})=\sum_I u_I({\vec x})\, e^I_{ij}({\vec
  x})$ is positive definite at each point ${\vec x}$ in $\Sigma$ and
can therefore be used as a reference metric for $\Sigma$.  Although
each metric $e^I_{ij}$ is smooth within its own domain
$\mathcal{S}_I$, it may not be smooth with respect to the Cartesian
coordinates of the other star-shaped domains that intersect
$\mathcal{S}_I$.  For this reason the combined metric $\bar g_{ij}$
will generally only be as smooth as the products $u_I(\vec
x)\,e^I_{ij}$.

At the present time we only know how to construct functions $u_I(\vec
x)$ that make the combined metric $\bar g_{ij}$ continuous (but not
$C^1$) across all the interface boundaries.  The metric $\bar g_{ij}$
can be modified in a systematic and fairly straightforward way,
however, to produce a new metric $\tilde g_{ij}$ whose extrinsic
curvature $\tilde K_{ij}$ vanishes along each multicube interface
boundary $\partial_\alpha\mathcal{B}_A$.  Continuity of the extrinsic
curvature is the geometrical condition needed to ensure the continuity
of the derivatives of the metric across interface boundaries.  The
modified metrics $\tilde g_{ij}$ constructed in this way can therefore
be used as $C^1$ reference metrics.  In the two-dimensional case, the
modification that converts $\bar g_{ij}$ into $\tilde g_{ij}$ can be
accomplished using a simple conformal transformation.  In higher
dimensions, a more complicated transformation is required.

The following sections present detailed descriptions of our procedure
for constructing reference metrics $\tilde g_{ij}$ on two-dimensional
multicube manifolds having arbitrary topologies.  In
Sec.~\ref{s:TwoDimensionalReferenceMetricsA0} an explicit method is
described for systematically constructing the overlapping star-shaped
domains ${\cal S}_I$; formulas are given for transforming between the
intrinsic Cartesian coordinates in each ${\cal S}_I$ and the global
Cartesian coordinates in ${\cal B}_A$; explicit representations are
given (in both local and global Cartesian coordinates) for the flat
metrics $e^I_{ij}(\vec x)$ in each domain ${\cal S}_I$; and examples
of useful $C^0$ partition of unity functions $u_I(\vec x)$ are given.
The resulting $C^0$ metrics are then modified in
Sec.~\ref{s:TwoDimensionalReferenceMetricsA1} by constructing an
explicit conformal transformation that produces a metric having
vanishing extrinsic curvature at each of the interface boundaries
$\partial_\alpha\mathcal{B}_A$.  The resulting metric is $C^1$ and can
therefore be used as a reference metric for these manifolds.

We test these procedures for constructing reference metrics on a
collection of compact, orientable two-dimensional manifolds in
Sec.~\ref{s:TwoDimensionalReferenceMetricsB}.  New multicube
representations of orientable two-dimensional manifolds having
arbitrary topologies are described in detail
in~\ref{s:2dMulticubeManifolds}.  These procedures have been
implemented in the Spectral Einstein Code (SpEC, developed by the SXS
Collaboration, originally at Caltech and Cornell~\cite{Kidder2000a,
  Scheel2006, Szilagyi:2009qz}).  Reference metrics are constructed
numerically in Sec.~\ref{s:TwoDimensionalReferenceMetricsB} for
two-dimensional multicube manifolds with genera $N_g$ between zero and
five; the scalar curvatures $\tilde R$ associated with these reference
metrics are illustrated; and numerical results are presented which
demonstrate that these two-dimensional reference metrics satisfy the
Gauss-Bonnet identity up to truncation level errors (which converge to
zero as the numerical resolution is increased).  We also show that the
continuous (but not $C^1$) reference metrics $\bar g_{ij}$ fail to
satisfy the Gauss-Bonnet identity numerically because of the curvature
singularities which occur on the interface boundaries in this case.

The scalar curvatures associated with the $C^1$ reference metrics
constructed in Sec.~\ref{s:TwoDimensionalReferenceMetrics} turn out to
be quite nonuniform.  Section~\ref{s:RicciFlow} explores the
possibility of using Ricci flow to smooth out the inhomogenities in
these metrics $\tilde g_{ij}$.  In particular we develop a slightly
modified version of volume-normalized Ricci flow with DeTurck gauge
fixing.  This version is found to perform better numerically with
regard to keeping the volume of the manifold fixed at a prescribed
value.  We describe our implementation of these new Ricci flow
equations in SpEC in Sec.~\ref{s:NumericalRicciFlow}.  We test this
implementation by evolving a round-sphere metric with random
perturbations on a six-region multicube representation of the
two-sphere manifold, $S^2$.  These tests show that our numerical Ricci
flow works as expected: the solutions evolve toward constant-curvature
metrics, the volumes of the manifolds are driven toward the prescribed
values, and the Gauss-Bonnet identities remain satisfied throughout
the evolutions.  In Sec.~\ref{s:SmootherReferenceMetrics} we use Ricci
flow to evolve the rather nonuniform $C^1$ reference metrics $\tilde
g_{ij}$ constructed in Sec.~\ref{s:TwoDimensionalReferenceMetrics},
using these $\tilde g_{ij}$ both as initial data and as the fixed
reference metrics throughout the evolutions.  We show that all these
evolutions approach constant curvature metrics, as expected for
two-dimensional Ricci flow. The volumes of these manifolds remain
fixed throughout the evolutions, and the Gauss-Bonnet identities are
satisfied for all the geometries tested (which include genera $N_g$
between zero and five).  These Ricci-flow-evolved metrics therefore
provide smoother and more uniform reference metrics for these
manifolds.

\section{Two-Dimensional Reference Metrics}
\label{s:TwoDimensionalReferenceMetrics}

This section develops a procedure for constructing reference metrics
on multicube representations of two-dimensional manifolds.  Continuous
reference metrics are created in
Sec.~\ref{s:TwoDimensionalReferenceMetricsA0} and then transformed in
Sec.~\ref{s:TwoDimensionalReferenceMetricsA1} into metrics whose
derivatives are also continuous across the multicube interface
boundaries.  The resulting $C^1$ reference metrics are tested in
Sec.~\ref{s:TwoDimensionalReferenceMetricsB} (on two-dimensional
manifolds with genera $N_g$ between zero and five) to ensure that they
satisfy the appropriate Gauss-Bonnet identities.

\subsection{Constructing Continuous Reference Metrics}
\label{s:TwoDimensionalReferenceMetricsA0}

The procedure for creating a continuous ($C^0$) reference metric $\bar
g_{ij}$ presented here has three basic steps.  First, a set of
star-shaped domains ${\cal S}_I$ for the multicube manifold is
constructed from a knowledge of the regions ${\cal B}_A$ and their
interface boundary identification maps $\partial_\alpha {\cal B}_A=
\Psi^{A\alpha}_{B\beta}(\partial_\beta{\cal B}_B)$.  The interiors of
these ${\cal S}_I$ have the topology of open balls in $\Rn$ and
together they form an open cover of the manifold $\Sigma$.  The
primary task in this first step of the procedure is to organize the
multicube structure in a way that allows us to determine which
star-shaped domain ${\cal S}_I$ is centered around
each vertex $\nu_{A\mu}$ of each multicube region ${\cal B}_A$, and to
determine how many regions ${\cal B}_A$ belong to each ${\cal S}_I$.
In the second step, intrinsic Cartesian coordinates and metrics are
constructed for each ${\cal S}_I$.  These intrinsic coordinates are
chosen to have smooth transformations with the global Cartesian
coordinates in each multicube region ${\cal B}_A$.  Metrics $e^I_{ij}$
for each star-shaped domain are introduced in this step to be the
Euclidean metric expressed in terms of the intrinsic Cartesian
coordinates in each ${\cal S}_I$.  In the third step, partitions of
unity $u_I({\vec x})$ are constructed that are positive for points
$\vec x$ inside ${\cal S}_I$, that vanish for points $\vec x$ outside
${\cal S}_I$, and that sum to unity at each point in the manifold:
$1=\sum_Iu_I(\vec x)$.  A global reference metric is then obtained by
taking weighted linear combinations of the flat metrics from each of
the domains ${\cal S}_I$: $\bar g_{ij}({\vec x}) = \sum_I u_I({\vec
  x})\, e^I_{ij}({\vec x})$.  At present we only know how to choose
the partition of unity functions $u_I(\vec x)$ in a way that makes
$\bar g_{ij}$ continuous across the boundary interfaces.

\subsubsection{Step One}
The first step is to compose and sort a list of all the vertices
${\nu}_{A\mu}$ in a given multicube structure. The index $\mu={\{
  1,...,2^n\}}$, where $n$ is the dimension of the manifold,
identifies the vertices of a particular multicube
region ${\cal B}_A$.  This list of vertices $\nu_{A\mu}$ can be sorted
into equivalence classes $\mathcal{V}_I$ whose members are identified
with one another by the interface boundary-identification maps, i.e.,
$\nu_{A\mu}$ and $\nu_{B\sigma}$ belong to the same $\mathcal{V}_I$
iff there exists a sequence of maps
$\Psi^{A\alpha}_{A_1\alpha_1}$,$\Psi^{A_1\alpha_1}_{A_2\alpha_2}$,\,\ldots,
$\Psi^{A_n\alpha_n}_{B\beta}$ with  ${\nu}_{A\mu}=\left(
\Psi^{A\alpha}_{A_1\alpha_1} \circ \Psi^{A_1\alpha_1}_{A_2\alpha_2}
\circ \ldots \circ \Psi^{A_n\alpha_n}_{B\beta} \right)(\nu_{B\sigma})$.

One star-shaped domain ${\cal S}_I$ is centered on each equivalence
class of vertices $\mathcal{V}_I$.  The domain $\mathcal{S}_I$
consists of copies of all the multicube regions $\mathcal{B}_A$ having
vertices that belong to the equivalence class $\mathcal{V}_I$.  For
two-dimensional manifolds, the primary computational task to be
completed in this first step is to determine the number $K_I$ of
vertices $\nu_{A\mu}$ that belong to each of the $\mathcal{V}_I$
classes.  The quantity $K_I$ represents the number of multicube
regions $\mathcal{B}_A$ clustered around the vertex $\mathcal{V}_I$ in
the star-shaped domain $\mathcal{S}_I$.  Our code performs this
counting process in two dimensions by using the fact that each vertex
$\nu_{A\mu}$ belongs to two different boundaries of the region ${\cal
  B}_A$.  The code arbitrarily picks one of these boundaries, say
$\partial_\alpha {\cal B}_A$, and follows the identification map
$\Psi^{B\beta}_{A\alpha}$ to the neighboring region ${\cal B}_B$.  The
mapped vertex $\nu_{B\sigma}=\Psi^{B\beta}_{A\alpha}(\nu_{A\mu})$
again belongs to two boundaries of the new region ${\cal B}_B$: the
mapped boundary $\partial_\beta {\cal B}_B$ and another one, say
$\partial_\gamma{\cal B}_B$.  The code then follows the map
$\Psi^{C\delta}_{B\gamma}$ across this other boundary to its
neighboring region ${\cal B}_C$ and to the new mapped vertex
$\nu_{C\rho} = \Psi^{C\delta}_{B\gamma}(\nu_{B\sigma})$.  Continuing
in this way, the code makes a sequence of transitions between regions
until it arrives back at the original vertex $\nu_{A\mu}$ of the
starting region ${\cal B}_A$. The code counts these transitions and
returns the number $K_I$ when the loop is closed.
Figure~\ref{f:StarShapedRegions} illustrates a two-dimensional
star-shaped domain with $K_I=5$.

\subsubsection{Step Two}
\begin{figure}[!htb]
\centering
\includegraphics[width=0.7\textwidth]{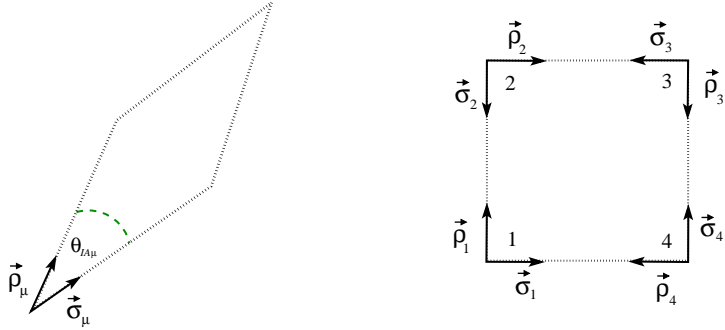}
\caption{Distorted and undistorted
  representations of a multicube region.  Left side shows one of the
  two-dimensional multicube regions ${\cal B}_A$ that has been
  distorted to allow it to fit together with the other regions in a
  particular star-shaped domain $\mathcal{S}_I$.  The vectors
  $\vec\rho_\mu$ and $\vec\sigma_\mu$ are tangent to the boundaries of
  ${\cal B}_A$.  Right side is a representation of this same ${\cal
    B}_A$, showing the associations of the vectors $\vec\rho_\mu$ and
  $\vec\sigma_\mu$ for its various possible
  vertices (labeled by the index $\mu$).
  \label{f:OverlapDomains3}} 
\end{figure}

The second step in this procedure is to construct local Cartesian
coordinates that cover each of the star-shaped domains
$\mathcal{S}_I$.  We do this by noting that each $\mathcal{S}_I$
consists of a cluster of cubes $\mathcal{B}_A$ whose vertices coincide
with the central point $\mathcal{V}_I$.  If these cubes are
appropriately distorted into parallelograms (by adjusting the angles
between their coordinate axes), they can be fitted together (without
overlapping and without leaving gaps between them) to form a domain in
$\Rn$ whose interior has the topology of an open ball.  Each
$\mathcal{S}_I$ can therefore be covered by a single coordinate chart,
which in two-dimensions can be written in the form $\bar x^i_I=(\bar
x_I,\bar y_I)$.  Figure~\ref{f:OverlapDomains3} illustrates both the
distorted (on the left) and the undistorted (on the right)
representations of a two-dimensional $\mathcal{B}_A$.

In two dimensions the distortions needed to allow the $\mathcal{B}_A$
to be fitted around a vertex point $\mathcal{V}_I$ are quite simple:
adjust the opening angles $\theta_{IA\mu}$ of the coordinate axes of
each cube so they sum to $2\pi$ around each vertex,
$\sum_{A\mu}\theta_{IA\mu}=2\pi$.  The optimal way to satisfy this
local flatness condition is to distort all of the two-dimensional
cubes that make up $\mathcal{S}_I$ in the same way, i.e., by setting
$\theta_{IA\mu}=2\pi/K_I$.  In higher dimensions the problem of
fitting the $\mathcal{B}_A$ together to form a smooth star-shaped
domain (without conical singularites and without gaps) is more
complicated.  The complication in higher dimensions comes from the
lack of uniqueness and a clear optimal choice, rather than being a
fundamental problem of existence.  We plan to study the problem of
finding a practical way to perform this construction in higher
dimensions in a future paper.

The simplest metric $\bar e^I_{ij}$ to assign to the star-shaped
domain ${\cal S}_I$ is the flat Euclidean metric expressed in terms of
the local coordinates of $\mathcal{S}_I$:
\begin{eqnarray}
ds^2 = \bar e^I_{ij}d\bar x^i_Id\bar x^j_I=d\bar x^2_I+d\bar y^2_I.
\label{e:LocalFlatMetric}
\end{eqnarray}
Each ${\cal B}_A$ that intersects ${\cal S}_I$ will inherit this flat
geometry via the coordinate transformation that connects them.  This
fact can be used to deduce the coordinate transformations between the
local Cartesian coordinates $\bar x^i_I=(\bar x_I,\bar y_I)$ of
$\mathcal{S}_I$ and the global coordinates $x^i_A=(x_A,y_A)$ of
$\mathcal{B}_A$.  The left side of Fig.~\ref{f:OverlapDomains3} shows
a region ${\cal B}_A$ in $\mathcal{S}_I$ that has been distorted into
a parallelogram having an opening angle $\theta_{IA\mu}$.  The vectors
$\vec \rho_\mu$ and $\vec \sigma_\mu$ in this figure represent unit
vectors (according to the local flat metric of ${\cal S}_I$) that are
tangent to the boundary faces of ${\cal B}_A$ at this vertex.  The
index $\mu$ identifies which of the vertices of ${\cal B}_A$ these
unit vectors belong to.  Since the opening angle at this particular
vertex is $\theta_{IA\mu}$, the inner product of these vectors is just
$\vec \rho_\mu\cdot\vec \sigma_\mu=\cos\theta_{IA\mu}$.  The vectors
$\vec \rho_\mu$ and $\vec \sigma_\mu$ are proportional to the
coordinate vectors $\partial_x$ and $\partial_y$ of the global
Cartesian coordinates used to describe points in the multicube region
${\cal B}_A$---exactly which coordinate vectors depends on which
vertex of ${\cal B}_A$ coincides with this point. The right side of
Fig.~\ref{f:OverlapDomains3} shows these vectors at each of the
vertices of ${\cal B}_A$, any of which could be the one that coincides
with the center of ${\cal S}_I$.  Table~\ref{t:TableI} gives the
relationships between $\vec \rho_\mu$ and $\vec \sigma_\mu$ and the
coordinate basis vectors in ${\cal B}_A$ for each vertex $\nu_{\mu}$.
Also listed in Table~\ref{t:TableI} are the vectors $\vec v_\mu$ that
give the location of each vertex relative to the center of its region
${\cal B}_A$.
\begin{table}[!htb]
  \centering
  \caption{The vectors $\vec \rho_\mu$ and $\vec \sigma_\mu$ are
      proportional to the basis vectors $\vec\partial_x$ and
      $\vec\partial_y$  at each vertex $\mu$ of the region
    $\mathcal{B}_A$.  This table gives the global Cartesian
      coordinate representations of $\vec \rho_\mu$ and $\vec
      \sigma_\mu$ at each vertex, the vertex-dependent constants
    $\epsilon_\mu$, and the locations $\vec \nu_\mu$ of the vertices
    with respect to the center of ${\cal B}_A$.
    \label{t:TableI} }
  \setlength\tabcolsep{12pt}
  \begin{tabular}{@{}lcccc@{}} \toprule
    $\mu$ & $\vec \rho_\mu$ & $\vec \sigma_\mu$ & $\epsilon_\mu$
    & $\vec v_\mu$ \\ \midrule
    1 & $(0,1)$ & $(1,0) $ & $+1$ & $\frac{1}{2}L(-1,-1)$ \\ [2pt]
    2 & $(1,0)$ & $(0,-1)$ & $-1$ & $\frac{1}{2}L(-1,+1)$ \\ [2pt]
    3 & $(0,-1)$& $(-1,0)$ & $+1$ & $\frac{1}{2}L(+1,+1)$ \\ [2pt]
    4 & $(-1,0)$& $(0,1)$  & $-1$ & $\frac{1}{2}L(+1,-1)$ \\ \bottomrule
  \end{tabular}
\end{table}

The inner products $\vec \rho_\mu \cdot \vec \rho_\mu$, $\vec
\sigma_\mu \cdot \vec \sigma_\mu$, and $\vec \rho_\mu\cdot \vec
\sigma_\mu$ are scalars that are independent of the coordinate
representation of the vectors.  Since $\vec \rho_\mu$ and $\vec
\sigma_\mu$ are unit vectors that are (up to signs) just the
coordinate basis vectors in the global Cartesian coordinates, it
follows that the components of the metric $e^I_{ij}$ in the
global coordinates of $\mathcal{B}_A$ must have the values $\vec
\rho_\mu\cdot\vec \rho_\mu =\vec \sigma_\mu\cdot \vec \sigma_\mu =
e^I_{xx}=e^I_{yy}=1$ and $\vec \rho_\mu\cdot\vec
\sigma_\mu=\cos\theta_{IA\mu} = \epsilon_\mu \,e^I_{xy}$, where
$\epsilon_\mu=\pm 1$ is the vertex-dependent constant defined in
Table~\ref{t:TableI}.  The flat metric $e^{I}_{ij}$ of the region
${\cal S}_I\cap{\cal B}_A$ therefore has the form 
\begin{eqnarray}
ds^2=e^{IA}_{ij}dx^i_Adx^j_A=dx^2_A
+2\epsilon_\mu\cos\theta_{IA\mu}\, dx_A\, dy_A + dy^2_A
\label{e:GlobalFlatMetric}
\end{eqnarray}
when expressed in terms of the global Cartesian coordinates
$x^i_A=(x_A,y_A)$ of ${\cal B}_A$.  This metric can also be 
written as
\begin{eqnarray}
ds^2=e^{IA}_{ij}dx^i_Adx^j_A=(dx_A+\epsilon_\mu\cos\theta_{IA\mu}\, dy_A)^2 
+ \sin^2\theta_{IA\mu}\, dy^2_A.
\label{e:GlobalFlatMetricII}
\end{eqnarray}
This is identical to the standard representation of $\bar e^I_{ij}$ in the
local coordinates of ${\cal S}_I$, Eq.~(\ref{e:LocalFlatMetric}), if
new coordinates $\tilde x_{IA}$ and $\tilde y_{IA}$ are defined as
\begin{eqnarray}
\tilde x_{IA} &=& x_A -c^x_A-v^x_\mu + \epsilon_\mu\cos\theta_{IA\mu}\,
(y_A -c^y_A-v^y_\mu),
\label{e:CoordinateChangex}
\\
\tilde y_{IA} &=& \sin\theta_{AI}\, (y -c^y_A-v^y_\mu).
\label{e:CoordinateChangey}
\end{eqnarray}
The constants $c_A^i$ represent the global Cartesian coordinates of
the center of region $\mathcal{B}_A$, and the constants $v_{\mu}^i$
represent the location of the $\mu$ vertex of the region relative to
its center.  These are included in the transformations in
Eqs.~(\ref{e:CoordinateChangex}) and (\ref{e:CoordinateChangey}) to
ensure that the point $\tilde x_{IA}=\tilde y_{IA} =0$ corresponds to
the point $\vec x = \vec c_A + \vec v_{\mu}$, which is the
${\nu}_{A\mu}$ vertex of ${\cal B}_A$ that coincides with the center
of ${\cal S}_I$.  These new coordinates $\tilde x_{IA}$ and $\tilde
y_{IA}$ are therefore equal to the local Cartesian coordinates of
${\cal S}_I$, $\bar x_I$ and $\bar y_I$, up to a rigid rotation:
\begin{eqnarray}
\bar x_{I}&=&\cos\psi_{IA}\,\tilde x_{IA}  + \sin\psi_{IA}\,\tilde y_{IA},
\label{e:CoordinateChangeIIx}
\\
\bar y_{I}&=&-\sin\psi_{IA}\,\tilde x_{IA} + \cos\psi_{IA}\,\tilde y_{IA},
\label{e:CoordinateChangeIIy}
\end{eqnarray}
for some angle $\psi_{IA}$.  The composition of
Eqs.~(\ref{e:CoordinateChangeIIx}) and (\ref{e:CoordinateChangeIIy})
with Eqs.~(\ref{e:CoordinateChangex}) and (\ref{e:CoordinateChangey})
therefore gives the transformation between the local Cartesian
coordinates of ${\cal S}_I$, $\bar x_I$ and $\bar y_I$, and the global
Cartesian coordinates, $x_A$ and $y_A$, of the multicube
representation of the manifold.

The metric $e^{IA}_{ij}$ given in
Eq.~(\ref{e:GlobalFlatMetric}) must be constructed for each vertex
$\nu_{A\mu}$ of each region ${\cal B}_A$ in terms of its global
Cartesian coordinates $x_A^i$.  These expressions depend only on the
opening angles $\theta_{IA\mu}$, which in turn depend only on the
parameter $K_I$.  The full coordinate transformations between the
global Cartesian coordinates $x_A$ and $y_A$ and the local coordinates
$\bar x_{I}$ and $\bar y_{I}$ given in
Eqs.~(\ref{e:CoordinateChangex})--(\ref{e:CoordinateChangeIIy}) are
not actually needed to evaluate the reference metrics.

\subsubsection{Step Three}
\begin{figure}
\centering
\includegraphics[width=0.52\textwidth]{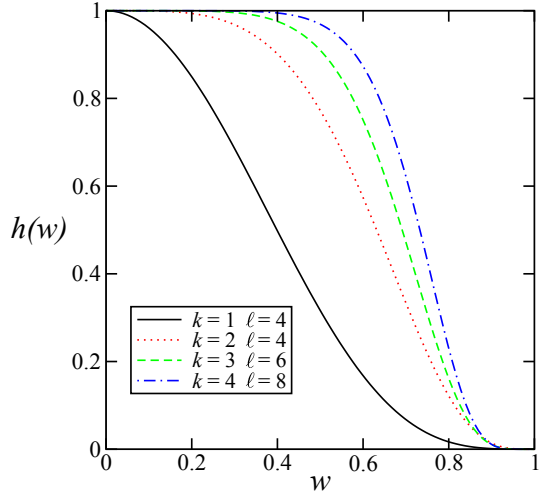}
\caption{Weight functions $h(w)$ defined in
  Eq.~(\ref{e:PartitionOfUnityhAlt}) are positive for $0\leq w < 1$
  and vanish for $w=1$. 
\label{f:PartitionOfUnityAlt} }
\end{figure}

\begin{figure}
\begin{picture}(0,150)(0,0)
\put(20,0){
  \includegraphics[width=0.38\textwidth]{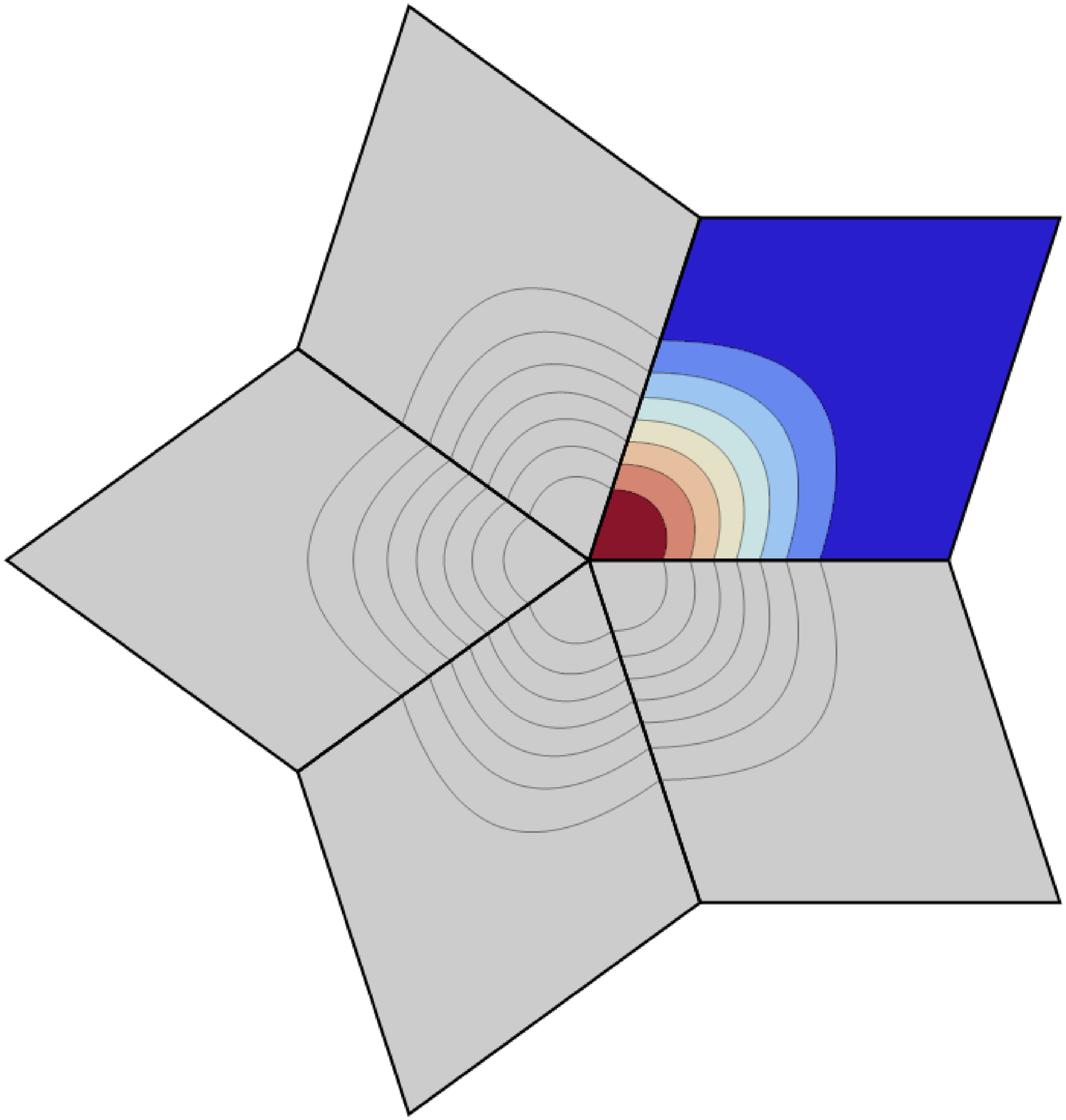}
}
\put(200,2){
  \includegraphics[width=0.42\textwidth]{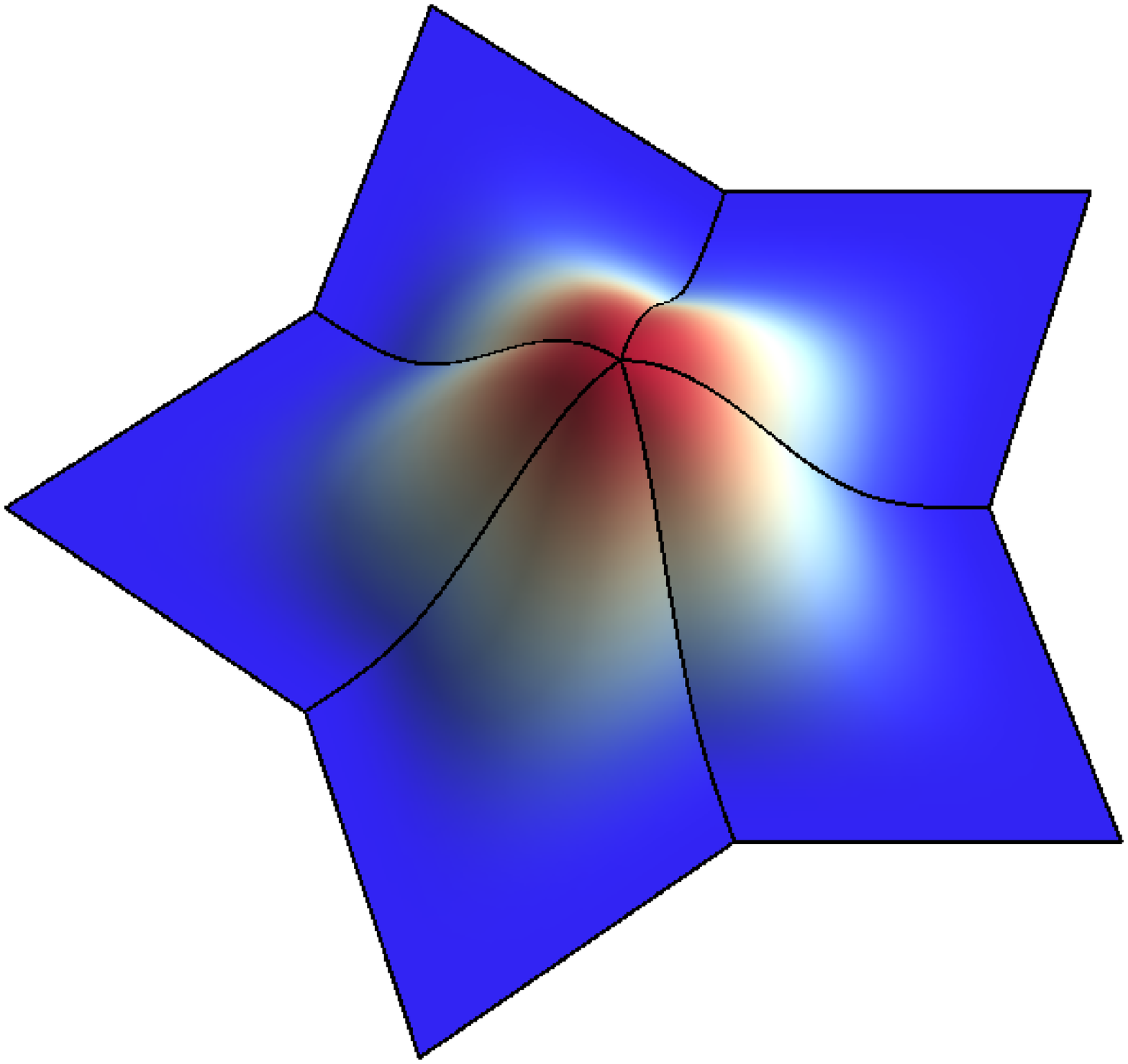}
}
\end{picture}
\caption{\label{f:WeightFunctions} Weight function $w_I(\vec x)$
  illustrated on a star-shaped domain $\mathcal{S}_I$ where five
  regions $\mathcal{B}_A$ meet.  Left illustration shows countours of
  $w_I(\vec x)$, which uses the $h(w)$ functions defined in
  Eq.~\ref{e:PartitionOfUnityhAlt} with $k=1$ and $\ell=4$.  Right
  illustration shows the same function in a three-dimensional
  rendering.  This example illustrates the fact that these $w_I(\vec
  x)$ are continuous but not $C^1$ across the region interface
  boundaries. }
\end{figure}

The third step in this procedure for constructing a reference metric
is to build a partition of unity $u_I(\vec x)$ that is adapted to the
star-shaped domains.  We do this by introducing a collection of weight
functions $w_I(\vec x)$ that are positive within a particular ${\cal
  S}_I$ and that fall to zero at its boundary.  We experimented with a
number of different weight functions and found that writing them as
simple separable functions of the global Cartesian coordinates of each
region $\mathcal{B}_A$ worked far better than anything else we
tried.  Thus we let
\begin{eqnarray}
w_I(\vec x)=h\left(\frac{x_A-c^x_A-v^x_{\mu}}{L}\right)
h\left(\frac{y_A-c^y_A-v^y_{\mu}}{L}\right),
\label{e:PartitionOfUnity}
\end{eqnarray}
where $L$ is the coordinate size of each region $\mathcal{B}_A$.
The functions $h(w)$ are chosen to have the value $h(0)=1$, which
corresponds to the vertex point at the center of the domain
$\mathcal{S}_I$, and the value $h(1)=0$ at the points which correspond
to the outer boundary of $\mathcal{S}_I$.  We find that the simple
class of functions
\begin{eqnarray}
h(w)=(1-w^{2k})^\ell,
\label{e:PartitionOfUnityhAlt}
\end{eqnarray}
with integers $k>0$ and $\ell>0$, works quite well.  Some of these
functions are illustrated in Fig.~\ref{f:PartitionOfUnityAlt}, with
integer values in the range that worked best in our numerical tests.
Figure~\ref{f:WeightFunctions} illustrates these weight functions
expressed in terms of the local Cartesian coordinates of one of the
star-shaped domains $\mathcal{S}_I$.  This figure shows clearly that
this choice of $u_I(\vec x)$ is continuous but not $C^1$ across the
interface boundaries.  We could also make these functions $C^1$ with
respect to the local coordinates in one of the $\mathcal{S}_I$,
however it is not possible to make them $C^1$ with respect to all of
the overlapping local star-shaped coordinates at the same time.

A partition of unity $u_I(\vec x)$ is constructed from
the weight functions $w_I(\vec x)$ by normalizing them:
\begin{eqnarray}
u_I(\vec x)=\frac{w_I(\vec x)}{H(\vec x)},
\label{e:PartitionOfUnityA}
\end{eqnarray}
where $H(\vec x)$ is defined by
\begin{eqnarray}
H(\vec x)=\sum_I w_I(\vec x).
\end{eqnarray}
This definition ensures that the $u_I(\vec x)$ satisfy the
normalization condition $\sum_Iu_I(\vec x)=1$ for every point $\vec x$
in the manifold.

A global reference metric is constructed by combining the metrics
$e^I_{ij}$ associated with each of the star-shaped domains
$\mathcal{S}_I$ and defined in Eq.~(\ref{e:GlobalFlatMetric}),
using the partition of unity defined in Eq.~(\ref{e:PartitionOfUnityA}):
\begin{eqnarray}
\bar g_{ij}({\vec x})=
\sum_I u_{I}({\vec x})\, e^{I}_{ij}({\vec x}).
\label{e:CompleteMetric}
\end{eqnarray}
This metric is positive definite, and it is continuous across all of the
multicube interface boundaries.  It can therefore be used as a
continuous reference metric.  

In an effort to reduce the spatial variation of the metric defined in
Eq.~(\ref{e:CompleteMetric}) and thus reduce the required numerical
resolution, we add additional terms of the form $u_A(\vec x)\,
e^A_{ij}$, where $e^A_{ij}$ are flat metrics with support in a single
multicube region $\mathcal{B}_A$.  Thus we let
\begin{eqnarray}
ds^2 = e^A_{ij}dx^i_Adx^j_A=dx^2_A+dy^2_A
\label{e:RegionFlatMetric}
\end{eqnarray}
be the flat Euclidean metric expressed in terms of the global
Cartesian coordinates $x_A$ and $y_A$.  We define new weight functions
$w_A(\vec x)$ associated with the individual multicube regions to be
\begin{eqnarray}
w_A(\vec x)=h\left(\frac{2(x_A-c^x_A)}{L}\right)
h\left(\frac{2(y_A-c^y_A)}{L}\right),
\label{e:RegionPartitionOfUnity}
\end{eqnarray}
which have the value $w_A(\vec c_A)=1$ at the center of the region
$\mathcal{B}_A$ and the value $w_A(\vec x)=0$ for points $\vec x$ on
its boundary.  These weight functions can be combined with those
assocated with the star-shaped domains,
Eq.~(\ref{e:PartitionOfUnity}), to form a new partition of unity.
We modify the normalization function $H(\vec x)$ to be
\begin{eqnarray}
H(\vec x)=\sum_I w_I(\vec x) + \sum_A w_A(\vec x).
\label{e:NewNormalization}
\end{eqnarray}
Then we redefine the functions $u_I(\vec x)$
using Eq.~(\ref{e:PartitionOfUnityA}) with this new $H(\vec x)$, and
we define functions $u_A(\vec x)$ using
Eqs.~(\ref{e:RegionPartitionOfUnity}) and (\ref{e:NewNormalization}):
\begin{eqnarray}
u_A(\vec x)=\frac{w_A(\vec x)}{H(\vec x)}.
\end{eqnarray}
A new metric is then formed by combining these
region-centered metrics with the star-shaped domain metrics constructed
above:
\begin{eqnarray}
\bar g_{ij}({\vec x})=
\sum_I u_{I}({\vec x})\, e^{I}_{ij}({\vec x})
+\sum_A u_{A}({\vec x})\, e^{A}_{ij}({\vec x}).
\label{e:CompleteMetricII}
\end{eqnarray}
The addition of the region-centered
metrics does not appear to have a significant impact on the required
numerical resolution.  Nevertheless, this is the two-dimensional
reference metric that we use (after conformally transforming as
described in the following section) in the numerical work described in
the later sections of this paper.

\subsection{Constructing $C^1$ Reference Metrics}
\label{s:TwoDimensionalReferenceMetricsA1}

The continuous metric $\bar g_{ij}$ has been constructed in a way that
ensures the geometry has no conical singularities at the vertices of
the multicube regions.  However, $\bar g_{ij}$ is not in general $C^1$
across the interface boundaries; e.g., the partition of unity that we
use is not $C^1$ there.  The geometry defined by $\bar g_{ij}$ will
therefore have curvature singularities along those interface
boundaries.  In order to remove these singularities, our next goal is
to modify $\bar g_{ij}$ by making it $C^1$, while at the same time
keeping it continuous, positive definite, and free of conical
singularities.  It should be possible, for example, to find a tensor
$\psi_{ij}$ that vanishes at the interface boundaries, and whose
normal derivatives are the negatives of those of $\bar g_{ij}$.  In
this case the new tensor $\tilde g_{ij}= \bar g_{ij}+\psi_{ij}$ and
its first derivatives should be continuous at the boundaries.  There
is in fact a great deal of freedom available in choosing $\psi_{ij}$.
In particular, it can be changed arbitrarily in the interior of a
region so long as its boundary values and derivatives remain
unchanged.  The idea is to use this freedom to keep $\psi_{ij}$ small
enough everywhere that $\tilde g_{ij}$ remains positive definite.  We
plan to find a practical way to do this for manifolds of arbitrary
dimension in a future work.  In this paper we focus on the
two-dimensional case, where a simple conformal transformation is all
that is needed to make the continuous metric $\bar g_{ij}$ $C^1$.  We
introduce the conformal factor $\psi_A$ for the metric in multicube
region $\mathcal{B}_A$:
\begin{eqnarray}
\tilde g^A_{ij}= \psi^4_A\, \bar g^A_{ij}.
\label{e:ConformalTrans}
\end{eqnarray}
The conformal factor $\psi_A$ is chosen to make the resulting metric
$\tilde g^A_{ab}$ and its derivatives continuous across interface
boundaries.

The extrinsic curvature $\bar K^{A\alpha}_{ij}$ of the
$\partial_\alpha\mathcal{B}_A$ boundary of cubic region $\mathcal{B}_A$
is defined by
\begin{eqnarray}
\bar K^{A\alpha}_{ij}=(\delta^k_i - \bar n_{A\alpha}^k \bar n_{A\alpha
  i})\bar \nabla_k \bar n_{A\alpha j},
\end{eqnarray}
where $\bar n_{A\alpha}^i$ is the unit normal to the boundary and
$\bar \nabla_k$ is the covariant derivative associated with the metric
$\bar g^A_{ij}$.  In two dimensions this can be rewritten as
\begin{eqnarray}
\bar K^{A\alpha}_{ij}=(\bar g^A_{ij}-\bar n_{A\alpha i}\bar n_{A\alpha
  j})\bar K_{A\alpha},
\end{eqnarray}
where $\bar K_{A\alpha}=\bar \nabla_k\bar n^k_{A\alpha}$ is the trace.
Since the normal vector $\bar n^i_{A\alpha}$ depends only on the metric
$\bar g_{ij}$, its divergence can be written explicitly in terms of
derivatives of the metric:
\begin{eqnarray}
\bar K_{A\alpha}&=&\bar \nabla_k \bar n_{A\alpha}^k =
\tfrac{1}{2}\left[\bar n^i_{A\alpha}(\bar g^{jk}
+ \bar n^j_{A\alpha} \bar n^k_{A\alpha})
-2\bar g^{ij}\bar n^k_{A\alpha} \right]
\partial_i \bar g_{jk}.\qquad
\end{eqnarray}

Under the conformal transformation given in
Eq.~(\ref{e:ConformalTrans}), the trace of the extrinsic curvature
$K_{A\alpha}$ transforms as follows:
\begin{eqnarray}
\tilde K_{A\alpha}=\psi^{-2}_A(\bar K_{A\alpha}+2\bar n^a_{A\alpha}\bar
\nabla_a\log \psi_A).
\end{eqnarray}
The idea is to choose the conformal factor $\psi_A$ so that it has the
value $\psi_A=1$ on each interface boundary
$\partial_\alpha\mathcal{B}_A$, with a normal derivative on each boundary
given by
\begin{eqnarray}
\bar n^a_{A\alpha}\bar \nabla_a\log \psi_A=-\tfrac{1}{2} \bar K_{A\alpha}.
\label{e:ConformalFactorBC}
\end{eqnarray}
These boundary conditions ensure that the metric $\tilde g_{ij}$
continues to be continuous everywhere and free of cone singularities
at the vertices of each cubic-block region, while also ensuring that
the extrinsic curvature at each interface boundary is zero.

There is no unique conformal factor satisfying the boundary conditions
$\psi_A=1$ and the normal-derivative condition given in
Eq.~(\ref{e:ConformalFactorBC}).  However, the following expression
for $\psi_A$ does satisfy these conditions:
\begin{eqnarray} 
\log\psi_A &=&
-f\left(\frac{x_A-c_A^x}{L}+\frac{1}{2}\right)
\frac{L\,\bar  K_{A-x}(y_A)}{2\,\bar n^x_{A-x}(y_A)}
+f\left(\frac{1}{2}-\frac{x_A-c_A^x}{L}\right)
\frac{L\,\bar K_{A+x}(y_A)}{2\,\bar n^x_{A+x}(y_A)}
\nonumber\\ &&-f\left(\frac{y_A-c_A^y}{L}+\frac{1}{2}\right)
\frac{L\,\bar K_{A-y}(x_A)}{2\,\bar n^y_{A-y}(x_A)}
+f\left(\frac{1}{2}-\frac{y_A-c_A^y}{L}\right)
\frac{L\,\bar K_{A+y}(x_A)}{2\,\bar n^y_{A+y}(x_A)}.
\label{e:ConformalFactor}
\end{eqnarray}
The required properties of the function $f(w)$
are that it has the values $f(0)=f(1)=0$ and the derivatives $f'(0)=1$
and $f'(1)=0$.  The simple choice $f(w)=w\, h(w)$ satisfies these
conditions, with $h(w)$ given in Eq.~(\ref{e:PartitionOfUnityhAlt}).
The expression for the conformal factor in
Eq.~(\ref{e:ConformalFactor}) has the property that $\log\psi_A=0$
everywhere on the boundary of the cubic-block region, while its
derivatives on the boundary satisfy Eq.~(\ref{e:ConformalFactorBC}).
The values of the extrinsic curvatures $\bar K_{A\alpha}$ and the
normal vectors $\bar n^i_{A\alpha}$ used in
Eq.~(\ref{e:ConformalFactor}) are those associated with the
continuous metric $\bar g_{ij}$ given in
Eq.~(\ref{e:CompleteMetricII}).

Continuity of the extrinsic curvature across interface boundaries is
the necessary and sufficient condition for the metric to be $C^1$ and
singularity-free at those interfaces (cf. the Israel junction
conditions~\cite{Israel1966}).  The metrics $\tilde g_{ij}$ defined in
Eq.~(\ref{e:ConformalTrans}), with conformal factor $\psi_A$ given by
Eq.~(\ref{e:ConformalFactor}), will be $C^1$ even across the multicube
interface boundaries, since their extrinsic curvatures vanish and are
continuous there.  The reference metrics $\tilde g_{ij}$ can thus be
used to define a $C^1$ differential structure, which defines the
continuity of tensor fields and their derivatives.  \ref{s:Uniqueness}
shows that this differential structure is unique in the sense that it
is the same as would be produced by any other $C^1$ reference metric
expressed in the same global multicube coordinates.


\subsection{Testing the Reference Metrics}
\label{s:TwoDimensionalReferenceMetricsB}

We have implemented the method outlined in
Secs.~\ref{s:TwoDimensionalReferenceMetricsA0} and
\ref{s:TwoDimensionalReferenceMetricsA1} for constructing a $C^1$
reference metric $\tilde g_{ij}$ in SpEC.  This section describes some
tests we have performed to verify that our code correctly constructs
reference metrics according to these procedures.  We begin by
constructing multicube representations of compact, orientable
two-dimensional manifolds having genera $N_g$ between zero and five.
\ref{s:2dMulticubeManifolds} gives detailed descriptions of these
multicube representations and also shows explicitly how they can be
generalized to compact, orientable two-dimensional manifolds of any
genus $N_g$.  These multicube representations consist of lists of the
regions $\mathcal{B}_A$ and their specific locations in $\Rn$,
together with a complete list of the specific interface boundary
identification maps $\Psi^{A\alpha}_{B\beta}$ that define how the
regions are to be connected together.

Any $C^1$ metric $g_{ij}$, including the reference metric $\tilde
g_{ij}$ from Eq.~(\ref{e:ConformalTrans}), must satisfy the
Gauss-Bonnet identity, which relates the scalar curvature $R$ to the
topology of any compact, orientable two-dimensional Riemannian
manifold:
\begin{eqnarray}
V\,||R ||= 8\pi(1-N_g),
\label{e:GaussBonet}
\end{eqnarray}
where $||R||$ is the spatially averaged scalar curvature,
\begin{eqnarray}
|| R || = \frac{\int R \sqrt{g}\, d^{\,2}x}{V},
\label{e:RAverage}
\end{eqnarray}
$V$ is the volume,
\begin{eqnarray}
V = \int\!\sqrt{g}\, d^{\,2}x,
\label{e:Volume}
\end{eqnarray}
and where $N_g$ is the genus of the manifold.  The Gauss-Bonnet
identity therefore provides a powerful test: The multicube manifold
must have the correct genus or the identity will fail.  And the metric
must be $C^1$ across all the interface boundaries, or curvature
singularities along those boundaries will cause the numerical
integrals used in the the identity to fail.

\begin{figure}[!htb]
\centering
\includegraphics[width=0.55\textwidth]{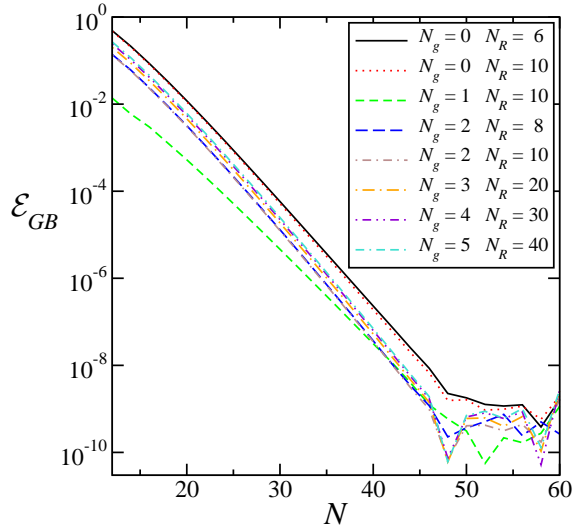}
\caption{\label{f:GBErrVsN}The error in the Gauss-Bonnet identity
  ${\cal E}_{GB}$, defined in Eq.~(\ref{e:e_gb}), as a function of
  resolution for two-dimensional multicube manifolds having different
  genera $N_g$ and different numbers of
  multicube regions $N_R$.}
\end{figure}

We use the quantity ${\cal E}_{GB}$, defined by
\begin{eqnarray}
{\cal E}_{GB} = \frac{\left|V\,|| R|| - 8\pi (1-N_g)\right|}{8\pi(1+N_g)}, 
\label{e:e_gb}
\end{eqnarray}
to monitor how well the Gauss-Bonnet identity is satisfied numerically
in our tests.  Figure~\ref{f:GBErrVsN} shows the values of ${\cal
  E}_{GB}$ computed for each of the multicube manifolds described in
\ref{s:2dMulticubeManifolds} using the $C^1$ reference metric $\tilde
g_{ij}$ defined in Eq.~(\ref{e:ConformalTrans}).  Each curve in
Fig.~\ref{f:GBErrVsN} represents ${\cal E}_{GB}$ for a particular
multicube manifold as a function of the numerical resolution $N$\ (the
number of grid points along each dimension of each multicube region
$\mathcal{B}_A$).  The manifolds are identified in
Fig.~\ref{f:GBErrVsN} by their genera $N_g$ and the numbers of regions
$N_R$ used in their particular representations.  These graphs show
that the Gauss-Bonnet identity is satisfied by the reference metrics
$\tilde g_{ij}$ with numerical errors that decrease exponentially as
the numerical resolution $N$ is increased.  The numerical errors arise
both in the numerical derivatives used in the computation of the
scalar curvature $R$ and in the numerical integrations used to
evaluate $|| R ||$.  A minimum error of $\mathcal{O}(10^{-9})$ is
reached at a resolution of about $N=46$, which corresponds to the
level of accumulated roundoff error in the calculation of ${\cal
  E}_{GB}$ at that resolution.

We have also tested the Gauss-Bonnet identity on this same collection
of multicube manifolds using the scalar curvatures computed from the
continuous reference metrics $\bar g_{ij}$ of
Eq.~(\ref{e:CompleteMetricII}) instead of the $C^1$ metrics $\tilde
g_{ij}$ of Eq.~(\ref{e:ConformalTrans}).  Using these $C^0$ reference
metrics, we find that $\mathcal{E}_{GB}$ is of order unity (with
values between about 0.5 and 2) for all of the tests illustrated in
Fig.~\ref{f:GBErrVsN}.  The Gauss-Bonnet identity fails in this case
because the curvatures associated with the $C^0$ reference metrics
have singularities along the multicube interface boundaries.  This
failure, which was expected in this case, reinforces the conclusion
that we have successfully implemented the procedure outlined in
Secs.~\ref{s:TwoDimensionalReferenceMetricsA0} and
\ref{s:TwoDimensionalReferenceMetricsA1} for constructing $C^1$
reference metrics on two-dimensional manifolds with arbitrary
topologies.

\section{Smoothing the Reference Metrics Using Ricci Flow}
\label{s:RicciFlow}

\begin{figure*}[!tb]
\centering
\subfigure{
  \includegraphics[width=\textwidth]{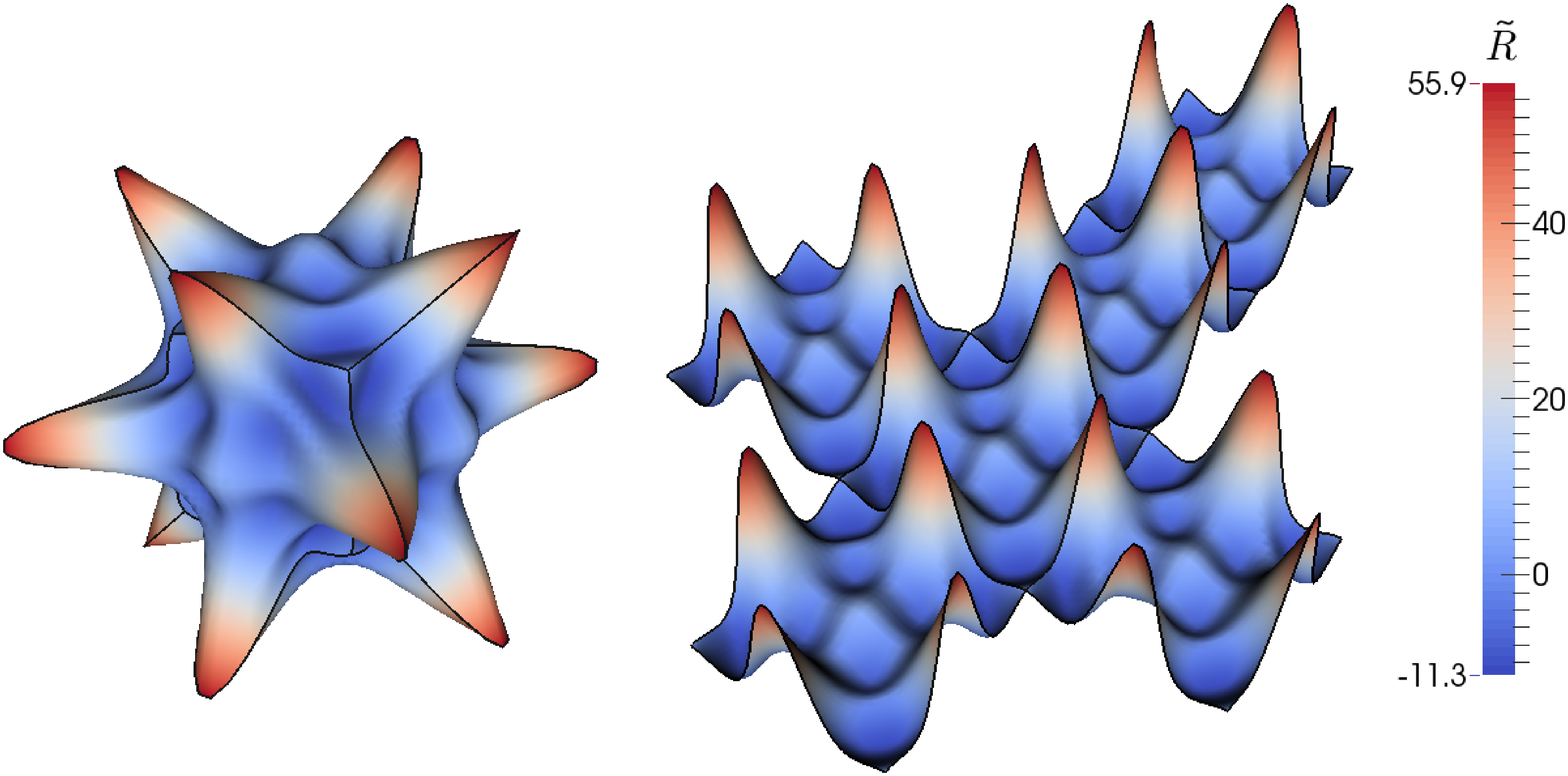}
}
\subfigure{
  \includegraphics[width=\textwidth]{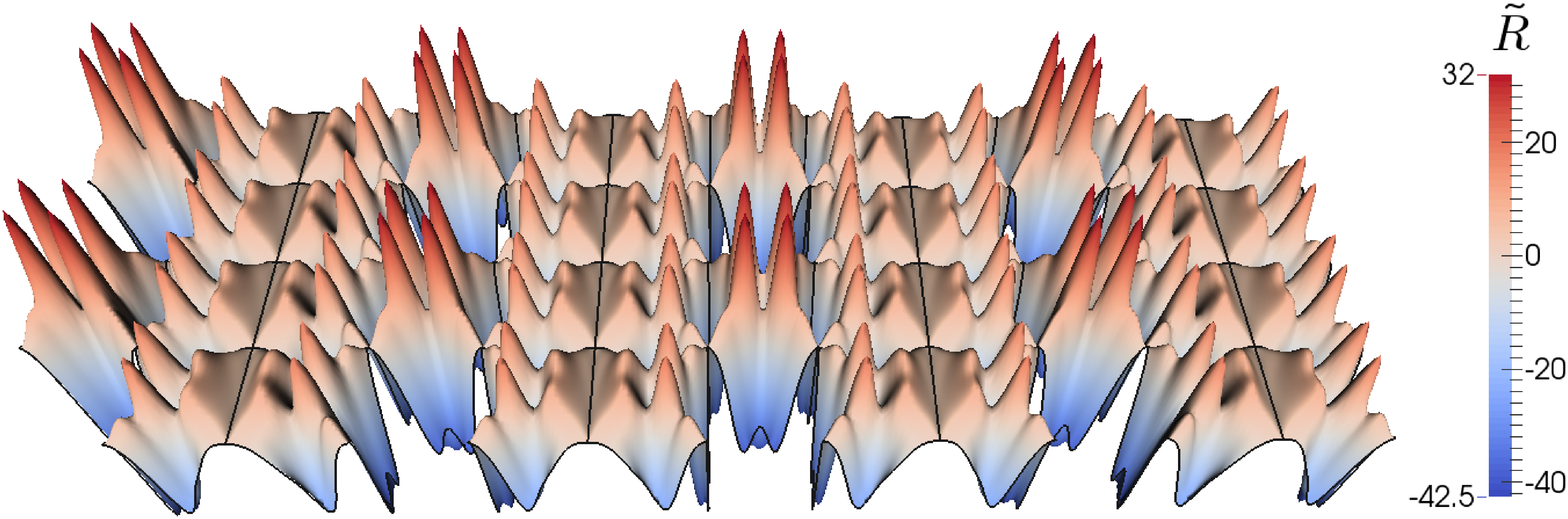}
}
\caption{\label{f:Genus05Paraview} Illustration of the scalar
    curvature $\tilde R$ of two multicube manifolds with $C^1$
  reference metrics $\tilde g_{ij}$ constructed via the procedure
  described in Sec.~\ref{s:TwoDimensionalReferenceMetrics}.  Both
  cases use a numerical resolution of $N=40$ grid points along each
  dimension of each multicube region. \textbf{Top:} The genus $N_g=0$,
  six-region case.  The left side shows the manifold mapped
  (non-isometrically) onto a 2-sphere, with radial warping
  proportional to the scalar curvature $\tilde R$.  The right side
  shows the same manifold in the multicube Cartesian coordinates, with
  warping in the $z$-direction proportional to $\tilde R$.
  \textbf{Bottom:} The genus $N_g=5$, forty-region multicube manifold
  in the multicube Cartesian coordinates, with warping in the
  $z$-direction proportional to the scalar curvature $\tilde R$. }
\end{figure*}

The $C^1$ reference metrics $\tilde g_{ij}$ introduced in
Secs.~\ref{s:TwoDimensionalReferenceMetricsA0} and
\ref{s:TwoDimensionalReferenceMetricsA1} satisfy the minimal
requirements needed to establish low-order differential structures on
two-dimensional manifolds.  These structures allow us to define the
continuity of tensors and their derivatives, which is all that is
required for solving the systems of second-order equations of most
interest in mathematical physics.  Unfortunately these metrics exhibit
a great deal of spatial structure and consequently require fairly high
numerical resolution to be represented accurately.
Figure~\ref{f:Genus05Paraview} illustrates the scalar curvature
$\tilde R$ associated with these reference metrics $\tilde g_{ij}$ for
the case of a six-region, $N_R=6$, representation of the genus $N_g=0$
multicube manifold (the two-sphere), and also for the case of a
forty-region, $N_R=40$, representation of the genus $N_g=5$ multicube
manifold (the five-handled sphere).
While these scalar curvatures appear to be continuous (even across the
region interface boundaries) they have very large spatial variations.
The goal of this section is to develop a method of transforming these
metrics into more uniform (and smoother) reference metrics.

The uniformization theorem implies that every orientable
two-dimensional manifold $\Sigma$ admits a metric having constant
scalar curvature~\cite{Chow2004}.  One approach to making the
reference metrics $\tilde g_{ij}$ more uniform, therefore, would be to
find a way to transform them into metrics having constant scalar
curvatures.  Fortunately there is a well-studied technique for doing
exactly that.  Volume-normalized Ricci flow is a parabolic evolution
equation for the metric whose solutions in two dimensions all evolve
toward metrics having spatially constant scalar
curvatures~\cite{Hamilton1988, Chow1991, Chen2006, Chow2004}.

The evolution equation we use for the volume-normalized Ricci flow of
a two-dimensional metric $g_{ij}$ is given by
\begin{eqnarray}
\partial_t g_{ij} &=& - 2 R_{ij} 
+||R(t)||\,g_{ij}
-\mu\frac{V(t)-V_0}{V(t)}\,g_{ij}
+ \nabla_i H_j + \nabla_j H_i
.
\label{e:RicciFlow}
\end{eqnarray}
The quantities $||R||$ and $V(t)$ in Eq.~(\ref{e:RicciFlow}) are the
volume-averaged scalar curvature and the volume of the manifold
defined in Eqs.~(\ref{e:RAverage}) and (\ref{e:Volume}), respectively.
The terms containing these quantities are added to control the volume
of the manifold.  The term proportional to $\mu$ in
Eq.~(\ref{e:RicciFlow}) is new to the best of our knowledge.  We have
found that it makes our numerical solutions of Eq.~(\ref{e:RicciFlow})
track the target volume $V_0$ more accurately. The DeTurck
gauge-fixing covector $H_i$ is defined by
\begin{eqnarray}
H_i=g_{ij}g^{k\ell}(\Gamma^j_{k\ell}-\tilde \Gamma^j_{k\ell}),
\end{eqnarray}
where $\Gamma^j_{k\ell}$ is the connection associated with the metric
$g_{ij}$, and $\tilde \Gamma^j_{k\ell}$ is any other fixed connection
on the manifold \cite{DeTurck1983}.  The DeTurck terms (those
containing $H_i$) are added to make Eq.~(\ref{e:RicciFlow}) strongly
parabolic, and thus to have a manifestly well-posed initial value
problem~\cite{Garfinkle2008}.

Contracting Eq.~(\ref{e:RicciFlow}) with the inverse metric $g^{ij}$
gives 
\begin{eqnarray}
\partial_t \log\sqrt{g} = - R+||R||
-\mu\frac{V(t)-V_0}{V(t)}+\nabla_iH^i.
\end{eqnarray}
Integrating this equation over any compact manifold provides the
evolution equation for the volume $V(t)$ of the manifold:
\begin{eqnarray}
\partial_t\left[V(t)-V_0\right]=-\mu\left[V(t)-V_0\right].
\end{eqnarray}
Without the term proportional to $\mu$, the volume of the manifold
would be fixed, $\partial_t V(t) = 0$, at the analytical level.  In
numerical simulations, however, discretization and roundoff error give
rise to slow, approximately linear drifts in the volume.  With
the damping term we have added, the volume of the manifold is driven
toward the target value $V_0$ at a rate determined by the constant
$\mu$.  In our numerical tests, we find that a value of $\mu=10$ works
well.

\subsection{Numerical Ricci Flow}
\label{s:NumericalRicciFlow}

We have implemented the volume-normalized Ricci flow equation with
DeTurck gauge fixing, Eq.~(\ref{e:RicciFlow}), in SpEC.  This code
evolves PDEs using pseudo-spectral methods to evaluate spatial
derivatives, and it performs explicit time integration at each
collocation point using standard ordinary differential equation
solvers (e.g., Runge-Kutta).  Boundary conditions are imposed at
multicube interface boundaries to enforce continuity of the metric
$g_{ij}$ and its normal derivative $\tilde n^k \tilde \nabla_k
g_{ij}$. The vector $\tilde n^k$ is the unit normal to the boundary
and $\tilde \nabla_k$ is the covariant derivative associated with the
reference metric $\tilde g_{ij}$.

Boundary conditions are imposed in SpEC using penalty methods.  The
desired boundary conditions are added to the evolution equations at
the boundary collocation points.  The evolution equations on the
$\partial_\alpha\mathcal{B}_A$ boundary, which is identified with the
$\partial_\beta\mathcal{B}_B$ boundary, for example, have the form
\begin{equation}
\partial_t g_{ij} = F_{ij} + \alpha\,\Bigl(g^A_{ij} - \langle
g^B_{ij}\rangle_A\Bigr) 
+\beta\, \tilde n_A^k \Bigl(\tilde \nabla_k g^A_{ij} -\langle \tilde \nabla_k
g^B_{ij}\rangle_A\Bigr),
\label{e:NumericalRicciFlow}
\end{equation}
where $F_{ij}$ represents the right side of Eq.~(\ref{e:RicciFlow}),
and $\alpha$ and $\beta$ are positive constant penalty factors.
The quantities $\langle g^B_{ij}\rangle_A$ and $\langle \tilde
\nabla_k g^B_{ij}\rangle_A$ represent the transformations of
$g^B_{ij}$ and $\tilde\nabla_k g^B_{ij}$ into the tensor basis of
region $\mathcal{B}_A$ using the interface boundary Jacobians:
\begin{eqnarray}
\langle g^B_{ij}\rangle_A &=& 
J^{*B\beta a}_{A\alpha i}J^{*B\beta b}_{A\alpha j}
g^B_{ab},\\
\langle \tilde \nabla_k g^B_{ij}\rangle_A 
&=& J^{*B\beta a}_{A\alpha i}J^{*B\beta b}_{A\alpha j}J^{*B\beta c}_{A\alpha k}
\,\tilde\nabla_c g^B_{ab}.
\end{eqnarray}
If the penalty factors $\alpha$ and $\beta$ are chosen properly, these
additional terms drive the evolution at the boundary in a way that
reduces any small boundary condition error~\cite{Hesthaven1996A}.
There is a range of constants $\alpha$ and $\beta$ that work
well---too small can lead to instability, while too large may make the
system overly stiff.  Empirically, we have found that the following
values work well in most cases:\footnote{We use the factor $N+1$ in
  Eq.~(\ref{e:PenaltyFactors}), instead of the simpler $N$, because
  it is natural to write $\alpha$ and $\beta$ as multiples of the
  inverse of the Legendre quadrature weight at the endpoints, $\omega
  = 2/N(N+1)$, since $\omega$
  enters the proofs of stability for these penalty methods.  In terms
  of $\omega$, we use $\alpha = N^2/\omega$ and
  $\beta = 1/\omega$. }
\begin{eqnarray}
\alpha = \tfrac{1}{2}N^3(N+1), \quad \beta = \tfrac{1}{2}N(N+1).
\label{e:PenaltyFactors}
\end{eqnarray}
In some cases the penalty factors (particularly $\alpha$) can be
decreased below the values given in Eq.~(\ref{e:PenaltyFactors})
without sacrificing stability.  Using smaller values allows a less
restrictive condition on the size of the maximum time step and
therefore allows more efficient numerical evolutions.  In rare cases,
we have found it necessary to increase $\beta$ above the value given
in Eq.~(\ref{e:PenaltyFactors}).  For example, in the low-resolution
$N=16$, ten-region, $N_R=10$, genus $N_g=0$ case, a
value of $\beta$ at least twice that given in
Eq.~(\ref{e:PenaltyFactors}) was needed for stability.  Hesthaven and
Gottlieb~\cite{Hesthaven1996A} have derived rigorous lower bounds on
the penalty factors needed for stable evolution of a simple,
second-order parabolic equation in one dimension.  They show that when
Robin-type boundary conditions are used (like those we use here),
penalty factors that scale like $\alpha\sim\mathcal{O}(N^2)$ and
$\beta \sim \mathcal{O}(N^2)$ are required.  Our results agree with
theirs for $\beta$, but we have found it necessary to use much larger
values of $\alpha$ that scale as $\alpha\sim\mathcal{O}(N^4)$ in most
cases.

We test the stability and robustness of our implementation of these
Ricci flow evolution equations on a six-region, $N_R=6$, multicube
representation of the two-sphere manifold, $S^2$, which is described
in detail in \ref{s:GenusZeroRegionSix}.  As initial data for these
tests we use the standard round-sphere metric with pseudo-random white
noise of amplitude $0.1$ added to each component of the metric
$g_{ij}$ at each collocation point.  The reference metric $\tilde
g_{ij}$ used in these tests is the usual smooth, unperturbed
round-sphere metric, which is given explicitly in global Cartesian
multicube coordinates in Ref.~\cite{Lindblom2013}.

We use several measures to determine whether our implementation of
numerical Ricci flow is working properly and whether it actually
drives the metric toward a constant-curvature state, as it is expected
to do in two dimensions.  First, we measure how well the numerical
Ricci flow evolves toward geometries having uniform scalar curvatures.
One possible dimensionless measure of this scalar-curvature uniformity
is the quantity $\tilde {\cal E}_R$, defined by
\begin{eqnarray}
\tilde {\cal E}_R^2 = 
\frac{\int (R - ||R||)^2 \sqrt{g}\, d^{\,2}x}{V||R||^2}. 
\end{eqnarray}
For the two-dimensional manifolds studied here, the volume-averaged
scalar curvature $||R||$ is given by the Gauss-Bonnet identity:
$||R||=8\pi(1-N_g)/V$.  The scalar-curvature uniformity measure can
therefore be rewritten in the form
\begin{eqnarray}
\tilde {\cal E}_R^2 = 
\frac{V\int (R - ||R||)^2 \sqrt{g}\, d^{\,2}x}{[8\pi(1-N_g)]^2}. 
\end{eqnarray}
This measure is singular for $N_g=1$, so we define an alternative
measure ${\cal E}_R$ as follows:
\begin{eqnarray}
{\cal E}_R^2 
= \frac{V\int (R - ||R||)^2 \sqrt{g}\, d^{\,2}x}{[8\pi(1+N_g)]^2}. 
\label{e:e_r}
\end{eqnarray}
This alternative measure is well defined for all compact, orientable
two-dimensional manifolds.  It differs from $\tilde {\cal E}_R$ by the factor
$|1-N_g|/(1+N_g)$, which is of order unity, except for the singular
case $N_g=1$.  We use the measure ${\cal E}_R$ to monitor the
uniformity of the scalar curvature in all of our Ricci flow
evolutions.  Second, we monitor the volume of the manifold to
determine whether the volume-normalized flow is working properly.  We
do this using the dimensionless quantity ${\cal E}_V$, defined by
\begin{eqnarray}
{\cal E}_V = \frac{|V(t)-V_0|}{V_0}, \label{e:e_v}
\end{eqnarray}
to measure the fractional change in the volume relative to the target
volume $V_0$.  Third, we use the quantity ${\cal E}_H$ to measure the
evolution of the DeTurck gauge-source covector:
\begin{eqnarray}
\label{e:e_h}
{\cal E}_H^2=\frac{\int g^{ij}H_iH_j\sqrt{g}\, d^{\,2} x}{
  \int\sum_{ij}\left(|g_{ij}|^2+\sum_{k}|\partial_kg_{ij}|^2
  \right)\sqrt{g}\, d^{\,2} x}.
\end{eqnarray}
And finally, we assess how well the geometries produced by this Ricci
flow satisfy the Gauss-Bonnet identity, using the quantity
$\mathcal{E}_{GB}$ defined in Eq.~(\ref{e:e_gb}).

\begin{figure}[!htb]
\centering 
\subfigure{
  \includegraphics[width=0.48\textwidth]{%
    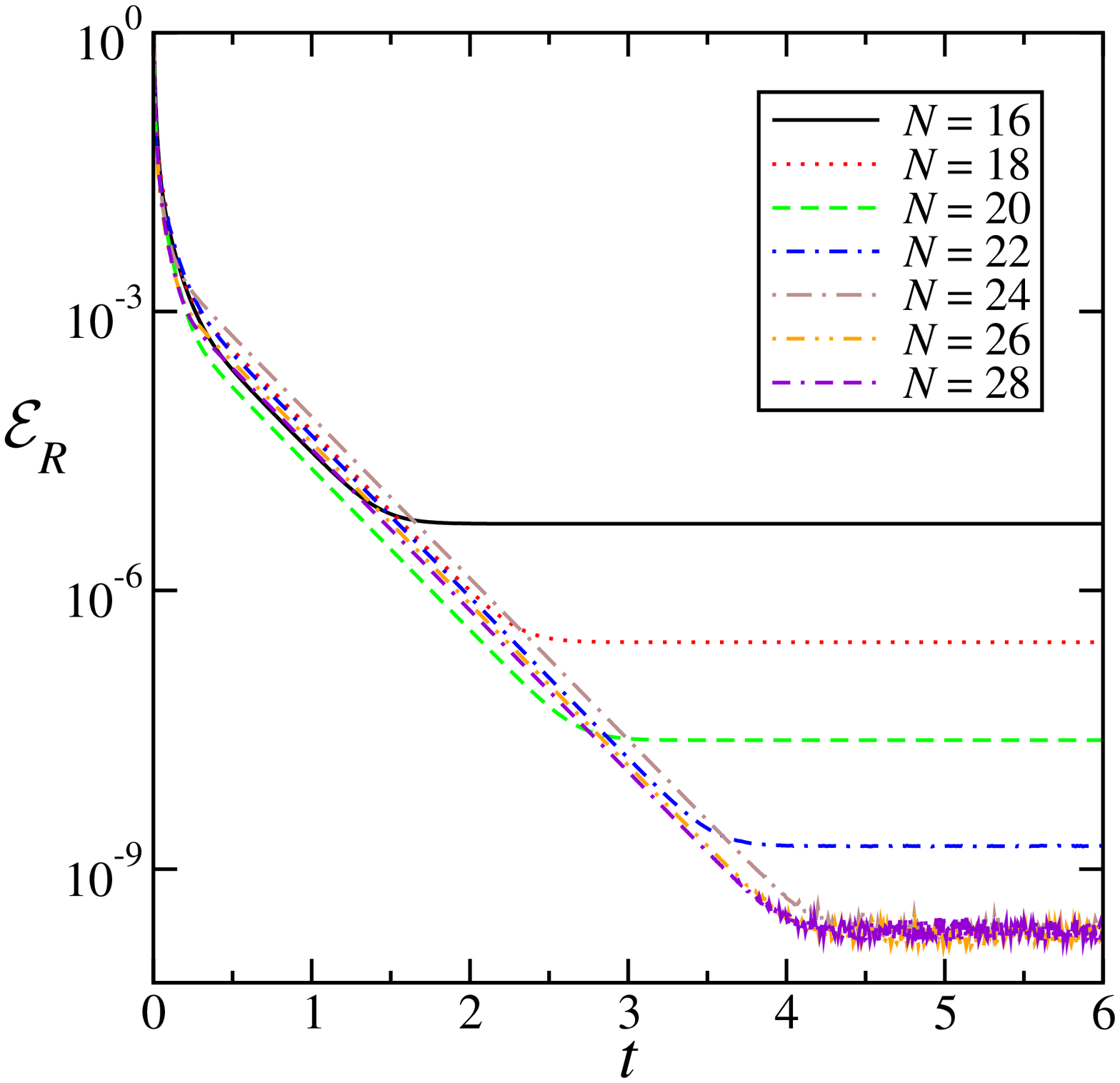}
} 
\subfigure{
  \includegraphics[width=0.48\textwidth]{%
    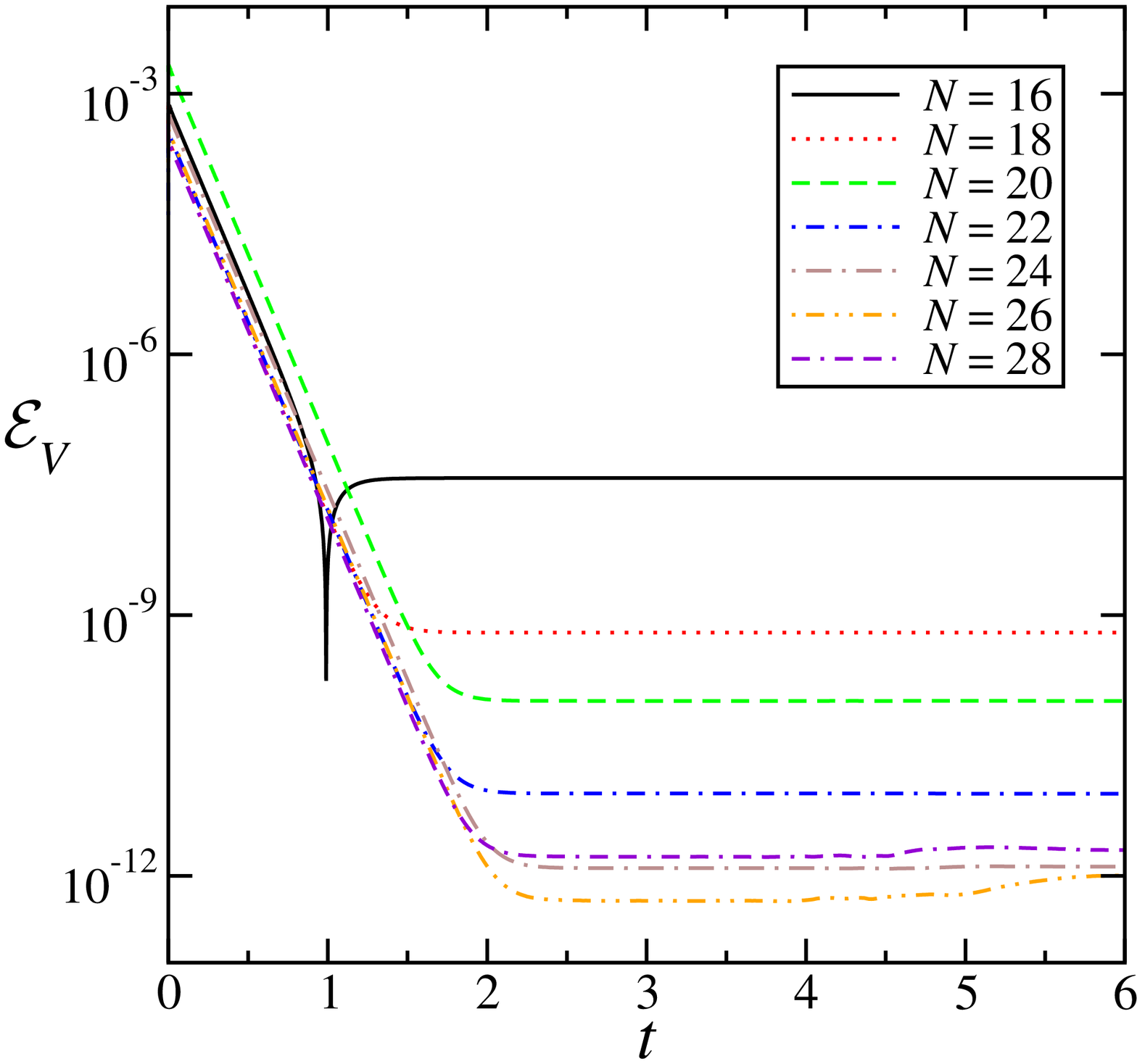}
} 
\subfigure{
  \includegraphics[width=0.48\textwidth]{%
    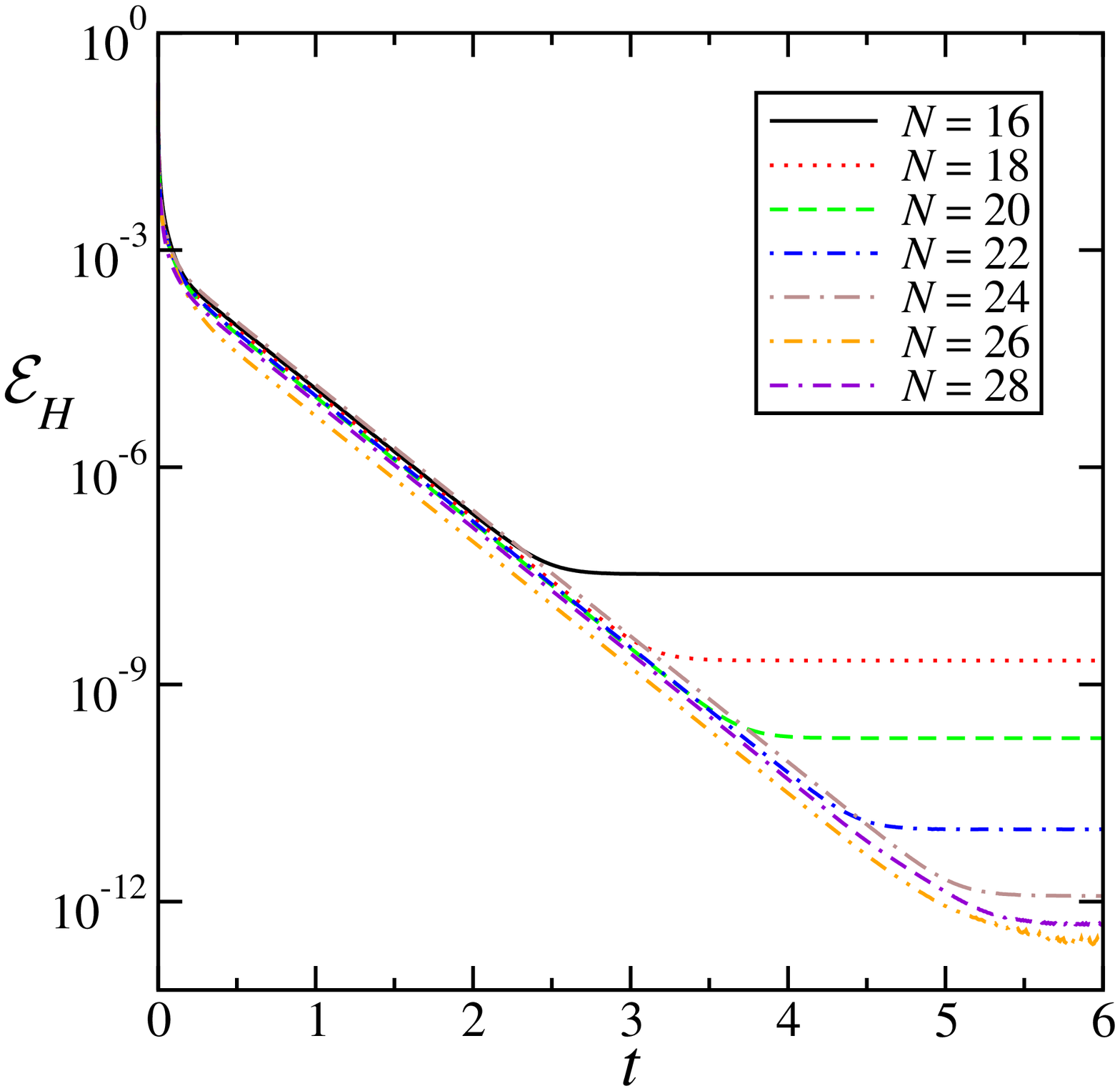}
}
\subfigure{ 
  \includegraphics[width=0.48\textwidth]{%
    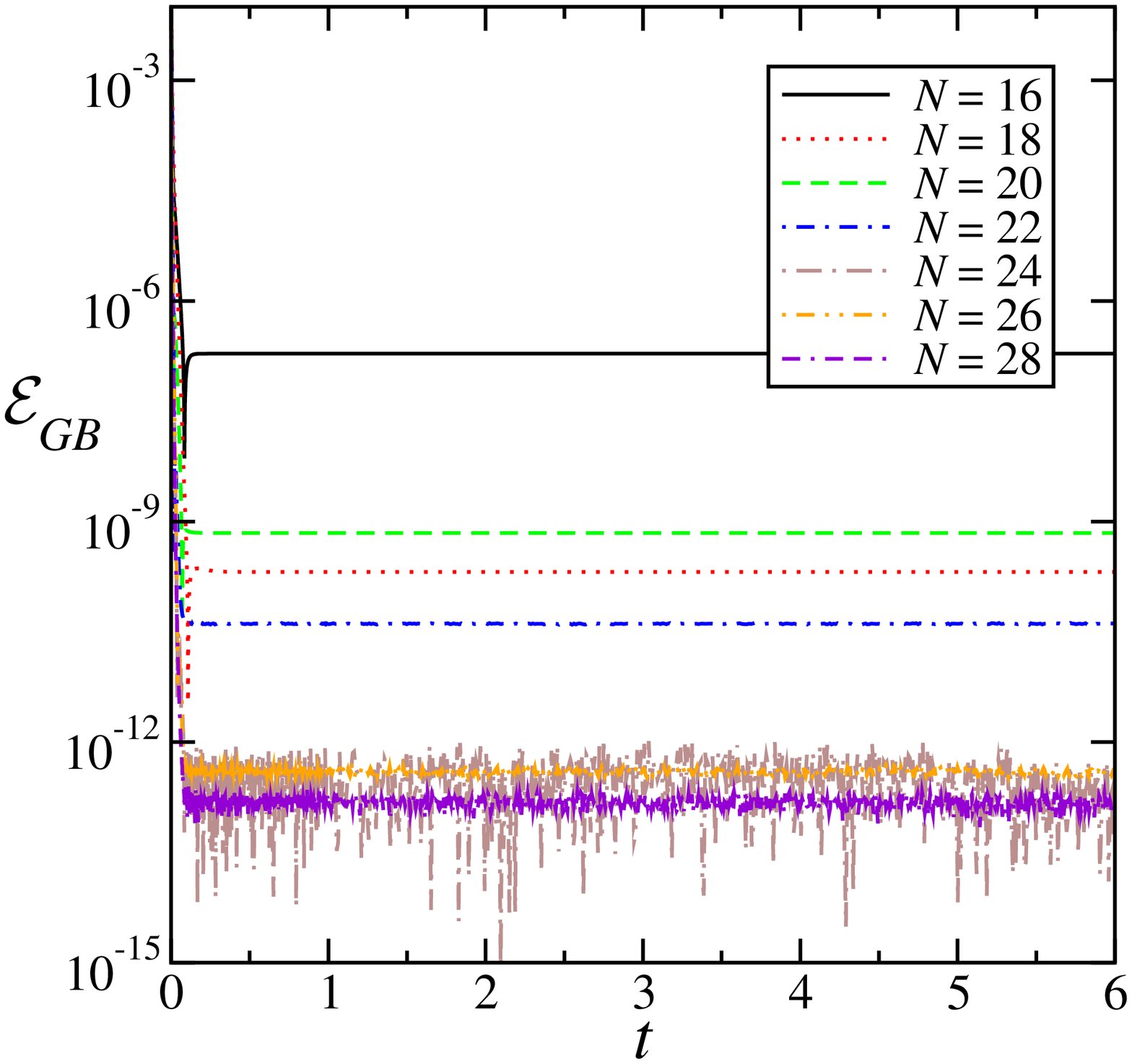}
} 
\caption{\label{f:RoundS2NoiseResults} Ricci flow evolutions of a
  six-region, $N_R=6$, multicube representation of the 2-sphere, with
  random noise added to the round-sphere metric as the initial data.
  Graphs show the evolutions of the scalar-curvature uniformity
  measure $\mathcal{E}_R$, the volume-normalization error
  $\mathcal{E}_V$, the DeTurck gauge-covector norm $\mathcal{E}_H$,
  and the Gauss-Bonnet identity error $\mathcal{E}_{GB}$.  These
  quantities are defined in Eqs.~(\ref{e:e_r}), (\ref{e:e_v}),
  (\ref{e:e_h}), and (\ref{e:e_gb}), respectively.  The reference
  metric used in these tests is the usual unperturbed round-sphere
  metric. The numerical resolution in each spatial dimension of each
  square region is denoted by $N$. }
\end{figure}

Figure~\ref{f:RoundS2NoiseResults} shows the results of our Ricci flow
evolutions using initial data constructed from the round-sphere metric
with random noise perturbations.  This figure plots the time
evolutions of the four error measures $\mathcal{E}_R$,
$\mathcal{E}_V$, $\mathcal{E}_H$, and $\mathcal{E}_{GB}$, defined in
Eqs.~(\ref{e:e_r}), (\ref{e:e_v}), (\ref{e:e_h}), and (\ref{e:e_gb}),
respectively, for evolutions performed with several different
numerical resolutions $N$.  As evidenced in these figures, the Ricci
flow evolutions are stable and convergent as the numerical resolution
$N$ is increased.  Nonuniformities in the random initial scalar
curvature, as measured by ${\cal E}_R$ and shown in the upper left
part of Fig.~\ref{f:RoundS2NoiseResults}, decay exponentially in
time as the geometry evolves toward the constant-curvature
round-sphere metric until the differences are dominated by truncation
level errors at each resolution.  The upper right
part of Fig.~\ref{f:RoundS2NoiseResults} shows that the
volume-controlling terms in Eq.~(\ref{e:RicciFlow}) are effective at
driving the volume of the manifold to the value $V_0$, as measured by
$\mathcal{E}_V$.  The target volume $V_0$ in these tests was taken to
be the volume measured by the smooth round-sphere reference metric,
rather than the volume of the initial random metric.  The lower left
part of Fig.~\ref{f:RoundS2NoiseResults} shows that the
gauge source one-form $H_i$, measured by ${\cal E}_H$, is effectively
driven to zero by the DeTurck term, and the lower right
part of Fig.~\ref{f:RoundS2NoiseResults} shows that the
Gauss-Bonnet error ${\cal E}_{GB}$ decays very quickly to truncation
level at each resolution.  Random noise was added to the initial data
in these tests at each grid point, so the precise structure of the
initial data is different at each resolution.  Therefore, numerical
convergence with increasing resolution $N$ at the initial and very
early times was not expected (or observed).

\subsection{Smoother Reference Metrics}
\label{s:SmootherReferenceMetrics}

We have used volume-normalized Ricci flow to construct smoother and
more uniform reference metrics for several multicube manifolds in two
dimensions.  In particular we have performed Ricci-flow smoothing of
the reference metrics for multicube representations of compact,
orientable two-dimensional manifolds with genera between $N_g=0$ (the
two-sphere) and $N_g=5$ (the five-handled two-sphere).  In each case,
initial data for the evolution are prepared by constructing the metric
$\tilde g_{ij}$ according to the procedure described in
Sec.~\ref{s:TwoDimensionalReferenceMetrics}.  These $\tilde g_{ij}$
use the polynomial generating functions $h(w)$ of
Eq.~(\ref{e:PartitionOfUnityhAlt}), with $k=1$ and $\ell=4$, both for
the partition of unity and for the functions $f(w)=w\,h(w)$ that
appear in the conformal factor in Eq.~(\ref{e:ConformalFactor}).
Although this choice of powers appears to give the best results, we
have found that other choices often work nearly as well.  We use the
metric $\tilde g_{ij}$ not only as initial data for these Ricci flow
evolutions, but also as the fixed reference metric, which defines the
continuity of all tensor fields and their derivatives throughout
the evolutions, including the Ricci-flow-evolved $g_{ij}(t)$.

We have performed Ricci flow evolutions on all the multicube manifolds
described in \ref{s:2dMulticubeManifolds}, and the results look
very similar to one another.  For this reason we describe only one of
these cases in detail, and then we summarize and compare the results
of our highest-resolution evolutions from all of the
cases.  We show detailed results
for our most complex case: a forty-region, $N_R=40$,
representation of a genus $N_g=5$ multicube
manifold (the five-handled two-sphere).  The scalar curvature for the
reference metric $\tilde g_{ij}$ in this case is illustrated in the
bottom part of Fig.~\ref{f:Genus05Paraview}.  The details of the
multicube structure for this case (and all our other cases) are given
in~\ref{s:2dMulticubeManifolds}.

\begin{figure}[!tb]
\centering 
\subfigure {
  \includegraphics[width=0.48\textwidth]{%
    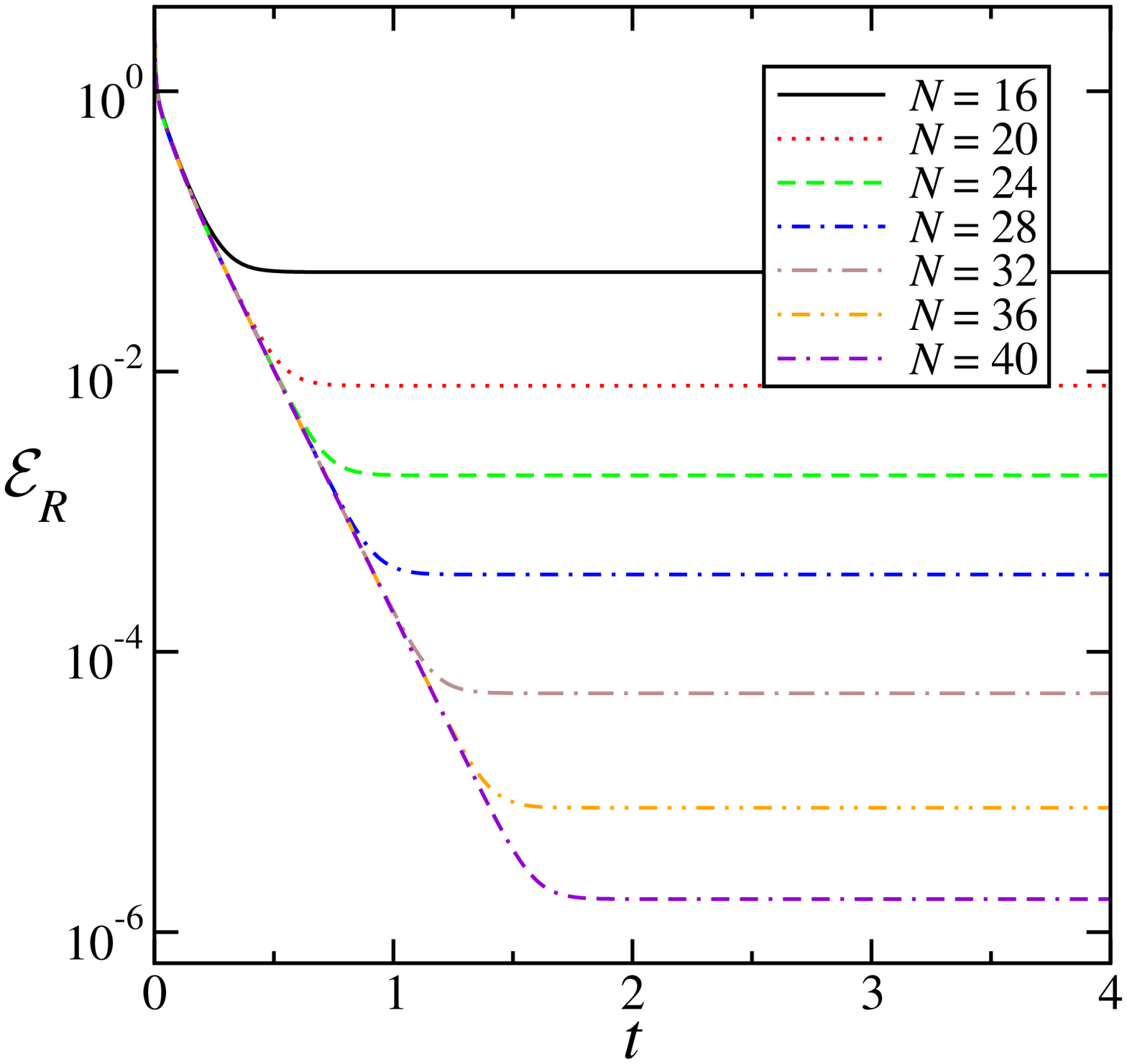}
} 
\subfigure {
  \includegraphics[width=0.48\textwidth]{%
    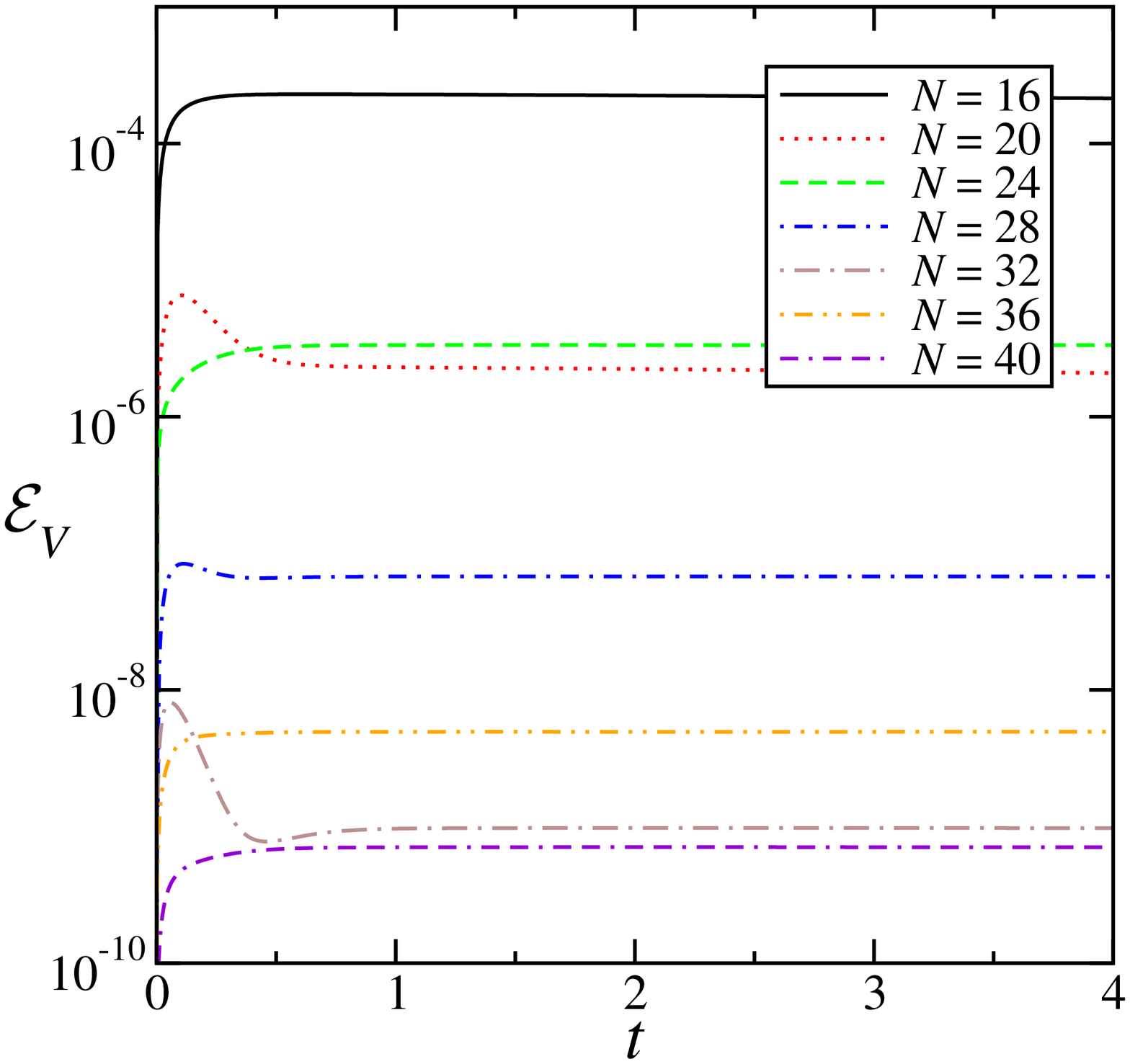}
} 
\subfigure {
  \includegraphics[width=0.48\textwidth]{%
    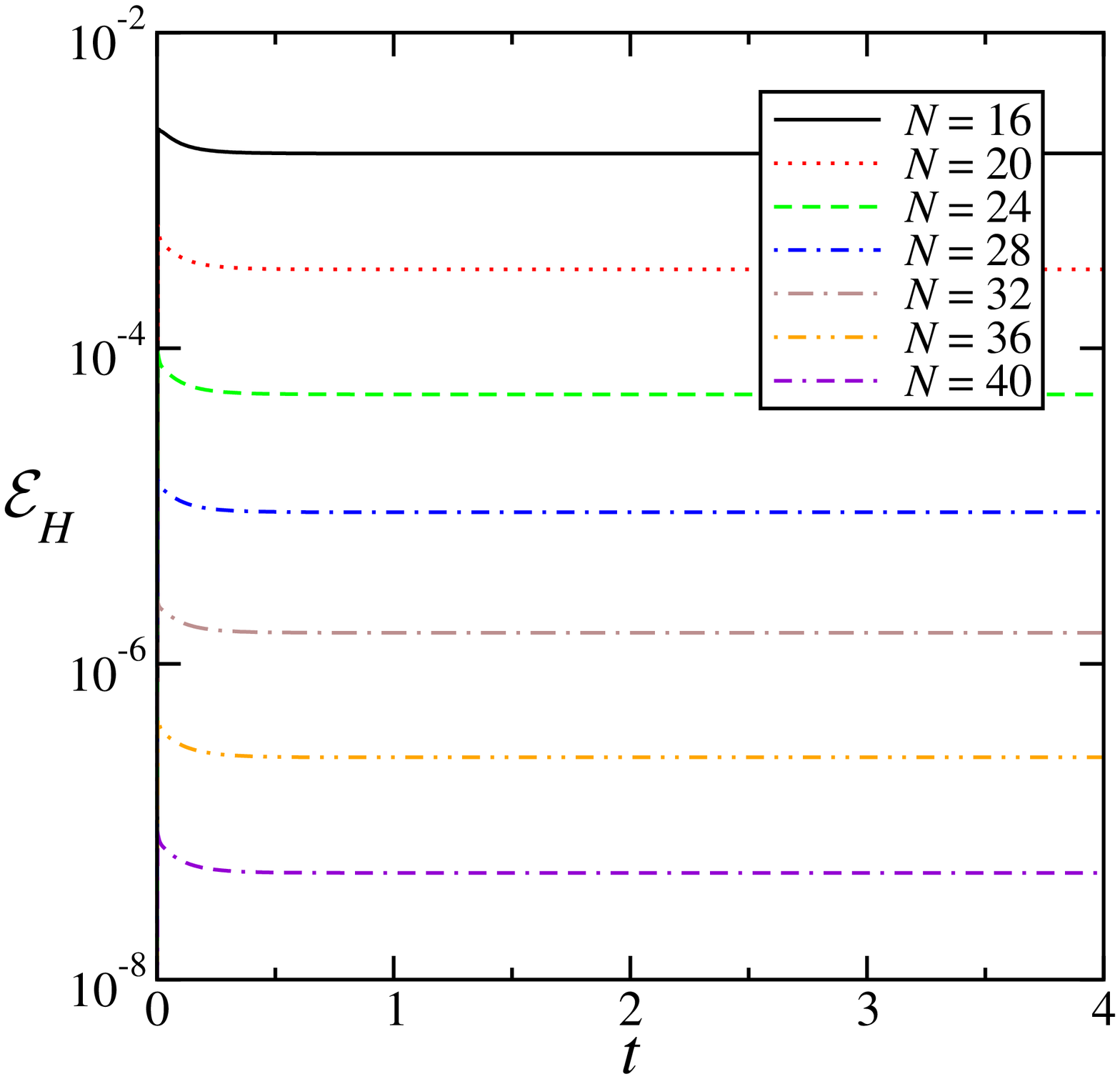}
}
\subfigure {
  \includegraphics[width=0.48\textwidth]{%
    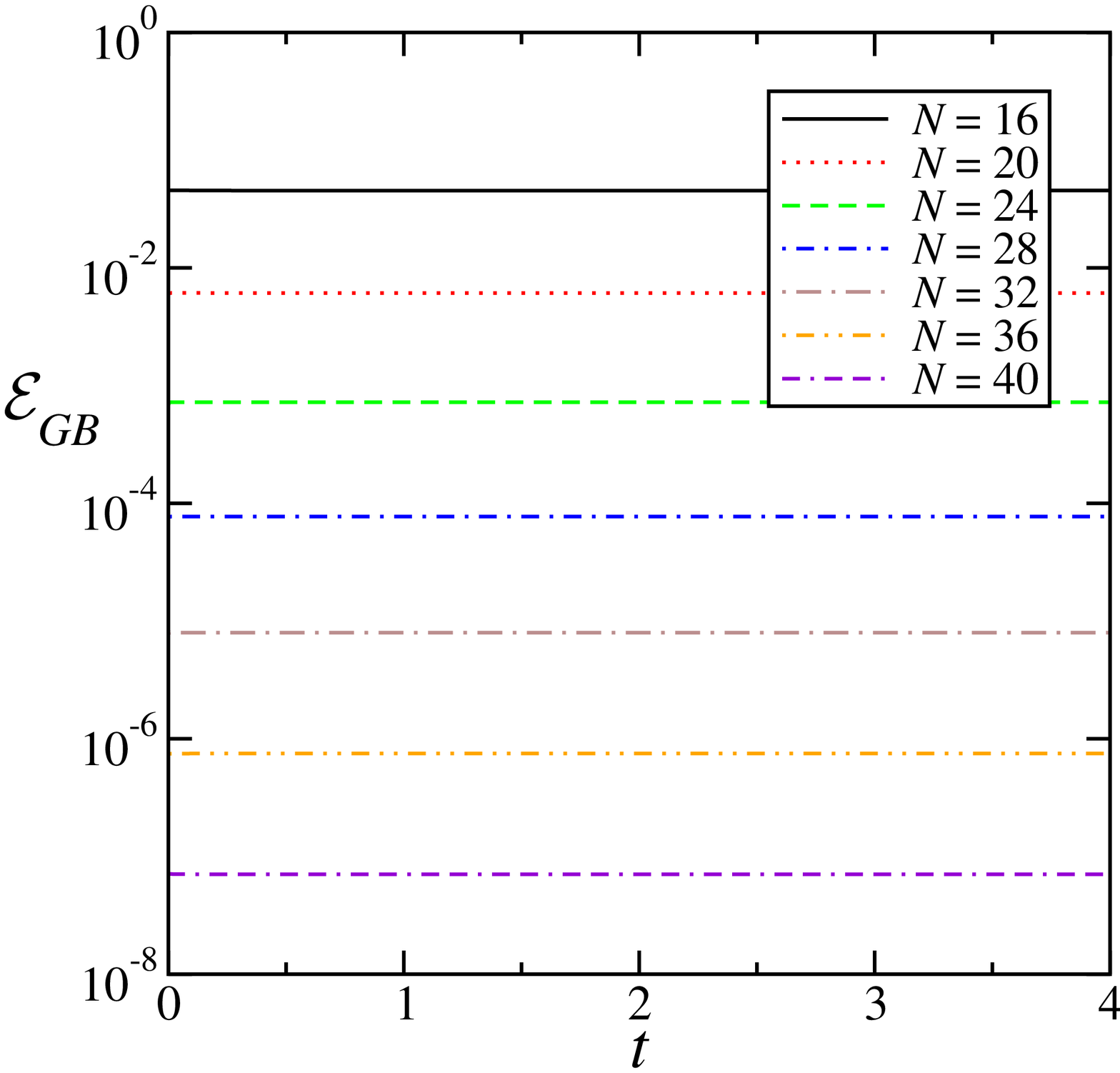}
} 
\caption{\label{f:Genus5Results} Ricci flow evolutions of a genus
  $N_g=5$, forty-region, $N_R=40$, multicube manifold.  Graphs show
  the evolutions of the scalar-curvature uniformity measure
  $\mathcal{E}_R$, the volume-normalization error $\mathcal{E}_V$, the
  DeTurck gauge-covector norm $\mathcal{E}_H$, and the Gauss-Bonnet
  identity error $\mathcal{E}_{GB}$.  These quantities are defined in
  Eqs.~(\ref{e:e_r}), (\ref{e:e_v}), (\ref{e:e_h}), and
  (\ref{e:e_gb}), respectively.  The reference metric, which is
  identical to the initial metric in this case, is constructed
  according to the procedure described in
  Sec.~\ref{s:TwoDimensionalReferenceMetrics}.  The numerical
  resolution in each spatial dimension of each multicube region is
  denoted by $N$. }
\end{figure}

Figure~\ref{f:Genus5Results} shows the results of these genus
$N_g=5$ evolutions for several different numerical
resolutions $N$.  
The graphs in Fig.~\ref{f:Genus5Results} indicate that the evolutions
are stable and convergent, demonstrating our ability to evolve PDEs on
arbitrary, complicated two-dimensional manifolds using the $C^1$
reference metrics developed in
Sec.~\ref{s:TwoDimensionalReferenceMetrics}.  These evolutions differ
from the random-metric evolutions shown in
Fig.~\ref{f:RoundS2NoiseResults} in several ways.  First, these
initial data are much smoother than the random metrics (which are
unresolved by construction).  Consequently, the Gauss-Bonnet error
$\mathcal{E}_{GB}$ is much smaller at early times.  Second, the
initial metric in these tests is identical to the reference metric,
and accordingly the error measures $\mathcal{E}_V$ and $\mathcal{E}_H$
are much smaller (about truncation level) at early times.  These error
measures remain close to these initial truncation-error levels
throughout the evolutions.  We also note that the more complicated
spatial structures of the reference metrics in these simulations
require somewhat higher numerical resolutions in order to obtain the
same level of truncation errors as the random-metric $S^2$ tests
described in Sec.~\ref{s:NumericalRicciFlow}.

\begin{figure}[!tb]
\centering 
\subfigure{
\includegraphics[width=0.48\textwidth]{%
    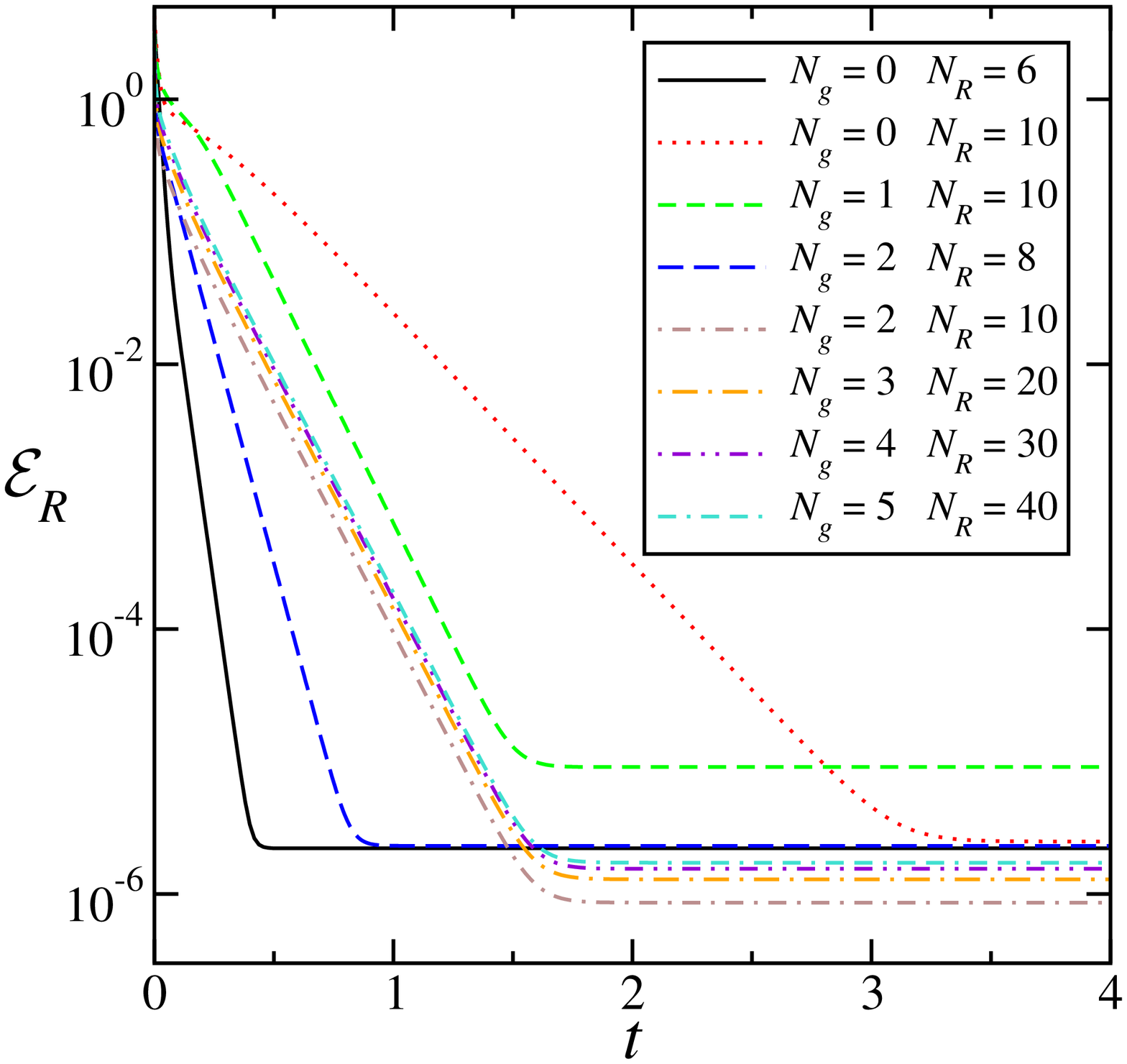}
} 
\subfigure{
  \includegraphics[width=0.48\textwidth]{%
    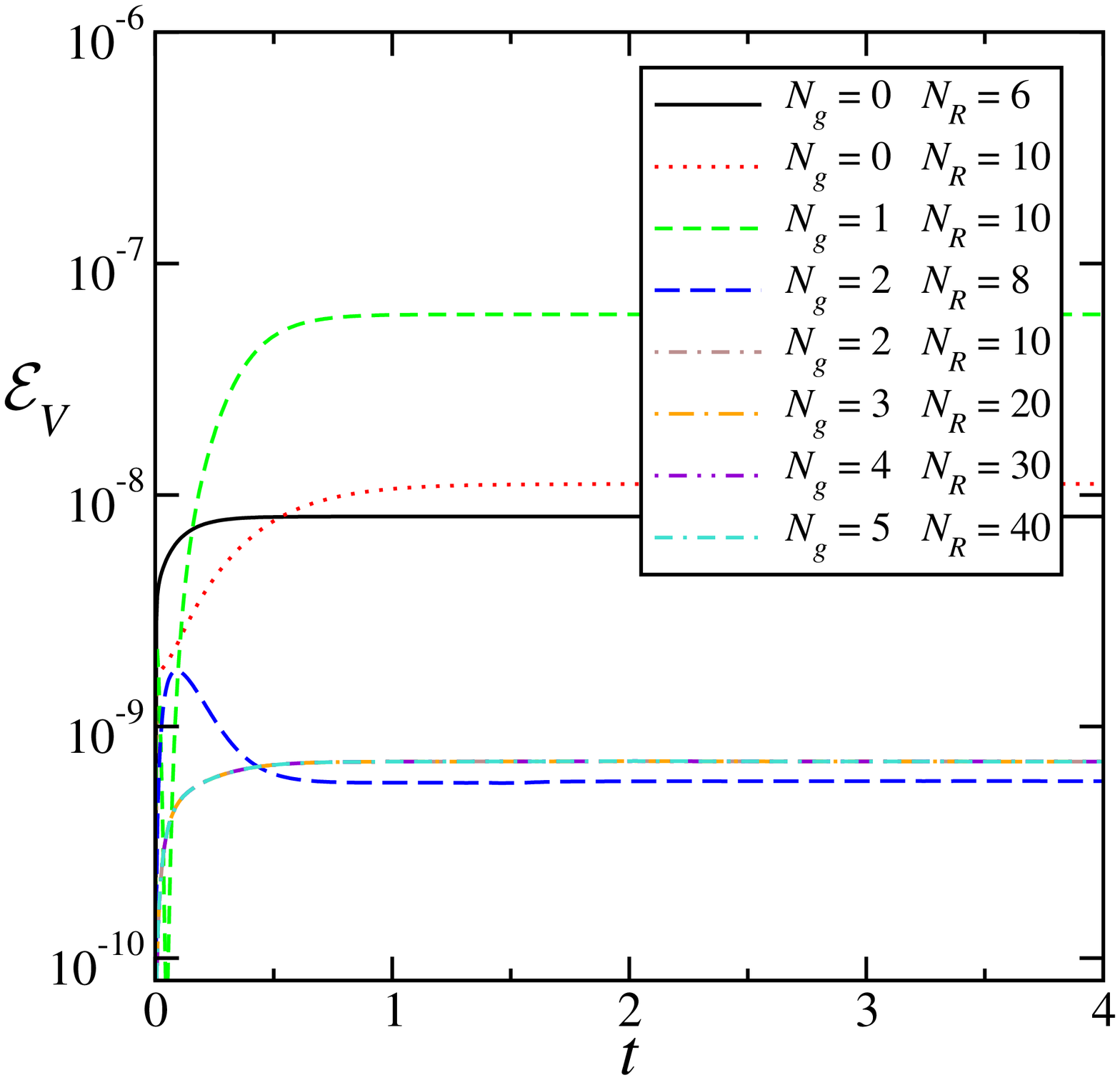}
} 
\subfigure{
  \includegraphics[width=0.48\textwidth]{%
    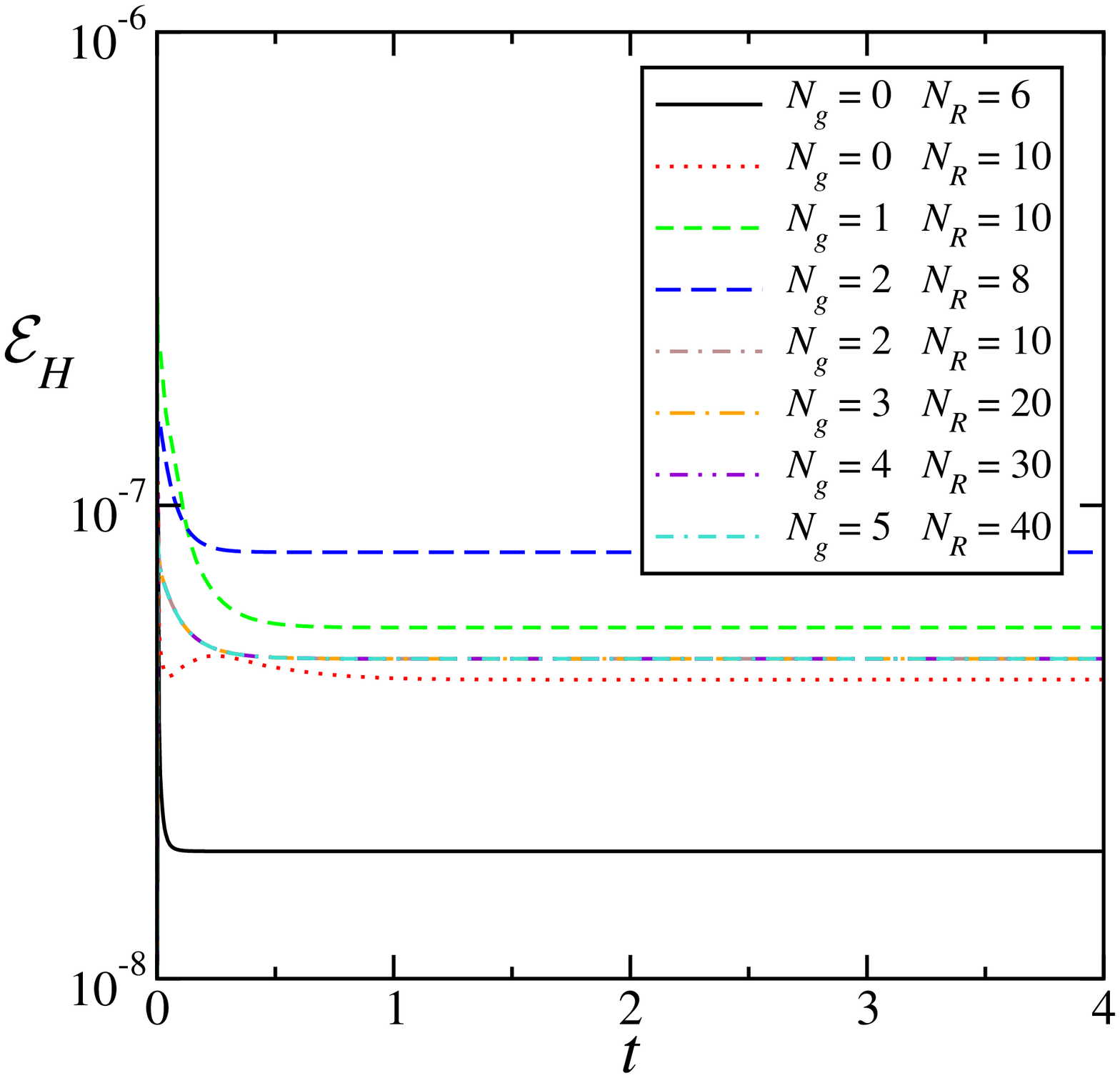}
}
\subfigure{
  \includegraphics[width=0.48\textwidth]{%
    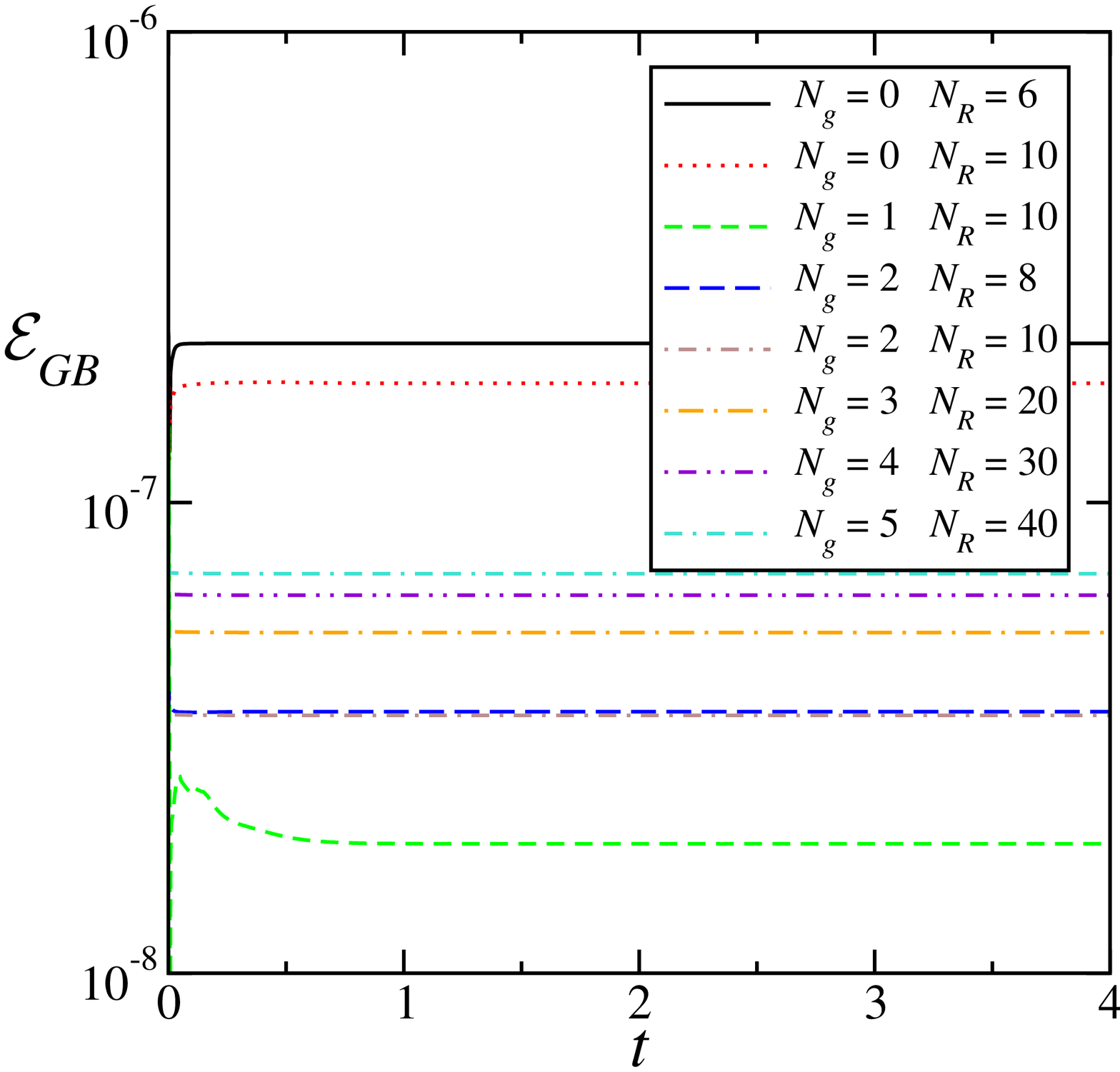}
} 
\caption{\label{f:AllCasesResults} High-resolution ($N=40$) results of
  Ricci flow evolutions on a variety of different multicube manifolds.
  The genus $N_g$ and the number of multicube regions $N_R$ of each
  case are indicated in the legends.  Graphs show the evolutions of
  the scalar-curvature uniformity measure $\mathcal{E}_R$, the
  volume-normalization error $\mathcal{E}_V$, the DeTurck
  gauge-covector norm $\mathcal{E}_H$, and the Gauss-Bonnet identity
  error $\mathcal{E}_{GB}$.  These quantities are defined in
  Eqs.~(\ref{e:e_r}), (\ref{e:e_v}), (\ref{e:e_h}), and
  (\ref{e:e_gb}), respectively.  In each case, the reference metric is
  identical to the initial metric and is constructed according to the
  procedure described in Sec.~\ref{s:TwoDimensionalReferenceMetrics}.
    }
\end{figure}

Figure~\ref{f:AllCasesResults} compares the highest-resolution Ricci
flow evolutions from each of the multicube manifolds described in
\ref{s:2dMulticubeManifolds} (up to and including the forty-region
representation of a genus 5 manifold).  
All of these cases are found to be stable and convergent, with
qualitatively similar results to the genus $N_g=5$ evolutions shown in
Fig.~\ref{f:Genus5Results}.  The only significant difference between
the cases is the rate at which nonuniformities in the scalar
curvatures decay.  The reference metrics that we construct on these
different multicube manifolds have nonuniformities on different length
scales, and these nonuniformities correspondingly decay at different
rates under the Ricci flow.  There are also differences in the levels
of the truncation errors for these cases at the same numerical
resolution.  The ten-region, $N_R=10$, representation of the genus
$N_g=1$ multicube manifold (the two-torus), for example, has the
highest level of truncation error among the examples we have studied.

\section{Discussion}
\label{s:Discussion}

This paper presents a method for constructing reference metrics on
multicube representations of manifolds having arbitrary
topologies. The method was implemented and successfully tested, as
described in Sec.~\ref{s:TwoDimensionalReferenceMetrics}, for a
variety of compact, orientable two-dimensional Riemannian manifolds
with genera between $0$ and $5$.  The reference metrics constructed in
this way are not smooth, but they have continuous derivatives, which
is sufficient to define the $C^1$ differential structures needed for
solving the systems of second-order PDEs of most interest in
mathematical physics.  We have demonstrated in Sec.~\ref{s:RicciFlow},
for example, that these $C^1$ reference metrics can be used
successfully to solve systems of second-order parabolic evolution
equations.

The reference metrics constructed using the methods in
Sec.~\ref{s:TwoDimensionalReferenceMetrics} have large spatial
variations, which are not easy to resolve numerically.  We demonstrate
in Sec.~\ref{s:RicciFlow} that these metrics can be made more uniform
by evolving them with Ricci flow.  The two-dimensional reference
metrics studied in our tests all evolve under Ricci flow to metrics
having constant scalar curvatures.

Ricci flow also has smoothing properties similar to the heat equation:
solutions to the Ricci flow equation on compact manifolds become
smooth, in fact real-analytic, for $t>0$ provided the initial
curvature is bounded (which is the case for our $C^1$ reference
metrics)~\cite{Bemelmans1984,Bando1987}.  Our numerical evolutions
show smoothing of the metrics that is consistent with this fact.  The
presence of the DeTurck gauge-fixing terms, however, somewhat
obfuscates this question of smoothness.  Our evolutions show that the
DeTurck gauge-fixing covector $H_i$ is zero, up to truncation level
errors, throughout the evolutions.  The connection $\Gamma^k_{ij}$ of
the metric $g_{ij}$ at the end of our Ricci flow evolutions could (in
principle) therefore retain some of the non-smooth features of the
reference connection $\tilde \Gamma^k_{ij}$, since
$H_i=0=g_{ij}g^{k\ell}(\Gamma^j_{k\ell}-\tilde \Gamma^j_{k\ell})$.
However, the vanishing of $H_i$ shows that the evolved metric
satisfies the original Ricci flow equation without the DeTurck terms,
and thus must be smooth by the aforementioned
theorems~\cite{Bemelmans1984,Bando1987}.  Hence any non-smoothness of
the connection must just reflect the (non-smooth) coordinate
transitions at the interface boundaries.

We made some effort to avoid even the potential effects of the
non-smoothness of the connection associated with the DeTurck terms by
modifying the basic Ricci flow Eq.~(\ref{e:RicciFlow}) in various
ways.  For example, we attempted to carry out numerical Ricci flow
evolutions without including the DeTurck terms at all, i.e., simply by
setting $H_i=0$ in Eq.~(\ref{e:RicciFlow}).  All of these evolutions
were unstable.  The DeTurck terms were added to the Ricci flow
equation to make it strongly parabolic and thereby manifestly
well-posed~\cite{Chow2004}.  Without the DeTurck terms, the basic
Ricci flow equations may simply be ill-suited for numerical solution.
We also tried modifying the DeTurck terms in a way that would attempt
to drive the solution to harmonic gauge, i.e., to a gauge in which
$0=g^{ij}\Gamma^k_{ij}$.  We did this by changing the definition of
$H_i$ to give the reference connection an explicit time dependence, as
in $H_i=g_{ij}g^{k\ell}(\Gamma^j_{k\ell}-e^{-\mu t}\,\tilde
\Gamma^j_{k\ell})$, for example.  Unfortunately all of these runs
failed as well.  While these runs appeared to be stable, the Ricci
flows in these cases did not evolve toward metrics having constant
scalar curvatures, and the DeTurck gauge-source covector $H_i$ did not
remain small during the evolutions.

We plan to continue to search for effective and efficient ways to
construct reference metrics on manifolds with arbitrary spatial
topologies.  In two dimensions the remaining questions are related to
finding better gauge conditions for the reference metrics.  In three
and higher dimensions the challenge will be to find efficient ways to
implement the general techniques developed here.


\section*{Acknowledgments} 
We thank J\"org Enders, Gerhard Huisken, James Isenberg, and Klaus
Kr\"oncke for helpful discussions about Ricci flow\Lee{, and Michael
  Holst and Ralf Kornhuber for helpful discussions on surface finite
  element methods}.  LL and NT thank the Max Planck Institute for
Gravitational Physics (Albert Einstein Institute) in Golm, Germany for
their hospitality during a visit when a portion of this research was
completed.  LL and NT were supported in part by a grant from the
Sherman Fairchild Foundation and by grants DMS-1065438 and PHY-1404569
from the National Science Foundation.  OR was supported by a
Heisenberg Fellowship and grant RI 2246/2 from the German Research
Foundation (DFG).  We also thank the Center for Computational
Mathematics at the University of California at San Diego for providing
access to their computer cluster (aquired through NSF DMS/MRI Award
0821816) on which all the numerical tests reported in this paper were
performed.

\appendix
\section{Uniqueness of the $C^1$ Multicube Differential Structure}
\label{s:Uniqueness}

The traditional definition of a $C^k$ differential structure on a
manifold consists of an atlas of coordinate charts having the property
that the transition maps between overlapping charts are $C^{k+1}$
functions.\footnote{We use the slightly non-standard terminology that
  a $C^k$ differential structure is needed to define $C^k$ tensor
  fields.  This choice implies that the transition maps between
  overlapping domains in the atlas must be $C^{k+1}$.}  Tensor fields
are defined to be $C^k$ with respect to this differential structure if
their components when represented in terms of this atlas are $C^k$
functions.  In a multicube representation of a manifold, we define the
continuity of tensor fields and their derivatives instead using the
Jacobians and the connection determined by a reference metric.  This
enables us to define these concepts without needing an overlapping
$C^{k+1}$ atlas.  The two definitions of differential structure are
equivalent on any manifold having both a multicube structure and a
$C^{k+1}$ atlas. In this appendix we consider the technical question
of the uniqueness of the multicube method of specifying the
differential structure.

The purpose of this appendix is to show that the $C^1$ differential
structure of a multicube manifold defined by a particular $C^1$
reference metric is independent of the choice of reference metric.  In
particular, we show that the definitions of continuity of tensor
fields and their covariant derivatives based on a $C^1$ reference
metric $\tilde g_{ab}$ are the same as those based on any other $C^1$
metric $\check{g}_{ab}$, i.e., any metric $\check{g}_{ab}$ that is
continuous and whose covariant gradient $\tilde\nabla_a
\check{g}_{bc}$ is continuous with respect to the differential
structure defined by $\tilde g_{ab}$.  Since any $C^k$ metric with
$k\geq 1$ is also $C^1$, this argument implies that the $C^1$
differential structure defined by the $C^1$ metric $\tilde g_{ab}$ is
also equivalent to the $C^1$ differential structure defined by any
$C^k$ metric $\check{g}_{ab}$.

We have shown~\cite{Lindblom2013} how the differential structure for a
multicube representation of a manifold may be specified by giving a
$C^1$ metric $\tilde g_{ab}$ represented
in the global Cartesian multicube coordinate
basis.\footnote{While the global Cartesian multicube coordinates are
  severely constrained (e.g., the faces of the cubic-block regions are
  required to be constant coordinate surfaces on which the values of
  the surface coordinates have particular fixed values), they are not
  fixed uniquely.  The remaining coordinate freedom is discussed at
  the end of this appendix, but for the first part of this discussion
  we assume that all tensor fields are represented in one particular
  choice of these global Cartesian multicube coordinates.}  This
method of defining the differential structure
constructs Jacobians $\tilde J^{A\alpha a}_{B\beta b}$ and their duals
$\tilde J_{A\alpha a}^{*B\beta b}$ that transform tensors from the
$\partial_\beta \mathcal{B}_B$ face of cubic region $\mathcal{B}_B$ to
the $\partial_\alpha \mathcal{B}_A$ face of cubic region
$\mathcal{B}_A$.  These Jacobians are determined by the metric $\tilde
g_{ab}$ and the rotation matrices $C^{A\alpha a}_{B\beta b}$ that
define the identification maps (cf.~\ref{s:2dMulticubeManifolds})
between neighboring regions.  The expressions for these Jacobians are
given by Lindblom and Szil\'agyi~\cite{Lindblom2013}:
\begin{eqnarray}
\tilde J^{A\alpha a}_{B\beta b}&=&C^{A\alpha a}_{B\beta c}
\left(\delta^c_b-\tilde n^c_{B\beta}\tilde n_{B\beta b}\right)-\tilde n^{a}_{A\alpha}
\tilde n_{B\beta b},
\label{e:TildeJacobianDef}\\
\tilde J_{A\alpha a}^{*B\beta b}&=& 
\left(\delta^c_a-\tilde n_{A\alpha a}\tilde n^c_{A\alpha}\right)
C^{B\beta b}_{A\alpha c}-\tilde n_{A\alpha a}\tilde n^b_{B\beta}.
\label{e:TildeInvJacobianDef}
\end{eqnarray}  
The vectors $\tilde n^a_{A\alpha}$ and $\tilde n^a_{B\beta}$ that
appear in these expressions represent the outward directed unit normal
vectors to the $\partial_\alpha \mathcal{B}_A$ face of region
$\mathcal{B}_A$ and the $\partial_\beta \mathcal{B}_B$ face of cubic
region $\mathcal{B}_B$, respectively.  These normals are unit vectors
with respect to the $\tilde g_{ab}$ metric, i.e., $1=\tilde
g_{Aab}\tilde n^a_{A\alpha}\tilde n^b_{A\alpha}=\tilde g_{Bab}\tilde
n^a_{B\beta}\tilde n^b_{B\beta}$.  These Jacobians, defined in
Eqs.~(\ref{e:TildeJacobianDef}) and (\ref{e:TildeInvJacobianDef}),
determine the way continuous tensor fields transform across interface
boundaries.  The reference metric also determines a covariant
derivative $\tilde \nabla_a$ that, together with the Jacobians,
defines how $C^1$ tensor fields transform across interface boundaries.
These definitions of continuity for tensor fields and their
derivatives determine the $C^1$ differential structure of the
manifold.  The question of the uniqueness of the $C^1$ differential
structure reduces therefore to the questions of the uniqueness of the
Jacobians $\tilde J^{A\alpha a}_{B\beta b}$, and of the uniqueness of
the continuity of the derivatives determined by the covariant
derivative $\tilde\nabla_a$.

The normal covectors $\tilde n_{A\alpha a}$ that appear in
Eqs.~(\ref{e:TildeJacobianDef}) and (\ref{e:TildeInvJacobianDef}) are
proportional to the gradients of the $x^{|\alpha|}_A$=constant
coordinate surfaces that define the particular boundary face of the
region (i.e., in this case the $\alpha$ face of region $A$):
\begin{eqnarray}
\tilde n_{A\alpha a}=\tilde N_{A\alpha}\partial_a x^{|\alpha|}_A.
\label{e:TildeUnitNormal}
\end{eqnarray}  
The index $\alpha$ can have either sign, e.g., to represent the $+x$
or the $-x$ coordinate boundary face.  The notation $x^{|\alpha|}_A$
indicates the coordinate associated with either case---i.e., both the
$+x$ and the $-x$ faces are surfaces of constant $x^x_A$.  The
proportionality constant $\tilde N_{A\alpha}$ in
Eq.~(\ref{e:TildeUnitNormal}) is determined by the requirement that
$\tilde n_{A\alpha a}$ is a unit covector with respect to the
reference metric $\tilde g_{Aab}$:
\begin{eqnarray}
\tilde N^{-2}_{A\alpha}=\tilde g^{ab}_A\partial_a x^{|\alpha|}_A\partial_b x^{|\alpha|}_A.
\end{eqnarray}
The sign of $\tilde N_{A\alpha}$ is chosen to ensure that $\tilde
n_{A\alpha a}$ is the outward directed normal. The normal vector is
defined as the dual to this normal covector: $\tilde
n^a_{A\alpha}=\tilde g^{ab}_A\tilde n_{A\alpha b}$. 

The Jacobians defined in Eqs.~(\ref{e:TildeJacobianDef}) and
(\ref{e:TildeInvJacobianDef}) transform these normals across interface
boundaries in the appropriate way:
\begin{eqnarray}
\tilde n^a_{A\alpha}&=&-\tilde J^{A\alpha a}_{B\beta b}\tilde n^b_{B\beta},
\label{e:NormalVectorTransform}\\
\tilde n_{A\alpha a}&=&-\tilde J_{A\alpha a}^{*B\beta b}\tilde n_{B\beta b}.
\label{e:NormalCoVectorTransform}
\end{eqnarray}
They also transform vectors $t^a_{B\beta}$ that are tangent to the
interface, $\tilde n_{A\alpha a} t^a_{A\alpha}=0$, by the
rotations $C^{A\alpha a}_{B\beta b}$ used to define the interface boundary
maps (cf.~\ref{s:2dMulticubeManifolds}):
\begin{eqnarray}
t^a_{A\alpha}=\tilde J^{A\alpha a}_{B\beta b} t^b_{B\beta}=
C^{A\alpha a}_{B\beta b} t^b_{B\beta}.
\label{e:TangentVectorTransform}
\end{eqnarray}
These Jacobians and dual Jacobians are inverses of each other as well
(cf. Ref.~\cite{Lindblom2013}):
\begin{eqnarray}
\delta^{a}_{b}=\tilde J^{A\alpha a}_{B\beta c}\tilde J_{A\alpha b}^{*B\beta c}.
\label{e:JacobianInverses}
\end{eqnarray}

Now consider a second positive-definite metric $\check g_{ab}$ that is
$C^1$ with respect to the differential structure defined by the metric
$\tilde g_{ab}$.  This second metric can be used to 
define alternate normal covectors $\check n_{A\alpha a}=\check
N_{A\alpha} \partial_a x_A^{|\alpha|}$ and vectors $\check
n^a_{A\alpha}=\check g^{ab}_A\check n_{A\alpha b}$, with $\check
N_{A\alpha}^{-2}= \check g^{ab}_A \partial_a x^{|\alpha|}_A\partial_b
x^{|\alpha|}_A$.  It follows from
Eq.~(\ref{e:NormalCoVectorTransform}) and the continuity of $\check
g_{ab}$ that the norm of $\tilde n_{A\alpha a}$ with respect to
$\check g_{ab}$ is continuous across interface boundaries:
\begin{eqnarray}
\check g^{ab}_A \tilde n_{A\alpha a}\tilde n_{A\alpha b}
=\check g^{ab}_B \tilde n_{B\beta a}\tilde n_{B\beta b}.
\label{e:InnerProductContinuity}
\end{eqnarray}
This norm can be rewritten as
\begin{eqnarray}
\check g^{ab}_A \tilde n_{A\alpha a}\tilde n_{A\alpha b}
= \tilde N_{A\alpha}^2\check g^{ab}_A \partial_a x^{|\alpha|}_A\partial_b x^{|\alpha|}_A
=\left(\frac{\tilde N_{A\alpha}}{\check N_{A\alpha}}\right)^2\!\!\!.\quad
\label{e:NormalRatioExpression}
\end{eqnarray} 
Equation (\ref{e:InnerProductContinuity}) therefore implies the
continuity of the ratio $\tilde N_{A\alpha}/\check N_{A\alpha}$ across
interface boundaries.  The alternate normal $\check n_{A\alpha a}$,
which can be written as $\check n_{A\alpha a}=(\check N_{A\alpha}/
\tilde N_{A\alpha})\tilde n_{A\alpha a}$, is therefore continuous (up
to a sign flip) across interface boundaries.  This also implies that
the alternate normal vector $\check n^a_{A\alpha}=\check
g^{ab}_A\check n_{A\alpha b}$ is continuous.  These alternate normals
must therefore satisfy the same continuity conditions (up to the sign
flips) across interface boundaries as any continuous tensor field:
\begin{eqnarray}
\check n^a_{A\alpha}&=&-\tilde J^{A\alpha a}_{B\beta b}\check n^b_{B\beta},
\label{e:AltNormalVectorTransform}\\
\check n_{A\alpha a}&=&-\tilde J_{A\alpha a}^{*B\beta b}\check n_{B\beta b}.
\label{e:AltNormalCoVectorTransform}
\end{eqnarray}

The normal vector $\tilde n^a_{A\alpha}$ together with a collection of
linearly independent tangent vectors, i.e., vectors $t^a_{A\alpha(k)}$
satisfying $0=t^a_{A\alpha(k)}\tilde n_{A\alpha a}$, can be used as a
basis of vector fields on the boundary.  Therefore any vector field,
including $\check n^a_{A\alpha}$, can be expressed as a linear
combination of the form
\begin{eqnarray}
\check n^a_{A\alpha}=Q\, \tilde n^a_{A\alpha} + \sum_k c_k\, t^a_{A\alpha(k)}.
\label{e:CheckNormalExpansion}
\end{eqnarray}
Contracting this expression with $\tilde n_{A\alpha a}$ and using
Eq.~(\ref{e:NormalRatioExpression}), it follows that $Q=\tilde
N_{A\alpha}/\check N_{A\alpha}$.  Note that the tangent vectors
$t^a_{A\alpha(k)}$, which are orthogonal to $\tilde n_{A\alpha a}$ by
definition, are also orthogonal to $\check n_{A\alpha a}$.  Therefore,
the alternate normal $\check n^a_{A\alpha}$ together with a linearly
independent collection of tangent vectors can also be used as a basis
of vectors on the boundary.

Next define alternate Jacobians $\check J^{A\alpha a}_{B\beta b}$ and
$\check J_{A\alpha a}^{*B\beta b}$ using the alternate metric $\check
g_{ab}$:
\begin{eqnarray}
\check J^{A\alpha a}_{B\beta b}&=&C^{A\alpha a}_{B\beta c}
\left(\delta^c_b-\check n^c_{B\beta}\check n_{B\beta b}\right)-\check n^{a}_{A\alpha}
\check n_{B\beta b},
\label{e:AltJacobianDef}\\
\check J_{A\alpha a}^{*B\beta b}&=&
\left(\delta^c_a-\check n_{A\alpha a}\check n^c_{A\alpha}\right)
C^{B\beta b}_{A\alpha c}-\check n_{A\alpha a}\check n^b_{B\beta}.
\label{e:AltInvJacobianDef}
\end{eqnarray}
These alternate Jacobians transform the alternate normal $\check
n^a_{A\alpha}$ and any tangent vector $t^a_{A\alpha(k)}$ in the
following way:
\begin{eqnarray}
\check n^a_{A\alpha}&=&-\check J^{A\alpha a}_{B\beta b}\,\check n^b_{B\beta},
\label{e:AltNormalVectorTransformA}\\
t^a_{A\alpha(k)}&=&\check J^{A\alpha a}_{B\beta b} \,t^b_{B\beta(k)}=
C^{A\alpha a}_{B\beta b} \,t^b_{B\beta(k)}.
\label{e:AltTangentVectorTransformA}
\end{eqnarray}
The alternative Jacobian and its dual are also inverse of each other:
\begin{eqnarray}
\delta^{a}_{b}=\check J^{A\alpha a}_{B\beta c}\check J_{A\alpha b}^{*B\beta c}.
\label{e:AltJacobianInverses}
\end{eqnarray}
The action of the alternate Jacobians $\check J^{A\alpha a}_{B\beta
  b}$ on the basis of vectors consisting of $\check n^a_{Aa}$ and a
collection of tangent vectors $t^a_{A\alpha(k)}$,
Eqs.~(\ref{e:AltNormalVectorTransformA}) and
(\ref{e:AltTangentVectorTransformA}), is identical to the action of
the original Jacobians $\tilde J^{A\alpha a}_{B\beta b}$ on this
basis, Eqs.~(\ref{e:TangentVectorTransform}) and
(\ref{e:AltNormalVectorTransform}).  It follows that the alternate
Jacobians must be identical to the originals:
\begin{eqnarray}
\check J^{A\alpha a}_{B\beta b}= \tilde J^{A\alpha a}_{B\beta b}.
\end{eqnarray}
Since the alternate dual Jacobians $\check J_{A\alpha a}^{*B\beta b}$
are the inverses of the alternate Jacobians, they must also be
identical to the original dual Jacobians (which are the inverses of
the original Jacobians).  We have shown therefore that the Jacobians
used to define the continuity of tensor fields across boundary
interfaces do not depend on which metric is used to construct them.
This argument depends only on the continuity of those metrics (not
their derivatives).

Now consider the uniqueness of the multicube definition of the
continuity of the derivatives of tensor fields.  Let $\tilde \nabla_a$
and $\check\nabla_a$ denote the covariant derivatives defined by the
$C^1$ reference metric $\tilde g_{ab}$ and the $C^1$ reference metric
$\check g_{ab}$, respectively.  Let $v^a$ and $w_a$ denote vector and
covector fields that are continuous across the interface boundaries,
as defined by the Jacobians constructed from either of the reference
metrics.  Assume that $\tilde \nabla_a v^b$ and $\tilde \nabla_a w_b$
are also continuous across interface boundaries.  The differences
between these tensors and those computed using the alternate covariant
derivative $\check\nabla_a$ are tensors:
\begin{eqnarray}
\tilde \nabla_a v^b - \check\nabla_a v^b = \Delta^b_{ac}v^c,\\
\tilde \nabla_a w_b - \check\nabla_a w_b = -\Delta^c_{ab}w_c.
\end{eqnarray}
The quantity $\Delta^b_{ac}=\tilde \Gamma^b_{ac}-\check
\Gamma^b_{ac}$, being the difference between connections, is also a
tensor.  It is continuous across interface boundaries as long as the
two metrics $\tilde g_{ab}$ and $\check g_{ab}$ used to construct it
are both $C^1$.  Continuity of the
derivatives $\tilde \nabla_a v^b$ and $\tilde \nabla_a w_b$ across
interface boundaries therefore implies the continuity of the
alternative derivatives $\check \nabla_a v^b$ and $\check \nabla_a
w_b$.

The equality of the Jacobians $\tilde J^{A\alpha a}_{B\beta b}$ and
$\check J^{A\alpha a}_{B\beta b}$, together with the continuity of the
covariant derivatives $\tilde \nabla_a$ and $\check \nabla_a$, implies
that the $C^1$ differential structure constructed from the $C^1$
metric $\tilde g_{ab}$ is equivalent to the one constructed from any
alternate $C^1$ metric $\check g_{ab}$.  In dimensions two and three
there is only one differential structure on a particular manifold
\cite{Lee2012}.  In those cases, this argument shows that the $C^1$
differential structures determined by any two $C^1$ metrics are
equivalent.  In higher dimensional manifolds, however, there can be
multiple inequivalent differential structures \cite{Lee2012}.  The
argument given here only establishes the independence of the multicube
differential structure constructed from reference metrics belonging to
the same differential structure in those cases.

The uniqueness of the Jacobians $J^{A\alpha \,a}_{B\beta\,b}$
discussed above assumed a particular fixed choice of global Cartesian
multicube coordinates.  Although these Cartesian multicube coordinates
are severely restricted, they are not unique.  The two assumptions
made about them are the following.  First, the faces
of each cubic-block region are assumed to be constant-coordinate
surfaces.  And second, the interface boundary maps
identify points in the manifold across boundaries in a particular way
(cf.~\ref{s:2dMulticubeManifolds}).  The global Cartesian multicube
coordinates on these manifolds can therefore be modified in any way
that leaves their interface boundary values and the identification of
points on the interface boundaries unchanged. The coordinates can be
modified smoothly in the interior of each cubic-block region, for
example, while keeping their values fixed on their faces. More
generally, the coordinates can be adjusted smoothly even on the
boundary faces as long as complementary adjustments are made to the
corresponding coordinates in the neighboring region.

Let $x_A^a$ denote one particular choice of coordinates on region $A$,
and let $\bar x^a_A$ denote another set of smoothly related
coordinates that satisfy the restrictions described above.  Also
assume that the Jacobians $\partial \bar x^a_A/\partial x^b_A$ are
everywhere nonsingular and nondegenerate.  Let $v^a_A$ and $w_{Aa}$
denote a smooth vector and covector fields in region $A$.  The
representations of these fields within this region using the $\bar
x^a_A$ coordinates are given by the standard expressions
\begin{eqnarray}
\bar v^a_A &=& \frac{\partial \bar x^a_A}{\partial x^b_A}v^b_A,
\label{e:VectorCoordinateChange}
\\
\bar w_{Aa} &=& \frac{\partial x^b_A}{\partial \bar x^a_A}w_{Ab}.
\label{e:CoVectorCoordinateChange}
\end{eqnarray}
Analogous changes of coordinates can be made in each of the
cubic-block regions.  The resulting Jacobians $\bar J^{A\alpha
  \,a}_{B\beta\,b}$ needed to transform tensor fields represented in
the $\bar x^a_A$ coordinates are related to those of the original
fixed coordinates $J^{A\alpha \,a}_{B\beta\,b}$ by the following
transformations:
\begin{eqnarray}
\bar J^{A\alpha \,a}_{B\beta\,b}=
J^{A\alpha \,c}_{B\beta\,d}
\frac{\partial \bar x^a_A}{\partial x^c_A}
\frac{\partial x^d_B}{\partial \bar x^b_B}.
\label{e:JacobianTransformation}
\end{eqnarray}
This multicube coordinate freedom does not require $\partial \bar
x^a_A/\partial x^b_A$ to be the identity $\delta^{a}_{b}$ on the faces
of the multicube regions, and consequently the Jacobians $\bar
J^{A\alpha \,a}_{B\beta\,b}$ need not be identical to $J^{A\alpha
  \,a}_{B\beta\,b}$.  Nevertheless, the formulas for the Jacobians,
Eqs.~(\ref{e:TildeJacobianDef}) and (\ref{e:TildeInvJacobianDef}),
have the same form in any particular multicube coordinate system.
When the individual elements (e.g., $n^a_{A\alpha}$) that enter these
equations for $J^{A\alpha \,a}_{B\beta\,b}$ are transformed to a
different coordinate basis using Eqs.~(\ref{e:VectorCoordinateChange})
and (\ref{e:CoVectorCoordinateChange}), the resulting $\bar J^{A\alpha
  \,a}_{B\beta\,b}$ is related to the original Jacobian by
Eq.~(\ref{e:JacobianTransformation}).  This equation represents the
coordinate freedom that exists in the expressions for the interface
Jacobians on multicube manifolds within a particular
differential structure.  Every two- and
three-dimensional manifold has a unique global differential structure,
and therefore Eq.~(\ref{e:JacobianTransformation}) represents all the
freedom that exists in the boundary interface Jacobians on those
manifolds.

\section{Two-Dimensional Multicube Manifolds}
\label{s:2dMulticubeManifolds}
\setcounter{table}{0} \setcounter{figure}{0} 
\renewcommand*{\theHtable}{\arabic{section}.\arabic{table}} 
\renewcommand*{\theHfigure}{\arabic{section}.\arabic{figure}} 
The purpose of this
appendix is to present explicit multicube representations of compact,
orientable two-dimensional manifolds with genera between zero and
three.  A straightforward procedure allows us to extend these examples
to arbitrary genus by gluing together copies of the $N_g=2$ multicube
structures.  The topologies of all these two-dimensional manifolds are
uniquely determined by their genus $N_g$, which can have non-negative
integer values.  The case $N_g=0$ is the two-sphere, $S^2$, and
$N_g=1$ is the two-torus, $T^2$.  Larger values of $N_g$ can be
thought of as two-spheres with $N_g$ handles attached.

A multicube representation of a manifold consists of a collection
of multicube regions $\mathcal{B}_A$ together with
maps $\Psi^{A\alpha}_{B\beta}$ that determine
how the boundaries $\partial_\alpha\mathcal{B}_A$ of these regions are
connected together.  We choose multicube regions $\mathcal{B}_A$ that
have uniform coordinate size $L$ and that are all aligned in $\Rn$
with the global Cartesian coordinate axes.  We position these
$\mathcal{B}_A$ in $\Rn$ in such a way that regions intersect (if at
all) only along boundaries that are identified with one another by one
of the $\Psi^{A\alpha}_{B\beta}$ maps.  For each multicube manifold,
we provide a table of vectors $\vec c_A$ that represent the global
Cartesian coordinates of the centers of each of the multicube regions
$\mathcal{B}_A$.  These tables serve as lists of the regions
$\mathcal{B}_A$ that are to be included in each particular multicube
representation.  We also provide tables of all of the interface
boundary identifications for each multicube representation.  A typical
entry in one of these tables is an expression of the form
$\partial_{+x}\mathcal{B}_2\leftrightarrow
\partial_{-y}\mathcal{B}_3$, which would indicate that the $+x$
boundary of multicube $\mathcal{B}_2$ is to be identified with the
$-y$ boundary of multicube $\mathcal{B}_3$.

The boundary identification maps used in our multicube manifolds are
simple linear transformations of the form
\begin{eqnarray}
x^i_A=c_A^i+f_\alpha^i + C^{A\alpha\,i}_{B\beta\,j}(x^j_B - c_B^j - f_\beta^j).
\label{e:TransitionMap}
\end{eqnarray}
This transformation takes points labeled by the global Cartesian
coordinates $x_B^j$ on the boundary $\partial_\beta\mathcal{B}_B$ to
points labeled by the global Cartesian coordinates $x_A^i$ on the
boundary $\partial_\alpha\mathcal{B}_A$.  The constants $c_A^i$
represent the location of the center of multicube region
$\mathcal{B}_A$, while the constants $f_\alpha^i$ represent the
position of the center of the $\alpha$ face relative to the center of
the region.  Since we have chosen the regions to have uniform sizes
and orientations, the constants $f_\alpha^i$ have the same form in
each multicube region:
\begin{eqnarray}
f_{\pm x}^i &=& \tfrac{1}{2}L(\pm 1,0),\\
f_{\pm y}^i &=& \tfrac{1}{2}L(0,\pm 1).
\end{eqnarray}
The matrix $\mathbf{C}^{A\alpha}_{B\beta}$ which appears in
Eq.~(\ref{e:TransitionMap}) is the combined rotation and reflection
matrix needed to reorient the $\partial_\beta\mathcal{B}_B$ boundary
with $\partial_\alpha\mathcal{B}_A$.  Our
specification of a particular multicube representation includes the
matrices $\mathbf{C}^{A\alpha}_{B\beta}$ for each interface boundary
identification map.  The list of possible matrices is quite small in
two-dimensions, consisting of the identity $\mathbf{I}$, various
combinations of 90-degree rotations $\mathbf{R}_\pm$, and reflections
$\mathbf{M}$.  Explicit representations of these matrices in terms of
the global Cartesian coordinate basis are given by
\begin{eqnarray}
\mathbf{I}=
\left(\begin{array}{cc}
1 &0\\
0&1
\end{array}\right),
\qquad\qquad
\mathbf{R}_\pm=
\left(\begin{array}{cc}
0 &\mp 1\\
\pm 1&0
\end{array}\right),
\qquad\qquad
\mathbf{M}=
\left(\begin{array}{cc}
-1 &0\\
0&1
\end{array}\right).
\end{eqnarray}
In the following sections we give the specific matrices
$\mathbf{C}^{A\alpha}_{B\beta}$ and their inverses
$\mathbf{C}_{A\alpha}^{B\beta}$ needed for each interface boundary
identification $\partial_\alpha\mathcal{B}_A\leftrightarrow
\partial_\beta\mathcal{B}_B$ of each multicube manifold.  The methods
and the notation used here are the same as those developed in
Ref.~\cite{Lindblom2013}.

\subsection{Six-Region, $N_R=6$, Representation of the Genus $N_g=0$ 
  Multicube Manifold}
\label{s:GenusZeroRegionSix}

\begin{figure}[!htb]
\centering
\includegraphics[width=0.7\textwidth]{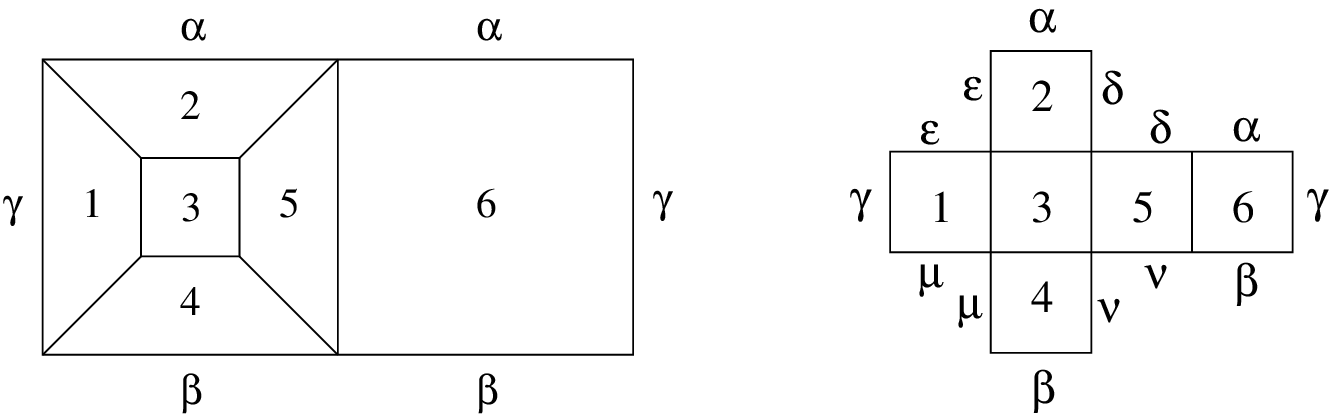}
\caption{Six-region, $N_R=6$, multicube representation of the genus
  $N_g=0$ manifold, the two-sphere, $S^2$.  Left figure shows a
  multicube representation using distorted squares to indicate as many
  interfacial connections as possible.  Greek letters indicate
  identifications between external edges.  Right figure shows the same
  multicube representation using uniformly sized, undistorted squares,
  including their relative locations in the background Euclidean
  space.
  \label{f:Genus0Region6} }
\end{figure}

The locations of the six square regions used to construct this
representation of $S^2$ are illustrated in Fig.~\ref{f:Genus0Region6}.
The values of the square-center location vectors $\vec c_A$ for this
configuration are summarized in
Table~\ref{t:TableGenus0Region6Locations}.  The inner edges of the
touching squares in the right side of Fig.~\ref{f:Genus0Region6} are
connected by identity maps.  The identifications of all the edges of
the regions are described in Table~\ref{t:TableGenus0Region6Maps}, and
the corresponding transformation matrices are given in
Table~\ref{t:TableGenus0Region6MapsA}.  This six-region representation
of $S^2$ is equivalent to the standard two-dimensional cubed-sphere
representation of $S^2$~\cite{Ronchi1996, Taylor1997, Dennis2003}.
\begin{table}[!htb]
\centering
\caption{Region center locations for the six-region, $N_R=6$, genus
  $N_g=0$ multicube manifold. 
  \label{t:TableGenus0Region6Locations}}
\setlength\tabcolsep{0.8em}
\begin{tabular}{@{}lll@{}} \toprule
  \multicolumn{3}{c}{$\vec c_A=(x,y)$} \\ \midrule
  $\vec c_1=(0,0)$ & $\vec c_2=(L,L)$ & $\vec c_3=(L,0)$ \\ [3pt]
  $\vec c_4=(L,-L)$ & $\vec c_5=(2L,0)$& $\vec c_6 =(3L,0)$ \\ \bottomrule
\end{tabular}
\end{table}
\begin{table}[!htb]
\centering
\caption{Region interface identifications $\partial_\alpha{\cal B}_A
  \leftrightarrow \partial_\beta {\cal B}_B$ for the six-region,
  $N_R=6$, representation of the genus $N_g=0$ manifold, the
  two-sphere, $S^2$.
  \label{t:TableGenus0Region6Maps}}
\setlength\tabcolsep{1.1em}
\begin{tabular}{@{}llll@{}} \toprule
  \multicolumn{4}{c}{$\partial_\alpha{\cal B}_A 
    \leftrightarrow \partial_\beta {\cal B}_B$} \\ \midrule
  $\partial_{+x}{\cal B}_1 \leftrightarrow \partial_{-x} {\cal B}_3$
  & $\partial_{-x}{\cal B}_1 \leftrightarrow \partial_{+x} {\cal B}_{6}$
  & $\partial_{+y}{\cal B}_1 \leftrightarrow \partial_{-x} {\cal B}_2$
  & $\partial_{-y}{\cal B}_1 \leftrightarrow \partial_{-x} {\cal B}_4$ 
  \\ [4pt]
  $\partial_{+x}{\cal B}_2 \leftrightarrow \partial_{+y} {\cal B}_5$
  & $\partial_{+y}{\cal B}_2 \leftrightarrow \partial_{+y} {\cal B}_6$
  & $\partial_{-y}{\cal B}_2 \leftrightarrow \partial_{+y} {\cal B}_3$
  & $\partial_{+x}{\cal B}_3 \leftrightarrow \partial_{-x} {\cal B}_5$
  \\ [4pt]
  $\partial_{-y}{\cal B}_3 \leftrightarrow \partial_{+y} {\cal B}_4$
  & $\partial_{+x}{\cal B}_4 \leftrightarrow \partial_{-y} {\cal B}_5$
  & $\partial_{-y}{\cal B}_4 \leftrightarrow \partial_{-y} {\cal B}_6$
  & $\partial_{+x}{\cal B}_5 \leftrightarrow \partial_{-x} {\cal B}_6$
  \\ \bottomrule
\end{tabular}
\end{table}
\begin{table}[!htb]
\centering
\caption{Transformation matrices ${\bf C}^{A\alpha}_{B\beta}$ for the
  interface identifications $\partial_\alpha{\cal B}_A \leftrightarrow
  \partial_\beta {\cal B}_B$ in the six-region, $N_R=6$,
  representation of the genus $N_g=0$ manifold, the two-sphere, $S^2$.
  All transformation matrices ${\bf C}^{A\alpha}_{B\beta}$ are assumed
  to be the identity $\mathbf{I}$, except those specified in this
  table.
  \label{t:TableGenus0Region6MapsA}}
\begin{tabular}{@{}lcc@{\hskip 2.3em}lcc@{}} \toprule
  $\partial_\alpha{\cal B}_A \leftrightarrow \partial_\beta {\cal B}_B$
  & ${\mathbf C}^{A\alpha}_{B\beta}$ 
  & ${\mathbf C}^{B\beta}_{A\alpha}$ 
  & $\partial_\alpha{\cal B}_A \leftrightarrow \partial_\beta {\cal B}_B$
  & ${\mathbf C}^{A\alpha}_{B\beta}$ 
  & ${\mathbf C}^{B\beta}_{A\alpha}$ \\ \midrule
  $\partial_{+y}{\cal B}_1 \leftrightarrow \partial_{-x} {\cal B}_2$
  & ${\mathbf R}_{+}$ & ${\mathbf R}_{-}$ 
  & $\partial_{-y}{\cal B}_1 \leftrightarrow \partial_{-x} {\cal B}_4$
  & ${\mathbf R}_{-}$ & ${\mathbf R}_{+}$ \\ [4pt]
  $\partial_{+x}{\cal B}_2 \leftrightarrow \partial_{+y} {\cal B}_5$
  & ${\mathbf R}_{+}$ & ${\mathbf R}_{-}$ 
  & $\partial_{+y}{\cal B}_2 \leftrightarrow \partial_{+y} {\cal B}_6$
  & ${\mathbf R}_{+}^2$ & ${\mathbf R}_{-}^2$ \\ [4pt]
  $\partial_{-y}{\cal B}_4 \leftrightarrow \partial_{-y} {\cal B}_6$
  & ${\mathbf R_+^2}$ & ${\mathbf R_-^2}$
  & $\partial_{+x}{\cal B}_4 \leftrightarrow \partial_{-y} {\cal B}_5$
  & ${\mathbf R_-}$ & ${\mathbf R_+}$ \\ \bottomrule
\end{tabular}
\end{table}

\subsection{Ten-Region, $N_R=10$, Representation of the Genus $N_g=0$ 
  Multicube Manifold}
\label{s:GenusZeroRegionTen}

\begin{figure}[!htb]
\centering
\includegraphics[width=0.8\textwidth]{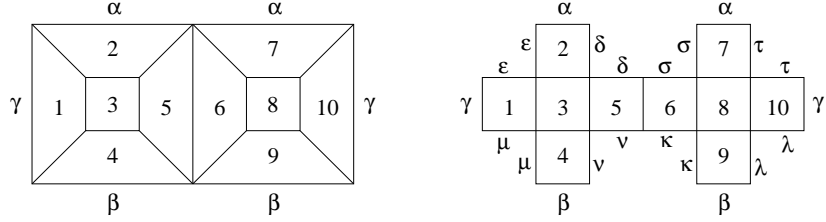}
\caption{Ten-region, $N_R=10$, multicube representation of the genus
  $N_g=0$ manifold, the two-sphere, $S^2$.  Left figure shows a
  multicube representation using distorted squares to indicate as many
  interfacial connections as possible.  Greek letters indicate
  identifications between external edges.  Right figure shows the same
  multicube representation using uniformly sized, undistorted squares,
  including their relative locations in the background Euclidean
  space.
  \label{f:Genus0Region10} }
\end{figure}

The locations of the ten square regions used to construct this
representation of $S^2$ are illustrated in
Fig.~\ref{f:Genus0Region10}.  The values of the square-center location
vectors $\vec c_A$ for this configuration are summarized in
Table~\ref{t:TableGenus0Region10Locations}.  The inner edges of the
touching squares in the right side of Fig.~\ref{f:Genus0Region10} are
assumed to be connected by identity maps.  The identifications of all
the edges of the regions are described in
Table~\ref{t:TableGenus0Region10Maps}, and the corresponding
transformation matrices are given in
Table~\ref{t:TableGenus0Region10MapsA}.  This ten-region
representation of $S^2$ is a simple generalization of the standard
two-dimensional cubed-sphere representation of $S^2$.  It is
constructed by splitting the four ``equatorial'' squares in the
standard six-region representation into eight squares with the new
interface boundaries running along the equator.
\begin{table}[!htb]
\centering
\caption{Region center locations for the ten-region, $N_R=10$, genus
  $N_g=0$ multicube manifold.
  \label{t:TableGenus0Region10Locations}}
\setlength\tabcolsep{0.8em}
\begin{tabular}{@{}lllll@{}} \toprule
  \multicolumn{5}{c}{$\vec c_A=(x,y)$} \\ \midrule
  $\vec c_1=(0,0)$ & $\vec c_2=(L,L)$ & $\vec c_3=(L,0)$ 
  & $\vec c_4=(L,-L)$ & $\vec c_5=( 2L,0)$ \\ [3pt]
  $\vec c_6=(3L,0)$ & $\vec c_7=(4L,L)$ & $\vec c_8=(4L, 0)$
  & $\vec c_9=(4L,-L)$ & $\vec c_{10}=(5L, 0)$ \\ \bottomrule
\end{tabular}
\end{table}
\begin{table}[!htb]
\centering
\caption{Region interface identifications $\partial_\alpha{\cal B}_A
  \leftrightarrow \partial_\beta {\cal B}_B$ for the ten-region, $N_R=10$,
  representation of the genus $N_g=0$ 
  manifold, the two-sphere, $S^2$.
  \label{t:TableGenus0Region10Maps}}
\setlength\tabcolsep{1.1em}
\begin{tabular}{@{}llll@{}} \toprule
  \multicolumn{4}{c}{$\partial_\alpha{\cal B}_A 
    \leftrightarrow \partial_\beta {\cal B}_B$} \\ \midrule
  $\partial_{+x}{\cal B}_1 \leftrightarrow \partial_{-x} {\cal B}_3$
  & $\partial_{-x}{\cal B}_1 \leftrightarrow \partial_{+x} {\cal B}_{10}$
  & $\partial_{+y}{\cal B}_1 \leftrightarrow \partial_{-x} {\cal B}_2$
  & $\partial_{-y}{\cal B}_1 \leftrightarrow \partial_{-x} {\cal B}_4$
  \\ [4pt]
  $\partial_{+x}{\cal B}_2 \leftrightarrow \partial_{+y} {\cal B}_5$
  & $\partial_{+y}{\cal B}_2 \leftrightarrow \partial_{+y} {\cal B}_7$
  & $\partial_{-y}{\cal B}_2 \leftrightarrow \partial_{+y} {\cal B}_3$
  & $\partial_{+x}{\cal B}_3 \leftrightarrow \partial_{-x} {\cal B}_5$
  \\ [4pt]
  $\partial_{-y}{\cal B}_3 \leftrightarrow \partial_{+y} {\cal B}_4$
  & $\partial_{+x}{\cal B}_4 \leftrightarrow \partial_{-y} {\cal B}_5$
  & $\partial_{-y}{\cal B}_4 \leftrightarrow \partial_{-y} {\cal B}_9$
  & $\partial_{+x}{\cal B}_5 \leftrightarrow \partial_{-x} {\cal B}_6$
  \\ [4pt]
  $\partial_{+x}{\cal B}_6 \leftrightarrow \partial_{-x} {\cal B}_8$
  & $\partial_{+y}{\cal B}_6 \leftrightarrow \partial_{-x} {\cal B}_7$
  & $\partial_{-y}{\cal B}_6 \leftrightarrow \partial_{-x} {\cal B}_9$
  & $\partial_{+x}{\cal B}_7 \leftrightarrow \partial_{+y} {\cal B}_{10}$
  \\ [4pt]
  $\partial_{-y}{\cal B}_7 \leftrightarrow \partial_{+y} {\cal B}_8$
  & $\partial_{+x}{\cal B}_8 \leftrightarrow \partial_{-x} {\cal B}_{10}$
  & $\partial_{-y}{\cal B}_8 \leftrightarrow \partial_{+y} {\cal B}_9$
  & $\partial_{+x}{\cal B}_9 \leftrightarrow \partial_{-y} {\cal B}_{10}$
  \\ \bottomrule
\end{tabular}
\end{table}
\begin{table}[!htb]
\centering
\caption{Transformation matrices ${\bf C}^{A\alpha}_{B\beta}$ for the
  interface identifications $\partial_\alpha{\cal B}_A \leftrightarrow
  \partial_\beta {\cal B}_B$ in the ten-region, $N_R=10$,
  representation of the genus $N_g=0$ manifold, the two-sphere, $S^2$.
  All transformation matrices ${\bf C}^{A\alpha}_{B\beta}$ are assumed
  to be the identity $\mathbf{I}$, except those specified in this
  table.
  \label{t:TableGenus0Region10MapsA}}
\begin{tabular}{@{}lcc@{\hskip 2.3em}lcc@{}} \toprule
  $\partial_\alpha{\cal B}_A \leftrightarrow \partial_\beta {\cal B}_B$
  &${\mathbf C}^{A\alpha}_{B\beta}$
  &${\mathbf C}^{B\beta}_{A\alpha}$
  &$\partial_\alpha{\cal B}_A \leftrightarrow \partial_\beta {\cal B}_B$
  &${\mathbf C}^{A\alpha}_{B\beta}$
  &${\mathbf C}^{B\beta}_{A\alpha}$ \\ \midrule
  $\partial_{+y}{\cal B}_1 \leftrightarrow \partial_{-x} {\cal B}_2$
  & ${\mathbf R}_{+}$ & ${\mathbf R}_{-}$ 
  & $\partial_{-y}{\cal B}_1 \leftrightarrow \partial_{-x} {\cal B}_4$
  & ${\mathbf R}_{-}$ & ${\mathbf R}_{+}$ \\ [4pt]
  $\partial_{+x}{\cal B}_2 \leftrightarrow \partial_{+y} {\cal B}_5$
  & ${\mathbf R}_{+}$ & ${\mathbf R}_{-}$ 
  & $\partial_{+y}{\cal B}_2 \leftrightarrow \partial_{+y} {\cal B}_7$
  & ${\mathbf R}_{+}^2$ & ${\mathbf R}_{-}^2$ \\ [4pt]
  $\partial_{-y}{\cal B}_4 \leftrightarrow \partial_{-y} {\cal B}_9$
  & ${\mathbf R_+^2}$ & ${\mathbf R_-^2}$
  & $\partial_{+x}{\cal B}_4 \leftrightarrow \partial_{-y} {\cal B}_5$
  & ${\mathbf R_-}$ & ${\mathbf R_+}$ \\ [4pt]
  $\partial_{+y}{\cal B}_6 \leftrightarrow \partial_{-x} {\cal B}_7$
  & ${\mathbf R}_{+}$ & ${\mathbf R}_{-}$
  & $\partial_{-y}{\cal B}_6 \leftrightarrow \partial_{-x} {\cal B}_9$
  & ${\mathbf R}_{-}$ & ${\mathbf R}_{+}$ \\ [4pt]
  $\partial_{+x}{\cal B}_7 \leftrightarrow \partial_{+y} {\cal B}_{10}$
  & ${\mathbf R}_{+}$ & ${\mathbf R}_{-}$
  & $\partial_{+x}{\cal B}_9 \leftrightarrow \partial_{-y} {\cal B}_{10}$
  & ${\mathbf R}_{-}$ & ${\mathbf R}_{+}$ \\ \bottomrule
\end{tabular}
\end{table}

\subsection{Ten-Region, $N_R=10$, Representation of the Genus $N_g=1$ 
  Multicube Manifold}
\label{s:GenusOneRegionTen}

\begin{figure}[!htb]
\centering
\includegraphics[width=0.8\textwidth]{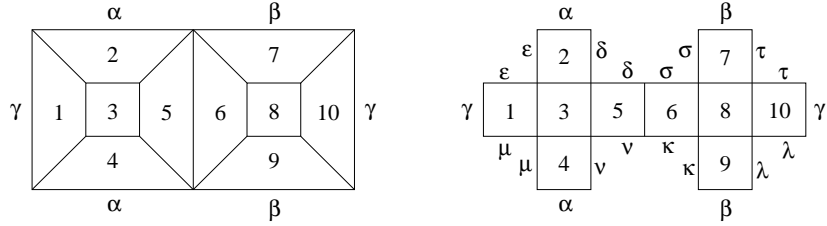}
\caption{Ten-region, $N_R=10$, multicube representation of the genus
  $N_g=1$ manifold, the two-torus, $T^2$.  Left figure shows a
  multicube representation using distorted squares to indicate as many
  interfacial connections as possible.  Greek letters indicate
  identifications between external edges.  Right figure shows the same
  multicube representation using uniformly sized, undistorted squares,
  including their relative locations in the background Euclidean
  space.
  \label{f:Genus1Region10} }
\end{figure}

The locations of the ten square regions used to construct this
representation of $T^2$ are illustrated in
Fig.~\ref{f:Genus1Region10}.  The values of the square-center location
vectors $\vec c_A$ for this configuration are summarized in
Table~\ref{t:TableGenus1Region10Locations}.  The inner edges of the
touching squares in the right side of Fig.~\ref{f:Genus1Region10} are
connected by identity maps.  The identifications of all the edges of
the regions are described in Table~\ref{t:TableGenus1Region10Maps},
and the corresponding transformation matrices are given in
Table~\ref{t:TableGenus1Region10MapsA}.  This ten-region
representation of $T^2$ is a simple generalization of the standard
one-region representation.  The outer edges of the squares in the left
illustration in Fig.~\ref{f:Genus1Region10} are identified with the
opposing outer edges using identity maps, just as in the standard
one-region representation of $T^2$.  This ten-region
representation merely subdivides the single-region representation into
ten regions, as shown in Fig.~\ref{f:Genus1Region10}.
\begin{table}[!htb]
\centering
\caption{Region center locations for the ten-region, $N_R=10$, genus 
  $N_g=1$ multicube manifold.
  \label{t:TableGenus1Region10Locations}}
\setlength\tabcolsep{0.8em}
\begin{tabular}{@{}lllll@{}} \toprule
  \multicolumn{5}{c}{$\vec c_A=(x,y)$} \\ \midrule
  $\vec c_1=(0, 0)$ & $\vec c_2=(L,L)$ & $\vec c_3=(L, 0)$
  & $\vec c_4=(L,-L)$ & $\vec c_5=(2L,0)$ \\ [3pt]
  $\vec c_6=(3L,0)$ & $\vec c_7=(4L,L)$ & $\vec c_8=(4L,0)$ 
  & $\vec c_9 =(4L,-L)$ & $\vec c_{10}=(5L,0)$ \\ \bottomrule
\end{tabular}
\end{table}
\begin{table}[!htb]
\centering
\caption{Region interface identifications $\partial_\alpha{\cal B}_A
  \leftrightarrow \partial_\beta {\cal B}_B$ for the ten-region,
  $N_R=10$, representation of the genus $N_g=1$ 
  manifold, the two-torus, $T^2$.
  \label{t:TableGenus1Region10Maps}}
\setlength\tabcolsep{1.1em}
\begin{tabular}{@{}llll@{}} \toprule
  \multicolumn{4}{c}{$\partial_\alpha{\cal B}_A 
    \leftrightarrow \partial_\beta {\cal B}_B$} \\ \midrule
  $\partial_{+x}{\cal B}_1 \leftrightarrow \partial_{-x} {\cal B}_3$
  & $\partial_{-x}{\cal B}_1 \leftrightarrow \partial_{+x} {\cal B}_{10}$
  & $\partial_{+y}{\cal B}_1 \leftrightarrow \partial_{-x} {\cal B}_2$
  & $\partial_{-y}{\cal B}_1 \leftrightarrow \partial_{-x} {\cal B}_4$
  \\ [4pt]
  $\partial_{+x}{\cal B}_2 \leftrightarrow \partial_{+y} {\cal B}_5$
  & $\partial_{+y}{\cal B}_2 \leftrightarrow \partial_{-y} {\cal B}_4$
  & $\partial_{-y}{\cal B}_2 \leftrightarrow \partial_{+y} {\cal B}_3$
  & $\partial_{+x}{\cal B}_3 \leftrightarrow \partial_{-x} {\cal B}_5$
  \\ [4pt]
  $\partial_{-y}{\cal B}_3 \leftrightarrow \partial_{+y} {\cal B}_4$
  & $\partial_{+x}{\cal B}_4 \leftrightarrow \partial_{-y} {\cal B}_5$
  & $\partial_{+x}{\cal B}_5 \leftrightarrow \partial_{-x} {\cal B}_6$
  & $\partial_{+x}{\cal B}_6 \leftrightarrow \partial_{-x} {\cal B}_8$
  \\ [4pt]
  $\partial_{+y}{\cal B}_6 \leftrightarrow \partial_{-x} {\cal B}_7$
  & $\partial_{-y}{\cal B}_6 \leftrightarrow \partial_{-x} {\cal B}_9$
  & $\partial_{+x}{\cal B}_7 \leftrightarrow \partial_{+y} {\cal B}_{10}$
  & $\partial_{+y}{\cal B}_7 \leftrightarrow \partial_{-y} {\cal B}_9$
  \\ [4pt]
  $\partial_{-y}{\cal B}_7 \leftrightarrow \partial_{+y} {\cal B}_8$
  & $\partial_{+x}{\cal B}_8 \leftrightarrow \partial_{-x} {\cal B}_{10}$
  & $\partial_{-y}{\cal B}_8 \leftrightarrow \partial_{+y} {\cal B}_9$
  & $\partial_{+x}{\cal B}_9 \leftrightarrow \partial_{-y} {\cal B}_{10}$
  \\ \bottomrule
\end{tabular}
\end{table}
\begin{table}[!htb]
\centering
\caption{Transformation matrices ${\bf C}^{A\alpha}_{B\beta}$ for the
  region interface identifications $\partial_\alpha{\cal B}_A
  \leftrightarrow \partial_\beta {\cal B}_B$ in the ten-region,
  $N_R=10$, representation of the genus $N_g=1$ manifold, the
  two-torus, $T^2$.  All transformation matrices ${\bf
    C}^{A\alpha}_{B\beta}$ are assumed to be the identity
  $\mathbf{I}$, except those specified in this table.
  \label{t:TableGenus1Region10MapsA}}
\begin{tabular}{@{}lcc@{\hskip 2.3em}lcc@{}} \toprule
  $\partial_\alpha{\cal B}_A \leftrightarrow \partial_\beta {\cal B}_B$
  &${\mathbf C}^{A\alpha}_{B\beta}$
  &${\mathbf C}^{B\beta}_{A\alpha}$
  &$\partial_\alpha{\cal B}_A \leftrightarrow \partial_\beta {\cal B}_B$
  &${\mathbf C}^{A\alpha}_{B\beta}$
  &${\mathbf C}^{B\beta}_{A\alpha}$
  \\ \midrule
  $\partial_{+y}{\cal B}_1 \leftrightarrow \partial_{-x} {\cal B}_2$
  & ${\mathbf R}_{+}$ & ${\mathbf R}_{-}$ 
  & $\partial_{-y}{\cal B}_1 \leftrightarrow \partial_{-x} {\cal B}_4$
  & ${\mathbf R}_{-}$ & ${\mathbf R}_{+}$ \\ [4pt]
  $\partial_{+x}{\cal B}_2 \leftrightarrow \partial_{+y} {\cal B}_5$
  & ${\mathbf R}_{+}$ & ${\mathbf R}_{-}$ 
  & $\partial_{+x}{\cal B}_4 \leftrightarrow \partial_{-y} {\cal B}_5$
  & ${\mathbf R_-}$ & ${\mathbf R_+}$ \\ [4pt]
  $\partial_{+y}{\cal B}_6 \leftrightarrow \partial_{-x} {\cal B}_7$
  & ${\mathbf R}_{+}$ & ${\mathbf R}_{-}$
  & $\partial_{-y}{\cal B}_6 \leftrightarrow \partial_{-x} {\cal B}_9$
  & ${\mathbf R}_{-}$ & ${\mathbf R}_{+}$ \\ [4pt]
  $\partial_{+x}{\cal B}_7 \leftrightarrow \partial_{+y} {\cal B}_{10}$
  & ${\mathbf R}_{+}$ & ${\mathbf R}_{-}$
  & $\partial_{+x}{\cal B}_9 \leftrightarrow \partial_{-y} {\cal B}_{10}$
  & ${\mathbf R}_{-}$ & ${\mathbf R}_{+}$ \\ \bottomrule
\end{tabular}
\end{table}

\subsection{Eight-Region, $N_R=8$, Representation of the Genus $N_g=1$ 
  Multicube Manifold}
\label{s:GenusOneRegionEight}

\begin{figure}[!htb]
\centering
\includegraphics[width=0.7\textwidth]{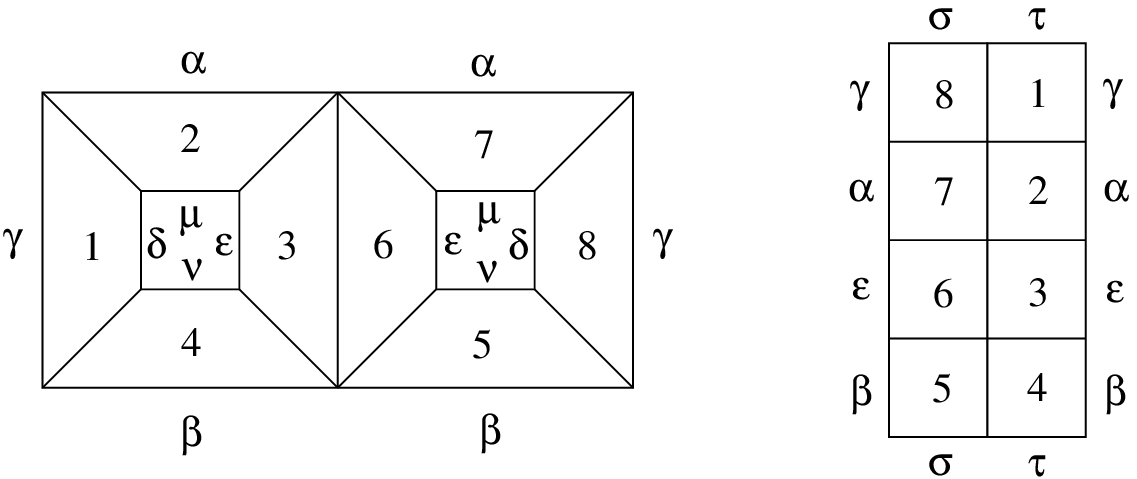}
\caption{Alternative eight-region, $N_R=8$, multicube representation
  of the genus $N_g=1$ manifold, the two-torus, $T^2$.  Left
  illustration shows a multicube representation using distorted
  squares to indicate as many interfacial connections as possible.
  Greek letters indicate identifications between external edges.
  Right illustration shows the same multicube representation using
  uniformly sized, undistorted squares, including their relative
  locations in the background Euclidean space.  The locations of the
  regions in the right illustration were chosen to show explicitly as
  many nearest neighbor identifications as possible.
  \label{f:Genus1Region8} }
\end{figure}

The locations of the eight square regions used to construct this
representation of $T^2$ are illustrated in Fig.~\ref{f:Genus1Region8}.
The values of the square-center location vectors $\vec c_A$ for this
configuration are summarized in
Table~\ref{t:TableGenus1Region8Locations}.  The inner edges of the
touching squares in Fig.~\ref{f:Genus1Region8} are connected by
identity maps.  The identifications of all the edges of the regions
are described in Table~\ref{t:TableGenus1Region8Maps}.  All of the
interface identification maps have transformation matrices
$\mathbf{C}^{A\alpha}_{B\beta}$ that are the identity matrix
$\mathbf{I}$, so they are not included in a table for this case. This
eight-region, $N_R=8$, representation of $T^2$ is constructed by
gluing a handle onto the ten-region representation of $S^2$ described
in \ref{s:GenusZeroRegionTen}.  The two inner regions (3 and 8 in
Fig.~\ref{f:Genus0Region10}) are removed, and the holes created in
this way are connected together to form a handle.  The outer edges in
this eight-region, $N_R=8$, representation of $T^2$ are therefore
connected together, as shown in the left side of
Fig.~\ref{f:Genus1Region8}, using the same identification maps as in
the ten-region representation of $S^2$ shown in the left side of
Fig.~\ref{f:Genus0Region10}.  The inner edges that make up the handle
in this new representation are identified as indicated by the Greek
letters in Fig.~\ref{f:Genus1Region8}.
\begin{table}[!htb]
\centering
\caption{Region center locations for the eight-region, $N_R=8$, genus 
  $N_g=1$ multicube manifold.
  \label{t:TableGenus1Region8Locations}}
\setlength\tabcolsep{0.8em}
\begin{tabular}{@{}llll@{}} \toprule
  \multicolumn{4}{c}{$\vec c_A=(x,y)$} \\ \midrule
  $\vec c_1=(L,2L)$ & $\vec c_2=(L,L)$ & $\vec c_3=(L,0)$  
  & $\vec c_4=(L,-L)$ \\ [3pt]
  $\vec c_5=(0,-L)$ & $\vec c_6=(0,0)$ & $\vec c_7=(0,L)$ 
  & $\vec c_8=(0,2L)$ \\ \bottomrule
\end{tabular}
\end{table}
\begin{table}[!htb]
\centering
\caption{Region interface identifications $\partial_\alpha{\cal B}_A
  \leftrightarrow \partial_\beta {\cal B}_B$ for the eight-region,
  $N_R=8$, representation of the genus $N_g=1$ 
  manifold, the two-torus, $T^2$.
  \label{t:TableGenus1Region8Maps}}
\setlength\tabcolsep{1.1em}
\begin{tabular}{@{}llll@{}} \toprule
  \multicolumn{4}{c}{$\partial_\alpha{\cal B}_A 
    \leftrightarrow \partial_\beta {\cal B}_B$} \\ \midrule
  $\partial_{+x}{\cal B}_1 \leftrightarrow \partial_{-x} {\cal B}_8$
  & $\partial_{-x}{\cal B}_1 \leftrightarrow \partial_{+x} {\cal B}_8$
  & $\partial_{+y}{\cal B}_1 \leftrightarrow \partial_{-y} {\cal B}_4$
  & $\partial_{-y}{\cal B}_1 \leftrightarrow \partial_{+y} {\cal B}_2$
  \\ [4pt]
  $\partial_{+x}{\cal B}_2 \leftrightarrow \partial_{-x} {\cal B}_7$
  & $\partial_{-x}{\cal B}_2 \leftrightarrow \partial_{+x} {\cal B}_7$
  & $\partial_{-y}{\cal B}_2 \leftrightarrow \partial_{+y} {\cal B}_3$
  & $\partial_{+x}{\cal B}_3 \leftrightarrow \partial_{-x} {\cal B}_6$
  \\ [4pt]
  $\partial_{-x}{\cal B}_3 \leftrightarrow \partial_{+x} {\cal B}_6$
  & $\partial_{-y}{\cal B}_3 \leftrightarrow \partial_{+y} {\cal B}_4$
  & $\partial_{+x}{\cal B}_4 \leftrightarrow \partial_{-x} {\cal B}_5$
  & $\partial_{-x}{\cal B}_4 \leftrightarrow \partial_{+x} {\cal B}_5$
  \\ [4pt]
  $\partial_{+y}{\cal B}_5 \leftrightarrow \partial_{-y} {\cal B}_6$
  & $\partial_{-y}{\cal B}_5 \leftrightarrow \partial_{+y} {\cal B}_8$
  & $\partial_{+y}{\cal B}_6 \leftrightarrow \partial_{-y} {\cal B}_7$
  & $\partial_{+y}{\cal B}_7 \leftrightarrow \partial_{-y} {\cal B}_8$
  \\ \bottomrule
\end{tabular}
\end{table}

\subsection{Eight-Region, $N_R=8$, Representation of the Genus $N_g=2$ 
  Multicube Manifold}
\label{s:GenusTwoRegionEight}

\begin{figure}[!htb]
\centering
\includegraphics[width=0.7\textwidth]{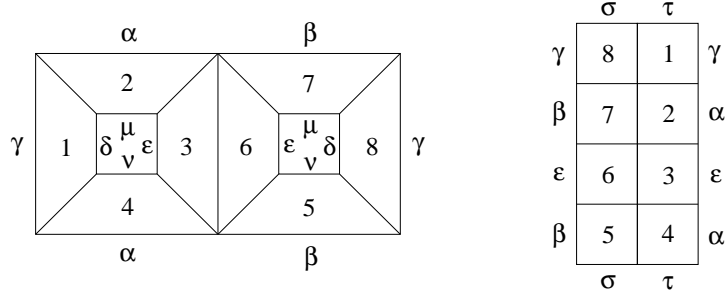}
\caption{Eight-region, $N_R=8$, multicube representation of the genus
  $N_g=2$ manifold, the two-handled sphere.  Left illustration shows a
  multicube representation using distorted squares that are arranged
  to indicate the association of this case with the $N_R=10$
  representation of the $N_g=1$ manifold.  Greek letters indicate
  identifications between external faces.  Right illustration shows
  the same multicube representation using uniformly sized, undistorted
  squares, including their relative locations in the background
  Euclidean space.  The locations of the regions in the right
  illustration were chosen to show explicitly as many nearest neighbor
  identifications as possible.
  \label{f:Genus2Region8} }
\end{figure}

The locations of the eight square regions used to construct this
representation of the genus $N_g=2$ manifold, the two-handled sphere,
are illustrated in Fig.~\ref{f:Genus2Region8}.  The values of the
square-center location vectors $\vec c_A$ for this configuration are
summarized in Table~\ref{t:TableGenus2Locations}.  The inner edges of
the touching squares in Fig.~\ref{f:Genus2Region8} are connected by
identity maps.  The identifications of all the edges of the regions
are described in Table~\ref{t:TableGenus2Maps}, and the corresponding
transformation matrices are given in Table~\ref{t:TableGenus2MapsA}.
This representation of the two-handled sphere is constructed by
starting with the ten-region representation of the two-torus shown in
Fig.~\ref{f:Genus1Region10}, removing the two internal regions (3 and
8 in Fig.~\ref{f:Genus1Region10}), and then connecting together the
holes created in this way to form the second handle.  The outer edges
in this eight-region representation of the genus $N_g=2$ manifold are
therefore connected together, as shown in the left side of
Fig.~\ref{f:Genus2Region8}, using the same identification maps as in
the ten-region representation of $T^2$ shown in the left side of
Fig.~\ref{f:Genus1Region10}.  The inner edges that make up the handle
in this new representation are identified as indicated by the Greek
letters in Fig.~\ref{f:Genus2Region8}.
\begin{table}[!htb]
\centering
\caption{Region center locations for the eight-region, $N_R=8$, genus 
  $N_g=2$ multicube manifold.
  \label{t:TableGenus2Locations}}
\setlength\tabcolsep{0.8em}
\begin{tabular}{@{}llll@{}} \toprule
  \multicolumn{4}{c}{$\vec c_A=(x,y)$} \\ \midrule
  $\vec c_1=(L,2L)$ & $\vec c_2=(L,L)$ & $\vec c_3=(L,0)$  
  & $\vec c_4=(L,-L)$ \\ [3pt] 
  $\vec c_5=(0,-L)$ & $\vec c_6=(0,0)$ & $\vec c_7=(0,L)$ 
  & $\vec c_8=(0,2L)$ \\ \bottomrule
\end{tabular}
\end{table}
\begin{table}[!htb]
\centering
\caption{Region interface identifications $\partial_\alpha{\cal B}_A
  \leftrightarrow \partial_\beta {\cal B}_B$ for the eight-region,
  $N_R=8$, representation of the genus $N_g=2$ 
  manifold, the two-handled sphere.
  \label{t:TableGenus2Maps}}
\setlength\tabcolsep{1.1em}
\begin{tabular}{@{}llll@{}} \toprule
  \multicolumn{4}{c}{$\partial_\alpha{\cal B}_A 
    \leftrightarrow \partial_\beta {\cal B}_B$} \\ \midrule
  $\partial_{+x}{\cal B}_1 \leftrightarrow \partial_{-x} {\cal B}_8$
  & $\partial_{-x}{\cal B}_1 \leftrightarrow \partial_{+x} {\cal B}_8$
  & $\partial_{+y}{\cal B}_1 \leftrightarrow \partial_{-y} {\cal B}_4$
  & $\partial_{-y}{\cal B}_1 \leftrightarrow \partial_{+y} {\cal B}_2$
  \\ [4pt]
  $\partial_{-x}{\cal B}_2 \leftrightarrow \partial_{+x} {\cal B}_7$
  & $\partial_{+x}{\cal B}_2 \leftrightarrow \partial_{+x} {\cal B}_4$
  & $\partial_{-y}{\cal B}_2 \leftrightarrow \partial_{+y} {\cal B}_3$
  & $\partial_{+x}{\cal B}_3 \leftrightarrow \partial_{-x} {\cal B}_6$
  \\ [4pt]
  $\partial_{-x}{\cal B}_3 \leftrightarrow \partial_{+x} {\cal B}_6$
  & $\partial_{-y}{\cal B}_3 \leftrightarrow \partial_{+y} {\cal B}_4$
  & $\partial_{-x}{\cal B}_4 \leftrightarrow \partial_{+x} {\cal B}_5$
  & $\partial_{-x}{\cal B}_5 \leftrightarrow \partial_{-x} {\cal B}_7$
  \\ [4pt]
  $\partial_{+y}{\cal B}_5 \leftrightarrow \partial_{-y} {\cal B}_6$
  & $\partial_{-y}{\cal B}_5 \leftrightarrow \partial_{+y} {\cal B}_8$
  & $\partial_{+y}{\cal B}_6 \leftrightarrow \partial_{-y} {\cal B}_7$
  & $\partial_{+y}{\cal B}_7 \leftrightarrow \partial_{-y} {\cal B}_8$
  \\ \bottomrule
\end{tabular}
\end{table}
\begin{table}[!htb]
\centering
\caption{Transformation matrices ${\bf C}^{A\alpha}_{B\beta}$ for the
  region interface identifications $\partial_\alpha{\cal B}_A
  \leftrightarrow \partial_\beta {\cal B}_B$ in the eight-region,
  $N_R=8$, representation of the genus $N_g=2$ manifold, the
  two-handled sphere. All transformation matrices ${\bf
    C}^{A\alpha}_{B\beta}$ are assumed to be the identity
  $\mathbf{I}$, except those specified in this table.
  \label{t:TableGenus2MapsA}}
\begin{tabular}{@{}lcc@{}} \toprule
  $\partial_\alpha{\cal B}_A \leftrightarrow \partial_\beta {\cal B}_B$
  &${\mathbf C}^{A\alpha}_{B\beta}$
  &${\mathbf C}^{B\beta}_{A\alpha}$
  \\ \midrule
  $\partial_{+x}{\cal B}_2 \leftrightarrow \partial_{+x} {\cal B}_4$
  & ${\mathbf R}_-^2$ & ${\mathbf R}_+^2$ \\ [4pt]
  $\partial_{-x}{\cal B}_5 \leftrightarrow \partial_{-x} {\cal B}_7$
  & ${\mathbf R}_+^2$ & ${\mathbf R}_-^2$ \\ \bottomrule
\end{tabular}
\end{table}

\subsection{Ten-Region, $N_R=10$, Representation of the Genus $N_g=2$ 
  Multicube Manifold}
\label{s:GenusTwoRegionTen}

\begin{figure}[!htb]
\centering
\includegraphics[width=0.8\textwidth]{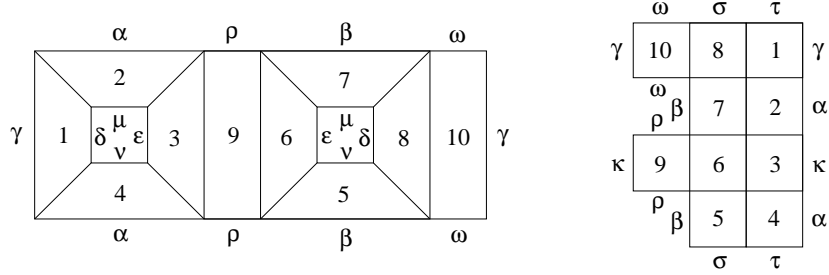}
\caption{Ten-region, $N_R=10$, multicube representation of the genus
  $N_g=2$ manifold, the two-handled sphere.  Left illustration shows a
  multicube representation using distorted squares that are arranged
  to indicate the association of this case with the $N_R=10$
  representation of the $N_g=1$ manifold.  Greek letters indicate
  identifications between external faces.  Right illustration shows
  the same multicube representation using uniformly sized, undistorted
  squares, including their relative locations in the background
  Euclidean space.  The locations of the regions in the right
  illustration were chosen to show explicitly as many nearest neighbor
  identifications as possible.
  \label{f:Genus2Region10} }
\end{figure}

The locations of the ten square regions used to construct this
representation of the genus $N_g=2$ manifold, the two-handled sphere,
are illustrated in Fig.~\ref{f:Genus2Region10}.  The values of the
square-center location vectors $\vec c_A$ for this configuration are
summarized in Table~\ref{t:TableGenus2Region10Locations}.  The inner
edges of the touching squares in Fig.~\ref{f:Genus2Region10} are
connected by identity maps.  The identifications of all the edges of
the regions are described in Table~\ref{t:TableGenus2Region10Maps},
and the corresponding transformation matrices are given in
Table~\ref{t:TableGenus2Region10MapsA}.  This representation of the
two-handled sphere is constructed by starting with the eight-region
representation shown in Fig.~\ref{f:Genus2Region8} and adding
additional squares to separate more distinctly the ends of the second
handle on the torus.  The outer edges in this ten-region
representation of the genus $N_g=2$ manifold are therefore connected
together as shown in Fig.~\ref{f:Genus2Region10}.  This representation
has the advantage that it reduces the maximum number of squares
meeting at a single vertex from eight to six.  The reference metric in
this case therefore requires less distortion of the flat metric pieces
that go into its construction.
\begin{table}[!htb]
\centering
\caption{Region center locations for the ten-region, $N_R=10$, genus
  $N_g=2$ multicube manifold.
  \label{t:TableGenus2Region10Locations}}
\setlength\tabcolsep{0.8em}
\begin{tabular}{@{}lllll@{}} \toprule
  \multicolumn{5}{c}{$\vec c_A=(x,y)$} \\ \midrule
  $\vec c_1=(L,2L)$ & $\vec c_2=(L,L)$ & $\vec c_3=(L,0)$  
  & $\vec c_4=(L,-L)$ & $\vec c_5=(0,-L)$ \\ [3pt]
  $\vec c_6=(0,0)$ & $\vec c_7=(0,L)$ & $\vec c_8=(0,2L)$ 
  & $\vec c_9=(-L,0)$ & $\vec c_{10}=(-L,2L)$ \\ \bottomrule
\end{tabular}
\end{table}
\begin{table}[!htb]
\centering
\caption{Region interface identifications $\partial_\alpha{\cal B}_A
  \leftrightarrow \partial_\beta {\cal B}_B$ for the ten-region,
  $N_R=10$, representation of the genus $N_g=2$ 
  manifold, the two-handled sphere.
  \label{t:TableGenus2Region10Maps}}
\setlength\tabcolsep{1.1em}
\begin{tabular}{@{}llll@{}} \toprule
  \multicolumn{4}{c}{$\partial_\alpha{\cal B}_A 
    \leftrightarrow \partial_\beta {\cal B}_B$} \\ \midrule
  $\partial_{+x}{\cal B}_1 \leftrightarrow \partial_{-x} {\cal B}_{10}$
  & $\partial_{-x}{\cal B}_1 \leftrightarrow \partial_{+x} {\cal B}_8$
  & $\partial_{+y}{\cal B}_1 \leftrightarrow \partial_{-y} {\cal B}_4$
  & $\partial_{-y}{\cal B}_1 \leftrightarrow \partial_{+y} {\cal B}_2$
  \\ [4pt]
  $\partial_{-x}{\cal B}_2 \leftrightarrow \partial_{+x} {\cal B}_7$
  & $\partial_{+x}{\cal B}_2 \leftrightarrow \partial_{+x} {\cal B}_4$
  & $\partial_{-y}{\cal B}_2 \leftrightarrow \partial_{+y} {\cal B}_3$
  & $\partial_{+x}{\cal B}_3 \leftrightarrow \partial_{-x} {\cal B}_9$
  \\ [4pt]
  $\partial_{-x}{\cal B}_3 \leftrightarrow \partial_{+x} {\cal B}_6$
  & $\partial_{-y}{\cal B}_3 \leftrightarrow \partial_{+y} {\cal B}_4$
  & $\partial_{-x}{\cal B}_4 \leftrightarrow \partial_{+x} {\cal B}_5$
  & $\partial_{-x}{\cal B}_5 \leftrightarrow \partial_{-x} {\cal B}_7$
  \\ [4pt]
  $\partial_{+y}{\cal B}_5 \leftrightarrow \partial_{-y} {\cal B}_6$
  & $\partial_{-y}{\cal B}_5 \leftrightarrow \partial_{+y} {\cal B}_8$
  & $\partial_{-x}{\cal B}_6 \leftrightarrow \partial_{+x} {\cal B}_9$
  & $\partial_{+y}{\cal B}_6 \leftrightarrow \partial_{-y} {\cal B}_7$
  \\ [4pt]
  $\partial_{+y}{\cal B}_7 \leftrightarrow \partial_{-y} {\cal B}_8$
  & $\partial_{-x}{\cal B}_{8} \leftrightarrow \partial_{+x} {\cal B}_{10}$
  & $\partial_{+y}{\cal B}_9 \leftrightarrow \partial_{-y} {\cal B}_9$
  & $\partial_{+y}{\cal B}_{10} \leftrightarrow \partial_{-y} {\cal B}_{10}$
  \\ \bottomrule
\end{tabular}
\end{table}
\begin{table}[!htb]
\centering
\caption{Transformation matrices ${\bf C}^{A\alpha}_{B\beta}$ for the
  region interface identifications $\partial_\alpha{\cal B}_A
  \leftrightarrow \partial_\beta {\cal B}_B$ in the ten-region,
  $N_R=10$, representation of the genus $N_g=2$ manifold, the
  two-handled sphere. All transformation matrices ${\bf
    C}^{A\alpha}_{B\beta}$ are assumed to be the identity
  $\mathbf{I}$, except those specified in this table.
  \label{t:TableGenus2Region10MapsA}}
\begin{tabular}{@{}lcc@{}} \toprule
  $\partial_\alpha{\cal B}_A \leftrightarrow \partial_\beta {\cal B}_B$
  &${\mathbf C}^{A\alpha}_{B\beta}$
  &${\mathbf C}^{B\beta}_{A\alpha}$
  \\ \midrule
  $\partial_{+x}{\cal B}_2 \leftrightarrow \partial_{+x} {\cal B}_4$
  & ${\mathbf R}_-^2$ & ${\mathbf R}_+^2$ \\ [4pt]
  $\partial_{-x}{\cal B}_5 \leftrightarrow \partial_{-x} {\cal B}_7$
  & ${\mathbf R}_+^2$ & ${\mathbf R}_-^2$ \\ \bottomrule
\end{tabular}
\end{table}

\subsection{Representations of Genus $N_g\geq 3$ Multicube Manifolds 
  Using $10(N_g-1)$ Regions}
\label{s:GenusLargerThanTwo}

\begin{figure}[!htb]
\centering
\includegraphics[width=2.2in]{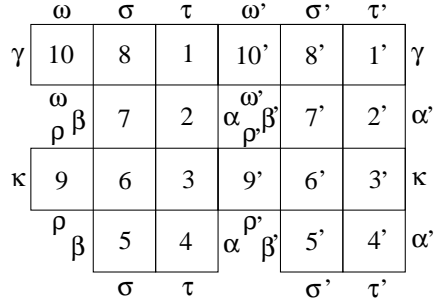}
\caption{Twenty-region, $N_R=20$, multicube representation of the
  genus $N_g=3$ manifold, the three-handled sphere.  The touching
  edges of adjacent squares in this figure are identified, while Greek
  letters indicate identifications between external edges.  This
  representation of the genus $N_g=3$ manifold was constructed by
  connecting together two copies of the $N_g=2$ manifold illustrated
  in Fig.~\ref{f:Genus2Region10}.
  \label{f:Genus3Region20} }
\end{figure}

Multicube representations of two-dimensional manifolds with genera
$N_g\geq 3$ can be constructed by gluing together copies of the genus
$N_g=2$ multicube manifold depicted in Fig.~\ref{f:Genus2Region10}.
This is done by breaking the interface identifications denoted
$\gamma$ and $\kappa$ in Fig.~\ref{f:Genus2Region10} and then
attaching in their place additional copies of the same multicube
structure, as shown in Fig.~\ref{f:Genus3Region20} for the genus
$N_g=3$ case.  Each copy of the genus $N_g=2$ multicube structure
added in this way increases the genus of the resulting manifold by
one.  The addition of one copy, as shown in
Fig.~\ref{f:Genus3Region20}, produces a multicube manifold of genus
$N_g=3$.  The values of the square-center location vectors $\vec c_A$
for this genus $N_g=3$ case are summarized in
Table~\ref{t:TableGenus3Region20Locations}.  The inner edges of the
touching squares in Fig.~\ref{f:Genus3Region20} are connected by
identity maps.  The identifications of all the edges of the twenty
square regions are described in Table~\ref{t:TableGenus3Region20Maps},
and the corresponding transformation matrices are given in
Table~\ref{t:TableGenus3Region20MapsA}. 

\begin{table}[!htb]
\centering
\caption{Region center locations for the twenty-region, $N_R=20$,
  genus $N_g=3$ multicube manifold, the three-handled sphere.
  \label{t:TableGenus3Region20Locations}}
\setlength\tabcolsep{0.8em}
\begin{tabular}{@{}lllll@{}} \toprule
  \multicolumn{5}{c}{$\vec c_A=(x,y)$} \\ \midrule
  $\vec c_1=(L,2L)$ & $\vec c_2=(L,L)$ 
  & $\vec c_3=(L,0)$ & $\vec c_4=(L,-L)$
  & $\vec c_5=(0,-L)$ \\ [3pt]
  $\vec c_6=(0,0)$ & $\vec c_7 =(0,L)$ 
  & $\vec c_8=(0,2L)$ & $\vec c_9=(-L,0)$ 
  & $\vec c_{10}=(-L,2L)$ \\ [3pt]
  $\vec c_{1'}=(4L,2L)$ & $\vec c_{2'}=(4L,L)$
  & $\vec c_{3'}=(4L,0)$ & $\vec c_{4'}=(4L,-L)$ 
  & $\vec c_{5'}=(3L,-L)$ \\ [3pt]
  $\vec c_{6'}=(3L,0)$ & $\vec c_{7'}=(3L,L)$ 
  & $\vec c_{8'}=(3L,2L)$ & $\vec c_{9'}=(2L,0)$ 
  & $\vec c_{10'}=(2L,2L)$ 
  \\ \bottomrule
\end{tabular}
\end{table}
\begin{table}[!htb]
\centering
\caption{Region interface identifications, $\partial_\alpha{\cal B}_A
  \leftrightarrow \partial_\beta {\cal B}_B$, for the twenty-region,
  $N_R=20$, representation of the genus $N_g=3$ 
  manifold, the three-handled sphere.
  \label{t:TableGenus3Region20Maps}}
\setlength\tabcolsep{1.1em}
\begin{tabular}{@{}llll@{}} \toprule
  \multicolumn{4}{c}{$\partial_\alpha{\cal B}_A 
    \leftrightarrow \partial_\beta {\cal B}_B$} \\ \midrule
  $\partial_{+x}{\cal B}_1 \leftrightarrow \partial_{-x} {\cal B}_{10'}$
  & $\partial_{-x}{\cal B}_1 \leftrightarrow \partial_{+x} {\cal B}_8$
  & $\partial_{+y}{\cal B}_1 \leftrightarrow \partial_{-y} {\cal B}_4$
  & $\partial_{-y}{\cal B}_1 \leftrightarrow \partial_{+y} {\cal B}_2$
  \\ [4pt]
  $\partial_{-x}{\cal B}_2 \leftrightarrow \partial_{+x} {\cal B}_7$
  & $\partial_{+x}{\cal B}_2 \leftrightarrow \partial_{+x} {\cal B}_4$
  & $\partial_{-y}{\cal B}_2 \leftrightarrow \partial_{+y} {\cal B}_3$
  & $\partial_{+x}{\cal B}_3 \leftrightarrow \partial_{-x} {\cal B}_{9'}$
  \\ [4pt]
  $\partial_{-x}{\cal B}_3 \leftrightarrow \partial_{+x} {\cal B}_6$
  & $\partial_{-y}{\cal B}_3 \leftrightarrow \partial_{+y} {\cal B}_4$
  & $\partial_{-x}{\cal B}_4 \leftrightarrow \partial_{+x} {\cal B}_5$
  & $\partial_{-x}{\cal B}_5 \leftrightarrow \partial_{-x} {\cal B}_7$
  \\ [4pt]
  $\partial_{+y}{\cal B}_5 \leftrightarrow \partial_{-y} {\cal B}_6$
  & $\partial_{-y}{\cal B}_5 \leftrightarrow \partial_{+y} {\cal B}_8$
  & $\partial_{-x}{\cal B}_6 \leftrightarrow \partial_{+x} {\cal B}_9$
  & $\partial_{+y}{\cal B}_6 \leftrightarrow \partial_{-y} {\cal B}_7$
  \\ [4pt]
  $\partial_{+y}{\cal B}_7 \leftrightarrow \partial_{-y} {\cal B}_8$
  & $\partial_{-x}{\cal B}_{8} \leftrightarrow \partial_{+x} {\cal B}_{10}$
  & $\partial_{+y}{\cal B}_9 \leftrightarrow \partial_{-y} {\cal B}_9$
  & $\partial_{+y}{\cal B}_{10} \leftrightarrow \partial_{-y} {\cal B}_{10}$
  \\ [4pt]
  $\partial_{+x}{\cal B}_{1'} \leftrightarrow \partial_{-x} {\cal B}_{10}$
  & $\partial_{-x}{\cal B}_{1'} \leftrightarrow \partial_{+x} {\cal B}_{8'}$
  & $\partial_{+y}{\cal B}_{1'} \leftrightarrow \partial_{-y} {\cal B}_{4'}$
  & $\partial_{-y}{\cal B}_{1'} \leftrightarrow \partial_{+y} {\cal B}_{2'}$
  \\ [4pt]
  $\partial_{-x}{\cal B}_{2'} \leftrightarrow \partial_{+x} {\cal B}_{7'}$
  & $\partial_{+x}{\cal B}_{2'} \leftrightarrow \partial_{+x} {\cal B}_{4'}$
  & $\partial_{-y}{\cal B}_{2'} \leftrightarrow \partial_{+y} {\cal B}_{3'}$
  & $\partial_{+x}{\cal B}_{3'} \leftrightarrow \partial_{-x} {\cal B}_{9}$
  \\ [4pt]
  $\partial_{-x}{\cal B}_{3'} \leftrightarrow \partial_{+x} {\cal B}_{6'}$
  & $\partial_{-y}{\cal B}_{3'} \leftrightarrow \partial_{+y} {\cal B}_{4'}$
  & $\partial_{-x}{\cal B}_{4'} \leftrightarrow \partial_{+x} {\cal B}_{5'}$
  & $\partial_{-x}{\cal B}_{5'} \leftrightarrow \partial_{-x} {\cal B}_{7'}$
  \\ [4pt]
  $\partial_{+y}{\cal B}_{5'} \leftrightarrow \partial_{-y} {\cal B}_{6'}$
  & $\partial_{-y}{\cal B}_{5'} \leftrightarrow \partial_{+y} {\cal B}_{8'}$
  & $\partial_{-x}{\cal B}_{6'} \leftrightarrow \partial_{+x} {\cal B}_{9'}$
  & $\partial_{+y}{\cal B}_{6'} \leftrightarrow \partial_{-y} {\cal B}_{7'}$
  \\ [4pt]
  $\partial_{+y}{\cal B}_{7'} \leftrightarrow \partial_{-y} {\cal B}_{8'}$
  & $\partial_{-x}{\cal B}_{8'} \leftrightarrow \partial_{+x} {\cal B}_{10'}$
  & $\partial_{+y}{\cal B}_{9'} \leftrightarrow \partial_{-y} {\cal B}_{9'}$
  & $\partial_{+y}{\cal B}_{10'} \leftrightarrow \partial_{-y} {\cal B}_{10'}$
  \\ \bottomrule
\end{tabular}
\end{table}
\begin{table}[!htb]
\centering
\caption{Transformation matrices ${\bf C}^{A\alpha}_{B\beta}$ for the
  region interface identifications $\partial_\alpha{\cal B}_A
  \leftrightarrow \partial_\beta {\cal B}_B$ in the twenty-region,
  $N_R=20$, representation of the genus $N_g=3$ manifold, the
  three-handled sphere. All transformation matrices ${\bf
    C}^{A\alpha}_{B\beta}$ are assumed to be the identity
  $\mathbf{I}$, except those specified in this table.
  \label{t:TableGenus3Region20MapsA}}
\begin{tabular}{@{}lcc@{\hskip 2.3em}lcc@{}} \toprule
  $\partial_\alpha{\cal B}_A \leftrightarrow \partial_\beta {\cal B}_B$
  &${\mathbf C}^{A\alpha}_{B\beta}$
  &${\mathbf C}^{B\beta}_{A\alpha}$
  &$\partial_\alpha{\cal B}_A \leftrightarrow \partial_\beta {\cal B}_B$
  &${\mathbf C}^{A\alpha}_{B\beta}$
  &${\mathbf C}^{B\beta}_{A\alpha}$
  \\ \midrule
  $\partial_{+x}{\cal B}_2 \leftrightarrow \partial_{+x} {\cal B}_4$
  & ${\mathbf R}_-^2$ & ${\mathbf R}_+^2$
  & $\partial_{-x}{\cal B}_5 \leftrightarrow \partial_{-x} {\cal B}_7$
  & ${\mathbf R}_+^2$ & ${\mathbf R}_-^2$ \\ [4pt]
  $\partial_{+x}{\cal B}_{2'} \leftrightarrow \partial_{+x} {\cal B}_{4'}$
  & ${\mathbf R}_-^2$ & ${\mathbf R}_+^2$
  & $\partial_{-x}{\cal B}_{5'} \leftrightarrow \partial_{-x} {\cal B}_{7'}$
  & ${\mathbf R}_+^2$ & ${\mathbf R}_-^2$ \\ \bottomrule
\end{tabular}
\end{table}

\clearpage
\bibliographystyle{model1-num-names}
\bibliography{../References/References}

\end{document}